\documentclass{emulateapj}

\newcommand{\ldl}{$\lambda/{\Delta}{\lambda}$}
\newcommand{\teff}{T$_{eff}$}

\newcommand{\meth}{CH$_4$}
\newcommand{\water}{H$_2$O}

\slugcomment{Submitted to ApJ, 30 March 2006; Accepted, 22 May 2006}

\shorttitle{T Dwarf Binaries}
\shortauthors{Burgasser et al.}

\begin{document}

\title{Hubble Space Telescope NICMOS Observations of T Dwarfs: Brown Dwarf Multiplicity
and New Probes of the L/T Transition}

\author{Adam J.\ Burgasser\altaffilmark{1},
J.\ Davy Kirkpatrick\altaffilmark{2},
Kelle L.\ Cruz\altaffilmark{3,4},
I.\ Neill Reid\altaffilmark{5},
Sandy K.\ Leggett\altaffilmark{6},
James Liebert\altaffilmark{7},
Adam Burrows\altaffilmark{7},
and
Michael E.\ Brown\altaffilmark{8}}

\altaffiltext{1}{Massachusetts Institute of Technology, Kavli Institute for Astrophysics and Space Research,
Building 37, Room 664B, 77 Massachusetts Avenue, Cambridge, MA 02139; ajb@mit.edu}
\altaffiltext{2}{Infrared Processing and Analysis Center, M/S
100-22, California Institute of Technology, Pasadena, CA 91125}
\altaffiltext{3}{Department of Astrophysics, 
American Museum of Natural History, Central Park West at 79th Street, New York, NY 10024}
\altaffiltext{4}{NSF Astronomy and Astrophysics Postdoctoral Fellow}
\altaffiltext{5}{Space Telescope Science Institute, 3700 San
Martin Drive, Baltimore, MD 21218}
\altaffiltext{6}{Joint Astronomy Centre, 660 North A'ohoku Place, Hilo, HI 96720}
\altaffiltext{7}{Steward Observatory, University of Arizona, 933 North Cherry Avenue, Tucson, AZ 85721}
\altaffiltext{8}{Division of Geological and
Planetary Sciences, M/S 105-21, California Institute of
Technology, Pasadena, California 91125}

\begin{abstract}
We present the results of a Hubble Space Telescope NICMOS imaging survey of 22 T-type field brown dwarfs.  Five are resolved as binary
systems
with angular separations of 0$\farcs$05--0$\farcs$35, and   
companionship is established on the basis of component
F110W-F170M colors (indicative of CH$_4$ absorption) and low probabilities of background contamination.  Prior ground-based observations show
2MASS 1553+1532AB to be a common proper motion binary.
The properties of these systems -- low multiplicity
fraction (11$^{+7}_{-3}$\% resolved, as
corrected for sample selection baises),
close projected separations ($\rho$ = 1.8--5.0~AU) and
near-unity mass ratios  ---
are consistent with previous results for field brown dwarf binaries.
Three of the binaries,
2MASS 0518-2828AB, SDSS 0423-0414AB and SDSS 1021-0304AB, 
have components that span the poorly-understood
transition between L dwarfs and T dwarfs.
Spectral decomposition analysis of
SDSS 1021-0304AB reveals a peculiar flux reversal in this system, 
with a T5 secondary that is $\sim$30\% brighter at 1.05
and 1.27~$\micron$ than the T1 primary.
This system, 2MASS 0518-2828AB and SDSS 1534+1615 all
demonstrate that the $J$-band brightening 
observed between late-type L to mid-type T dwarfs
is an intrinsic feature
of this spectral transition, albeit less pronounced than previously surmised.
We also find that the resolved binary fraction 
of L7 to T3.5 dwarfs is twice that
of other L and T dwarfs, an anomaly that
can be explained by a relatively rapid evolution of brown dwarfs
through the L/T transition, perhaps
driven by dynamic (nonequilibrium) depletion
of photospheric condensates.
\end{abstract}


\section{Introduction}

Multiple star systems are of fundamental importance in the study of
stellar populations, and by inference much of galactic and
extragalactic astrophysics.
These systems remain the predominant outlet for the direct measurement
of individual stellar masses, either through the detection of orbital
motion or microlensing techniques 
\citep{an02}.
Eclipsing binaries also enable measurement of stellar radii.
The properties of and interactions between the components of multiple star
systems are fundamental to the phenomena
of cataclysmic variables, X-ray binaries, Type Ia supernovae, planetary
nebulae and several classes of peculiar stars.
Indeed, the creation of multiple systems is
inherent in the star formation process itself.  Measurement of
multiplicity statistics ---  the binary fraction, mass ratio
distribution and separation distribution --- provide key
empirical constraints on star formation theory.
The formation and character of planet-forming debris disks around young stars
can be modulated by the presence of companions.
Finally, coeval binary systems provide a unique
control environment for studying the detailed
physical properties of individual stars, yielding insight
on the general characteristics of a stellar class.

Multiple systems have been particularly useful in the study of
brown dwarfs, stars with insufficient mass to sustain core hydrogen
fusion \citep{kum62,hay63}. 
Indeed, many of the first brown dwarfs to be identified
are members of nearby multiple systems \citep{bec88,nak95,opp95,reb98}.  
Over the past few years,
high resolution imaging and spectroscopic 
surveys of very low mass (VLM; M $\lesssim$ 0.1M$_{\sun}$) 
stars and brown dwarfs
in the field and in young stellar clusters
have revealed roughly 75 binaries
(cf.\ Burgasser et al.\ 2006b\footnote{A current list of known VLM binaries is maintained by N.\ Siegler at the Very Low Mass Binaries Archive, \url{http://paperclip.as.arizona.edu/$\sim$nsiegler/VLM\_binaries/}.}), 
with intriguing results.
The resolved VLM 
binary fraction (the frequency of binary systems
in a given sample of stars) is $\sim$10-20\%, 
significantly lower than the binary fractions of
solar-type stellar systems ($\sim$65\%; e.g.\ Abt \& Levy 1976;
Duquennoy \& Mayor 1991) and early-type M stars
($\sim$30--40\%; e.g.\ Fischer \& Marcy 1992;
Reid \& Gizis 1997; Delfosse et al.\ 2004), indicating a decline in 
the binary fraction with later spectral types \citep{bou06}.
The resolved binary fraction is likely a lower limit
to the true binary fraction due to the existence of unresolved, 
closely separated systems \citep{max05}.  This possibility is
an important consideration for low-mass systems, for
while the distribution of separations
of F- through M-type stellar pairs is quite broad, ranging
over 0.1 AU to 0.1 pc, $>$90\% of all known VLM binaries 
have projected separations $<$20 AU \citep{me06ppv},
with maximum separations scaling
with total system mass \citep[however, see Luhman 2004 and Billeres et al.\ 2005]{clo03,me03hst}.
The mass ratio distribution of resolved VLM binaries is also distinct,
peaking sharply at $q \equiv {\rm M}_2/{\rm M}_1 \approx 1$
\citep{rei01,bou03,me03hst}, in contrast to the relatively flat
mass ratio distributions of stellar systems (e.g., Mazeh et al.\ 1992).
These properties have led researchers to suggest that VLM stars and brown dwarfs may
form via a different mechanism than stars (e.g., Bate, Bonnell \& Bromm 2002),
although this idea remains controversial (e.g., Luhman 2004).
More concretely, astrometric and spectroscopic followup of VLM binaries
have provided the first brown dwarf mass measurements
\citep{bas99,lan01,bou04,bra04,zap04,sta06},
important empirical constraints for theoretical evolutionary
models.

In order to constrain the binary properties of brown dwarfs
in greater detail,
to identify new systems useful for mass measurements, and to search
for very low luminosity brown dwarf companions,
we have conducted a high resolution imaging survey of 22 T dwarfs
using NICMOS
on the {\em Hubble Space Telescope} (hereafter, {\em HST}).
T dwarfs are the lowest luminosity (L $\lesssim$ 3$\times$10$^{-5}$ L$_{\sun}$)
and coldest ({\teff} $\lesssim$ 1400~K; Golimowski et al.\ 2004)
brown dwarfs currently known.  They are distinguished by the presence of
strong {\water} and {\meth} absorption bands in their near infrared
spectra \citep{me02a,geb02} and the absence of photospheric condensates
that dominate warmer L dwarf atmospheres \citep{mar96,tsu96,all01}.
We have identified five binaries in our sample,
of which three have well-resolved components allowing
detailed characterization of their empirical properties.

Observations are described in $\S$~2, including the sample
composition, observing strategy and data reduction.
In $\S$~3 we present photometric results, including
color/spectral type, photometric conversion
and bolometric correction relations.  In $\S$~4 we describe point
spread function (PSF) fits to
our resolved sources, and determine sensitivity limits for putative
faint companions.  Detailed analysis of 
individual systems is given in $\S$~5.
In $\S$~6 we provide an updated
assessment of the multiplicity properties of field brown dwarfs,
including the overall binary fraction, separation distribution and mass
ratio distribution.  In $\S$~7 we examine what currently known 
brown dwarf binaries
reveal about the poorly understood transition between L dwarfs and T dwarfs.
Results are summarized in $\S$~8.

\section{Observations\label{sec:observations}}

\subsection{T Dwarf Targets}

Observations presented here incorporate data from two {\em HST} programs, 
GO-9833 and GO-10247 conducted during Cycles 12 and 13, respectively.
The first program targeted 22 T dwarfs identified in the 
Sloan Digitial Sky Survey \citep[hereafter SDSS]{yor00}
and the Two Micron All Sky Survey \citep[hereafter 2MASS]{skr06},
spanning the full range of T spectral types (T0 to T8),
including the lowest luminosity brown dwarf so far identified, 
2MASS 0415-0935\footnote{We use abbreviated notation for sources 
in our observed sample throughout
the text; e.g., 2MASS hhmm$\pm$ddmm, where the suffix is
the J2000 sexigesimal
Right Ascension (hours and minutes) and declination (degrees and minutes).
Full source names and coordinates are provided in Table~1.}
\citep{me02a,vrb04,gol04}.  Twelve of these sources have
measured parallaxes from \citet{dah02,tin03}; and \citet{vrb04};
and 17 have proper motion measurements (although several 
did not at the time {\em HST} images were obtained; see below).
Program GO-10247 targeted the peculiar T1 dwarf
2MASS 0518-2828 \citep{cru04}, a
source suspected of being multiple due to its peculiar near infrared
spectrum.  A compilation of the observed properties of all of the
sources is provided in Table~1.

\subsection{Imaging and Data Reduction}

Table 2 provides a log of our {\em HST} observations.
Each target was imaged over one orbit in the three filters
F090M (except SDSS 1254-0122), F110W and F170M using 
the highest-resolution  camera NIC1 (pixel scale 0$\farcs$043,
field of view 11$\arcsec$$\times$11$\arcsec$). 
2MASS 0518-2828 was also observed with the F145M and F160W
filters. The F110W and F170M filters 
sample the peak spectral flux
of T dwarfs (around 1.2 $\micron$) and the 
1.6 $\micron$ {\meth} band, respectively, as illustrated in Figure~1.  
As the near infrared {\meth} bands are primary classification
diagnostics for T dwarfs \citep{me06a}, 
F110W-F170M color can provide a rough estimate of
spectral type (cf.~$\S$~3.2) as well as a discriminant
for bona-fide, low-temperature brown dwarf companions. 
The F090M filter samples the red wing of the pressure-broadened
0.77~$\micron$ \ion{K}{1} doublet \citep{bur00,all03,bur03}, and provides
an additional discriminant against background sources.

All data were acquired in MULTIACCUM mode.  Multiple
exposures in the F110W and F170M filters were obtained
in a spiral dither pattern with 
steps of 1$\farcs$3 ($\sim$ 30 NIC1 pixels).
Total integration times in these two filters
ranged over 791-912 s and 1519-1600 s, respectively, for the majority
of our sample.  Exceptions include
2MASS 0348-6022, which was observed for a longer period due to its location in
the {\em HST} continuous viewing zone; and 2MASS 0518-2828, where
shorter exposures were taken to allow observations in 
five filters over one orbit.
Short (48-88 s), single F090M exposures 
were obtained for the GO-9833 targets.
Multiple F090M, F145M and F160W exposures were obtained for 2MASS 0518-2828
using the same dither pattern
as the F110W and F170M observations.

Several of our targets were not well centered on the NIC1 camera
due largely to their uncertain or unknown proper motions at the time of the observations.
The most extreme case is
that of 2MASS 0727+1710.  At the time of the {\em HST} observations
the proper motion of 2MASS 0727+1710 had not been measured, so the
telescope was pointed at the 1997.83 epoch position as measured by 2MASS.  
Unfortunately, this source has one of the largest proper motions in 
our sample, 1$\farcs$297$\pm$0$\farcs$005 \citep{vrb04},
and the resulting 8$\farcs$3 motion between the 2MASS and {\em HST} imaging epochs
was sufficient to move 2MASS 0727+1710 out of the NIC1 field of view.
Observations of SDSS 0151+1244 were also offset due to source
motion, and the object was imaged in the corner of the NIC1
camera's field of view (1$\farcs$2 from the closest edge of the array), 
limiting the area sampled for companions.  
The remaining sources were detected sufficiently close to the center of the
NIC1 array ($>$2$\farcs$5 from the array
edge) to provide adequate sampling of separations within the resolution
of the original discovery surveys 
($\sim1{\farcs}5$--$2\arcsec$ for SDSS and 2MASS).

Images were reduced by standard pipeline processing 
(CALNICA, Bushouse et al.\ 1997) using
updated calibration images and photometric keywords as of August 2004.
CALNICA reduction includes analog-to-digital correction, 
subtraction of bias and dark current frames, linearity correction,
correction for readout artifacts (the ``bars'' anomaly), division by an
appropriate flat field image, photometric calibration, cosmic
ray identification, and combination of MULTIACCUM frames
into a single calibrated image.  Post-CALNICA processing was limited
to the cleaning of cosmic rays and persistent bad pixels by nearest-neighbor interpolation,
and the mosaicking of the F110W and F170M dithered image sets (and all five filter sets 
of 2MASS 0518-2828) using the CALNICB routine.

\subsection{Resolved Sources}

Subsections (2$\farcs$5$\times$2$\farcs$5) of the reduced 
F090M, F110W and F170M mosaic images for 
each source are shown in Figure~\ref{fig_image1}.
North/east orientations are indicated by arrows, and the images are
scaled logarithmically to highlight low flux features.  Note the clearly
resolved PSFs in most of the F110W and F170M images, resulting
in significant structure outside of the core of the PSF including
first-order Airy rings and diffraction spikes at wider separations.
Three sources immediately stand out as obvious doubles. 
The previously reported binary SDSS 0423-0414 \citep{me0423}
shows two overlapping PSFs roughly oriented along a NNE/SSW axis,
with the northern component appearing to be slightly fainter in the F110W and F170M bands.
SDSS 1021-0304 also appears to be a close double aligned along a ENE/WSW axis,
with the western component appearing to be fainter at both F090M and F170M, but
not F110W.  2MASS 1553+1532 is a well-resolved
pair aligned along a N/S axis, with the southern component
appearing to be slightly fainter at F090M (it is marginally detected in this
band), F110W and F170M.  This source had previously been reported as a
candidate binary by \citet{me02aas}.

In addition to these
three sources, PSFs of 2MASS 0518-2828 and 2MASS 0926+5847 are slightly
elongated in the {\em HST} images and therefore also appear to be double. 
These sources are shown in more detail in Figure~\ref{fig_contour},
which displays contour plots of the central 0$\farcs$9$\times$0$\farcs$9 
regions of
the F090M and F110W images around 2MASS 0518-2828 and SDSS 0926+5847,
respectively, and equivalent data for the unresolved source
2MASS 1503+2525.  2MASS 0518-2828 is slightly elongated along a N/S axis
(PSF full width at half maximum of 1.86 pixels, as compared to 1.57 pixels for
2MASS 1503+2525),
with the shape of its southern extension indicating a fainter component.
All three F090M images obtained for 2MASS 0518-2828 show the same elongation
in the same orientation, lending confidence to its reliability.
The elongation is less obvious in the F110W and F145M images of this source, 
and marginally detected in the F160W and F170M images, presumably because it is 
obscured by the broader PSF
at these wavelengths.
SDSS 0926+5847 is clearly elongated along a NW/SE axis
(PSF FWHM of 2.67 pixels at F110W, versus 2.06 pixels for 2MASS 1503+2525), 
and appears
to be more symmetric, suggesting near-equal brightness components.
Again, the same elongation is seen in each of the F110W and F170M exposures
(the source
is only marginally detected at F090M).  We therefore conclude that both
systems are resolved doubles.  PSF fits for all of the doubles are 
presented in $\S$~4.
The remaining 17 targets appear to be 
single at the resolution of the NIC1 camera.

\section{NICMOS Photometry}

\subsection{Measurements}

Aperture photometry for all of the sources in our sample
were measured from the individual calibrated images using the 
IRAF\footnote{IRAF is distributed by the National Optical
Astronomy Observatories, which are operated by the Association of
Universities for Research in Astronomy, Inc., under cooperative
agreement with the National Science Foundation.} PHOT routine.
Various aperture radii ranging from 2-20 pixels 
(0$\farcs$086-0$\farcs$86) about the source flux peak were examined,
with a common background annulus of 20-30 pixels (0$\farcs$86-1$\farcs$3).  
Integrated source count rates were converted to photometric magnitudes on the 
Arizona Vega system ($M_{Vega}$ = 0.02)
using the photometric keyword parameter PHOTFNU and
Vega fluxes of 2157.3, 1784.9 and 946.1 Jy at F090M, F110W and F170M, respectively
\citep{sch05}.
Individual magnitudes from dithered exposure frames
were averaged to derive a single photometric measurement for each source.
F145M and F160W magnitudes for 2MASS 0518-2828 were similarly measured
using Vega fluxes of 1197.1 and 1042.6 Jy, respectively.

Aperture corrections in the F110W and F170M filters were determined from observations
of the three highest signal-to-noise (S/N) single
sources in our sample, 2MASS 0348+6022, SDSS 1254-0122 and 2MASS 1503+2525.
Comparison of integrated flux profiles as a function of aperture
size between these sources demonstrates
excellent agreement, with deviations of less than 0.01 mag 
for apertures wider than 4 pixels.  For F090M observations,
we adopted aperture corrections measured 
from observations of 2MASS 0518-2828, even though
this source is slightly resolved (all other sources
have insufficient S/N in this filter).
Table~3 lists the aperture corrections obtained
for each filter, corrected from an 11.5 pixel (0$\farcs$49)
reference aperture to an infinite aperture using values
from \citet{sch05}.

Photometric measurements are listed in Table~4.
For unresolved sources and the marginally resolved
pairs SDSS 0926+5847 and 2MASS 0518-2828,
we report 5-pixel (0$\farcs$22) aperture
photometry corrected to an infinite aperture using the values
in Table 3.  For the resolved doubles SDSS 0423-0414
and SDSS 1021-0304, we report
15-pixel (0$\farcs$65) aperture photometry encompassing both components
with no aperture correction.  
This aperture size was chosen as it includes $>$90\% of the
light in all three filters and minimizes photometric noise.  
For the well-resolved
double 2MASS 1553+1532, we report corrected 3-pixel (0$\farcs$13)
aperture photometry for each component separately.  
Uncertainties include contributions from the scatter of individual 
measurements (typically 1-2\%) and in the aperture
corrections ($<$1\% for F110W and F170M, 5\% for F090M), as well as 5\% absolute calibration
uncertainties and 1\% zeropoint drift \citep{sch05}.
The 5\% calibration uncertainties, which dominate the error
budgets for F110W and F170M magnitudes, are highly
correlated and reduce to 3\% 
for NICMOS colors; e.g., F110W-F170M.  These values are reported separately
in Table~4.  For 2MASS 0518-2828, we also measured F145M = 15.86$\pm$0.05 and 
F160W = 15.20$\pm$0.05.

Formal limiting magnitudes for each source field and filter were
determined by PSF simulation. Scaled PSFs of 2MASS 1503+2525
were added onto blank regions
of the individual F090M, F110W and F170M exposures 
and checked for visual
detection.  Reliable detections were possible for
peak flux scalings of 7, 5 and 5 times the background noise
at F090M, F110W and F170M, respectively.  These limits are given in Table~2.
There is a slight correlation of these limiting magnitudes 
with telescope pointing angle with respect to the Moon, likely the
result of increased background emission.
Several of our sources were detected in the F090M
exposures at magnitudes
below the formal detection limits, and have appropriately poorer S/N.

\subsection{T Dwarf Colors}

The original motivation for the filter set employed in this study 
was to provide
adequate color discrimination of bona-fide companions from 
coincident background
sources, and to determine photometric classifications.
Figure~\ref{fig_colorvsspt1} compares F110W-F170M colors
to spectral type for sources in our sample and unresolved
late-type L dwarfs observed in the
{\em HST} NICMOS program of \citet{rei06}.
Spectral types are based on optical spectroscopy for the L dwarfs
and near infrared spectroscopy for the T dwarfs.
We also show synthetic colors measured from low resolution
near infrared spectra of late-type L and T dwarfs from \citet{me06a}.
Earlier than type T1, F110W-F170M color is relatively
constant at $\sim$1.8 mag, albeit with significant dispersion ($\pm$0.3 mag)
that is larger than the photometric uncertainties.
For subclasses T1 and later, 
there is a tight correlation between spectral type and color.  
A linear fit to the photometric data 
for unresolved sources yields
\begin{equation}
SpT = 7.26 - 3.44(F110W-F170M)
\end{equation}
(where SpT(T1) = 1, SpT(T5) = 5, etc.), with an 
RMS scatter of 0.4 subclasses.  Thus, F110W-F170M is
a reliable proxy for spectral type in the T dwarf regime.
F090M-F110W color also provides a gross discriminant
of T spectral type, as shown in
Figure~\ref{fig_colorvsspt2}.  These colors redden
from 1.25 to 2.55 over spectral types T1 to T8, due
largely to increased absorption by \ion{K}{1}.
However, photometric
uncertainties are much larger for the F090M data, reducing its
utility.  We therefore focus on the 
F110W-F170M colors for our analysis.

Despite the apparent utility of {\em HST} colors to distinguish and
classify T dwarfs, 
the vast majority of photometric data for these objects
are from ground-based studies based principally on 
the $J$ (1.2~$\micron$), $H$ (1.6~$\micron$) and $K$ 
(2.0~$\micron$) telluric opacity windows.
To put our photometry into context 
with existing data,
we compared $F110W$ and $F170M$ magnitudes
to $J$ and $H$ photometry, respectively, on the 2MASS and 
Mauna Kea Observatory (MKO; \citet{sim02,tok02}) systems.
MKO data were collated from \citet{leg02}; \citet{kna04}; and references 
therein\footnote{A compilation of these data is maintained by S.\ 
Leggett at \url{http://www.jach.hawaii.edu/$\sim$skl/LTdata.html}.}.
Figure~\ref{fig_mkovshstvscolor} compares $J$-F110W and $H$-F170M
colors to F110W-F170M color for T dwarfs in our sample.
We also plot MKO/NICMOS synthetic colors
derived from low resolution near infrared spectroscopy
(from Burgasser et al.\ 2006a) as a comparison.
There is a marked difference between 2MASS and MKO
$J$-F110W colors.  The former are roughly constant 
($\sim$-0.75 mag) for -0.2 $\leq$ F110W-F170M $\leq$ 1.7, 
but there is significant scatter 
($\sim$0.2~mag) due primarily to the large uncertainties
associated with faint 2MASS T dwarf photometry (typically 0.04-0.10 mag).  In contrast,
MKO $J$-F110W colors show a tight correlation with F110W-F170M color
(MKO photometric
uncertainties are typically 0.03-0.05 mag) and 
a 0.2-0.3 magnitude offset from 2MASS $J$-F110W colors
due to differences in the filter profiles (cf.\ Stephens \& Leggett 2004).
A polynomial fit to the MKO/NICMOS photometric data for unresolved
sources yields the relation
\begin{eqnarray}
J_{MKO}-F110W & = & -1.15 + 0.025(F110W-F170M) \nonumber \\
 &  &  + 0.035(F110W-F170M)^2 
\end{eqnarray}
with a scatter of 0.04~mag.  
2MASS and MKO $H$-F170M colors are similar
(the result of their equivalent $H$-band filter
profiles), and both span a wider range
than $J$-F110W colors due to
strong 1.6~$\micron$ {\meth} absorption
in the later-type T dwarfs. 
$H$-F170M color is also
correlated with F110W-F170M color, 
and a fit to MKO photometry for unresolved sources yields 
\begin{eqnarray}
H_{MKO}-F170M & = & -0.75 + 0.89(F110W-F170M)  \nonumber \\
 &  & - 0.27(F110W-F170M)^2 
\end{eqnarray}
with a scatter of 0.05~mag.  
Interestingly, the combined light colors of SDSS 1021-0304
fall well off both of these lines
As discussed in $\S$~5.3, these discrepancies are due to the
unique composition of the SDSS 1021-0304 system.
Note that equivalent relations for 2MASS/NICMOS photometry are
presented in \citet{rei06}.

In Figure~\ref{fig_bc} we plot derived F110W and F170M
bolometric corrections (BC) for T dwarfs in our sample as a function
of F110W-F170M color.  These were computed from ground-based
MKO photometry and $K$-band
BCs from \citet{gol04} as, e.g., 
$BC_{F110W} = BC_K + K_{MKO} - F110W$.
Again, there is good correlation between the
NICMOS BCs and F110W-F170M color, 
and a polynomial fit to the
F110W photometry of unresolved sources yields
\begin{eqnarray}
BC_{F110W} & = & 1.43 + 0.11(F110W-F170M)  \nonumber \\
 &  & - 0.24(F110W-F170M)^2
\end{eqnarray}
with a scatter of 0.07~mag.

\section{PSF Fitting}

\subsection{Method}

The properties of the five resolved doubles in our sample
were determined by fitting individual images to PSF
models using an algorithm similar to that described in \citet{me03hst}.
One important modification in this study 
was the use of model PSFs generated by the Tiny Tim 
program\footnote{See \url{http://www.stsci.edu/software/tinytim/tinytim.html}.} 
\citep{kri95}.
Tiny Tim was specifically designed to generate PSFs for {\em HST} imaging data,
and includes mirror zonal errors and filter passband effects in the model
PSF shape.  We generated several grids of Tiny Tim PSFs for the
F090M, F110W and F170M filters appropriate
for the post-cryocooler NICMOS NIC1 detector, sampling 169 positions
across the chip in row/column increments of 16 pixels.  
For each calibration image, our PSF fitting routine 
employed the model PSF located closest to the position of the target
on the chip.  
We also used near infrared
spectral data for SDSS 1021-0304 from \citet{me06a} as input to the 
Tiny Tim program in order to model the appropriate spectral
response across the passband.
Model PSFs were sampled at 10 times the native pixel resolution of NIC1 for
subpixel shifting.    

PSF fits were made for 2$\farcs$5$\times$2$\farcs$5 
subsections of each image centered on the target sources,
and initial guesses for the pixel positions and fluxes 
of the two components were made using a simple peak detection algorithm
(for the primary)
and single-PSF subtraction (for the secondary).  
Model images were then generated using two
Tiny Tim PSFs resampled to the resolution of the data and scaled to the 
estimated peak fluxes of the detected sources.  
Our routine then iteratively searched for the optimal solution to the
primary position, secondary position, primary flux and secondary flux,
in that order, by computing the residuals between the 
model image and the data.  
Positional shifts of 0.1 pixels were made by shifting the 
oversampled model PSFs in integer units, and then downsampling to the
resolution of the data.  Fluxes were varied in steps of 1\% (0.01 mag).
The PSF fits were done recursively for convergence; i.e., if 
better solutions to any of the four binary parameters were found,
the routine retested all of the parameters starting with the primary
position until no reduction in residuals could be made.  

Figure~\ref{fig_psffit} illustrates the quality of these fits
for one of the F170M images of the SDSS 1021-0304 pair.
Shown are surface plots of the original data on a logarithmic vertical scale
(to bring up the background noise), the best-fit PSF
model, the result of subtracting the primary PSF model
from the data, and the result of subtracting the full PSF model
from the data.  The primary-subtracted image shows
a well-resolved secondary component, 
with clear detection of that component's first-order Airy ring.  
The final subtraction is extremely clean and average residuals
(typically $<$1\% of the peak source flux for all fits) are
at the level of background noise.

Final estimates of the flux ratio, separation and position angles of each double
were determined as the mean of individual determinations from each calibrated
image, with some vetting of very poor fits caused largely
by cosmic ray hits close to the target source.   Uncertainties
include scatter in the individual fits and systematic
uncertainties of 0.01 mag and 0$\farcs$004 in flux ratio and separation,
respectively, as prescribed by the fitting routine.
These values are listed in Table~\ref{tab_binparam}

For the wider doubles SDSS 0423-0414, SDSS 1021-0304 and 2MASS 1553+1532,
we estimate that additional systematic effects in the fitting process
are insignificant given the well-resolved nature of these sources and minimal 
residuals.  Indeed, 
resolved aperture photometry for the 2MASS 1553+1532 pair are consistent
with the PSF results within measured uncertainties. 
For the closer binaries 2MASS 0518-2828 and SDSS 0926+5847,
whose angular separations 
are less than 2 NIC1 pixels, systematic effects may be more important.  
To examine this possibility, we experimented with
PSF fits to test data constructed to mimic the measured
properties of the 2MASS 0518-2828 and SDSS 0926+5847 pairs 
in the F110W and F170M bands.
A total of 300 test images were constructed for each target/filter
simulation using random pairings 
of 2$\farcs$5$\times$2$\farcs$5 subsections of images
for the brightest unresolved sources in our sample (2MASS 0348-6022, SDSS 1254-0122
and 2MASS 1503+2525, for a total of 23 PSF images at F110W and 13 PSF images at F170M).
The selected PSF images were shifted by subpixel resampling
to replicate separations and position angles randomly drawn from Gaussian 
distributions centered at the measured values of the binary under investigation,
and with distribution widths twice the measured uncertainties.  The secondaries
in each test image were scaled to a flux ratio randomly 
drawn from a uniform distribution (in magnitude space)
spanning 0 mag to the measured magnitude difference plus 
three times the measured uncertainty.
These test images were then run through the same PSF fitting algorithm
as described above to derive experimental values.  
Systematic effects were ascertained by selecting only
those test cases where the experimental values agreed with the measured
values for the binaries, and then computing the mean and standard deviation of
the associated input parameters.   For 2MASS 0926+5847,
these simulations indicate that the underlying flux ratios
for this system ($\Delta$F110W$^{(sim)}$ = 0.4$\pm$0.2, $\Delta$F170M$^{(sim)}$ = 0.4$\pm$0.3)
are closer to unity, as suggested by visual inspection of the images themselves 
(cf.\ Figure~\ref{fig_contour}). Similar systematic flux ratio offsets are indicated 
for 2MASS 0518-2828 ($\Delta$F110W$^{(sim)}$ = 0.8$\pm$0.5, $\Delta$F170M$^{(sim)}$ = 0.9$\pm$0.6).
The flux ratio offsets are largely
due to the PSF fitting algorithm attempting to fit
both components with a single PSF, while the secondary PSF fits
the largest peak in the residuals.  
There were no indications of systematic offsets in the separations or position angles
of these systems, however.
The ``systematics-corrected'' flux ratios resulting from these simulations are 
given in Table~\ref{tab_binparam}.

\subsection{Results}

The angular separations of our sources range from 
0$\farcs$051$\pm$0$\farcs$012 (2MASS 0518-2828)
to 0$\farcs$349$\pm$0$\farcs$005 (2MASS 1553+1532). 
Assuming spectrophotometric distance estimates
for 2MASS 0518-2828, SDSS 0926+5847 and 2MASS 1553+1532 
of 34$\pm$6, 38$\pm$7 and 12$\pm$2 pc based on their combined light
2MASS $J$-band magnitudes (corrected for equal-brightness components) and 
the $M_J$/spectral type relations of \citet{tin03},
projected separations range over 1.8-5.0~AU.
These values are 
consistent with the small separations typically 
found for resolved brown dwarfs \citep{me06ppv}.

Relative magnitudes and combined light
photometry were used to determine component F110W and F170M
magnitudes and colors; these
are also listed in Table~\ref{tab_binparam}. For SDSS 0423-0414, SDSS 1021-0304 and
2MASS 1553+1532, the secondary colors are consistent with T dwarf 
spectral types of T2, T5 and T7, respectively.  
Component colors for SDSS 0926+5847 (after correction for systematic effects)
indicate spectral types of T4$\pm$0.5, consistent with the composite spectral type of T4.5.
Component colors for 2MASS 0518-2828 
have much larger uncertainties, and we can only ascertain that they
are consistent with spectral types $\sim$T3 and earlier.
In all cases, we can rule
out that the secondaries are hotter background stars, 
since both component and composite
F090M magnitudes would
be significantly brighter than observed.
Furthermore, the likelihood of a background source lying near 
any of the target sources is very small.  In our entire sample, only five additional point sources
were detected at F110W in the 11${\arcsec}^2$ NIC1 field of view,
with magnitudes of 19.5-22.5.  Assuming that the background
surface density scales as $10^{0.6F110W}$ (i.e., scaling as $d^3$), then
the probability of a background source with F110W $\lesssim$ 18.0 mag (bracketing the 
estimated magnitudes of the detected secondaries)
lying within 1$\arcsec$ of any target source is $5{\times}10^{-6}$, and can be ruled
out at the $>$4$\sigma$ confidence level.  
We therefore conclude that all five secondaries are physically bound T dwarf
companions.

Utilizing our derived F110W bolometric correction/color relation
(Eqn.~4), we can determine the relative
bolometric luminosities of the binary components as 
\begin{eqnarray}
{\Delta}M_{bol} & \equiv & M_{bol}(B) - M_{bol}(A)  \nonumber \\
 & = & {\Delta}F110W + BC_{F110W}(B) - BC_{F110W}(A).
\end{eqnarray}
These values are given in Table~\ref{tab_binparam}.  We also list
absolute $M_{bol}$ values for the components of
SDSS 0423-0414 and SDSS 1021-0304, which have parallax distance measurements.
In all cases, we verify that the secondaries are less luminous than the primaries,
as expected.  Relative effective temperatures, $T_B/T_A$, were determined from the relative
bolometric luminosities assuming identical component radii, so that 
T$_B$/T$_A$ = $(L_B/L_A)^{1/4}$.  Again,  
secondary {\teff}s are less than primary {\teff}s for all five systems,
although in no case are differences more than $\sim$20\%.
These ratios are consistent with estimated component {\teff}s based on their
spectral types (from F110W-F170M color and spectral decomposition; see $\S$~5.3)
and the {\teff}/spectral type relation of \citet{gol04}, taking into
account the 124~K scatter in the latter relation.  For SDSS 0423-0414 and SDSS 1021-0304,
we derived component {\teff}s from their individual luminosities and assumed
radii of 0.095$\pm$0.010~R$_{\sun}$, appropriate for 0.5-5~Gyr brown dwarfs
in the {\teff} range of late-type L and T dwarfs \citep{bur97}.  These effective
temperatures are consistently 100-200~K lower than those based on the
\citet{gol04} relation. Although the deviations are comparable to the
uncertainties in both this relation and our {\teff} determinations,
the \citet{gol04} results may be overestimated in this spectral type regime 
due to contamination by these previously unresolved (and hence overluminous) binaries.

Finally, we derived mass ratios for the five binary systems, assuming coevality,
using the mass-luminosity power-law relation of \citet{bur01}, L $\propto$
M$^{2.64}$, implying
\begin{equation}
q \approx 10^{\frac{-{\Delta}M_{bol}}{6.6}}.
\end{equation}
System mass ratios are all 0.7 or greater, similar to most currently known brown dwarf pairs.
Individual component masses were estimated from the evolutionary models
of \citet{bur97} using the component {\teff}s (for 2MASS 0518-2828, SDSS 0926+5847
and 2MASS 1553+1532) or $M_{bol}$s (for SDSS 0423-0414 and SDSS 1021-0304),
and assuming an age range of 0.5-5~Gyr, typical for local 
disk dwarfs \citep{rei00b}.  These component masses are roughly consistent
with the estimated mass ratios. 
Orbital period estimates were derived assuming circular orbits and semimajor
axes $a \approx 1.26\rho$ \citep{fis92}.  These periods
range over 10-50~yr,  
with the
SDSS 0423-0414, 2MASS 0518-2828 and SDSS 0926+5847 systems (P $\lesssim$ 20~yr)
appearing to be
the best targets for dynamical mass measurements.

\subsection{Search Limits for Very Faint Companions}

In order to search for even fainter companions to unresolved T dwarfs in our
sample, we repeated the PSF fitting analysis described above
on the F110W calibrated images using a 
single PSF model for the primary.  The F110W images were chosen because
this filter samples the peak of the 
near infrared spectral flux of brown dwarfs down to 
{\teff} $\approx$ 500~K \citep{bur03b}, and therefore provides
the most sensitive probe for low mass companions.
We examined residual
images (subtraction of the PSF model from the data) 
by eye for faint point sources that persist in the same location
relative to the original primary.
Out of the entire sample, 
only one ``candidate'' companion was identified, a faint source
(F110W $\approx$ 22.0) located 0$\farcs$97 northeast of the T1 SDSS 0151+1244.  
This source has a high probability of being spurious, however; it does not appear in either the F090M or F170M images, and
due to the pointing offset of the {\em HST} observations of SDSS 0151+1244
(see $\S$~2.2) was only detected 
in one F110W exposure.  While it is therefore likely to be
a residual cosmic ray or ghost, it warrants follow-up confirmation
imaging since its brightness relative to SDSS~0151+1244 is consistent 
with a {\teff} $\sim$ 450~K brown dwarf companion.

Sensitivity limits for putative faint companions were quantified
by examining the F110W radial profiles for each of the unresolved sources
before and after PSF subtraction.  Examples of 
the brightest (2MASS 1503+2525)
and faintest (SDSS 0207+0000) sources are shown in Figure~\ref{fig_psfsub}.
PSF subtraction results in residuals that are $\sim$4-4.5 mag 
fainter than the peak source flux in the core and first Airy ring 
($\rho \lesssim 0{\farcs}2$), irrespective
of the brightness of the primary.  At larger separations, there is
little or no improvement in 
sensitivity beyond the inherent decrease in the primary flux, 
and residuals are largely
background limited at $\rho \gtrsim 0{\farcs}4$.
Faint source sensitivity limits (assuming 3$\sigma$ detections)
for each of the unresolved sources, including mass ratio
limits assuming $\Delta$F110W $\sim$ ${\Delta}M_{bol}$ and using Eqn.~6
are given in Table~\ref{tab_senslimits}
These detection limits can be characterized as follows:
\begin{itemize}
\item No detections for $\rho \lesssim 0{\farcs}04$,
\item $\Delta$F110W $\sim$ 3-3.5 mag ($q \gtrsim$ 0.3-0.4) for $0{\farcs}04 \lesssim \rho \lesssim 0{\farcs}2$,
\item $\Delta$F110W varying from $\sim$3-3.5 mag to the background limit ($\sim$4-6.5 mag; $q \gtrsim$ 0.1-0.3) for 
$0{\farcs}2 \lesssim \rho \lesssim 0{\farcs}4$, and
\item $\Delta$F110W background limited for $\rho \gtrsim 0{\farcs}4$.
\end{itemize}
Assuming all of our sources have masses below the hydrogen burning
minimum mass ($\sim$0.075~M$_{\sun}$; 
Chabrier et al.\ 2000; Burrows et al.\ 2001),
these observations rule out companions down 
to just above the deuterium burning limit 
($\sim$0.012~M$_{\sun}$; Burrows et al.\ 2001) for 
most of our targets.

Finally, we comment on the 5 faint 
sources (F110W = 19.5--22.5) detected at wider
separations ($\rho \gtrsim 4\arcsec$)
in the NIC1 images. These can be ruled out as low temperature
companions based on their magnitudes and F110W-F170M colors, 
typically 1.2-1.4 mag, inconsistent with co-spatial
mid- and late-type T dwarf
companions but typical for background M stars.\footnote{The most 
interesting source is located 7$\farcs$63 northeast
of SDSS 1254-0122 at 12$^h$54$^m$54$\fs$14 $-$01$^d$22$\arcmin$42$\farcs$84.
With F110W-F170M = 2.31$\pm$0.14, this source 
is likely to be a highly reddened background star or faint 
unresolved galaxy.} 
We conclude that no bona-fide companions
are present around any of our targets
with $\rho \gtrsim 0{\farcs}4$ and $q \gtrsim$ 0.1-0.3.

\section{Analysis of Individual Sources}

\subsection{SDSS 0423-0414}

The identification of SDSS 0423-0414 as a binary 
system was previously reported
in \citet{me0423}.  The parameters reported here
supersede those of the previous paper, although
all measurements are
consistent within the reported uncertainties.
This source, similar in composition
to the recently resolved binary 2MASS J22521073-1730134 
\citep{rei2252},
is an unusual system,
as its combined light optical spectrum \citep[Kirkpatrick et al.\ in prep.]{cru03}
exhibits both 6563~{\AA} H$\alpha$ emission, an indicator
of magnetic activity; and 6708~{\AA} \ion{Li}{1} absorption, present in
brown dwarfs with masses below the Li-burning minimum mass ($\sim$0.065~M$_{\sun}$;
\citet{reb92}).  Both signatures are rare in very late-type L dwarfs and T dwarfs,
as the strength and frequency of H$\alpha$ emission plummets across
the L dwarf regime \citep{giz00,kir00,moh03,wes04},
while \ion{Li}{1} absorption becomes increasingly
difficult to detect against a progressively fainter 
continuum suppressed by pressure-broadened \ion{Na}{1} and \ion{K}{1}
lines \citep{bur03}. The only other sources known 
to exhibit the same combination of 
features are the L2 Kelu~1 \citep{rui97}, which has also been
resolved as a binary system \citep{liu05,gel06}; 
the L0.5 2MASS J20575409-0252302 \citep{cru03}, which has not \citep{rei06};
and the L0 2MASS J11544223-3400390 (Kirkpatrick et al., in prep.), which 
has not yet been imaged at high angular resolution.  

This raises the question: from which component
or components do these spectral features arise?
Decomposition of the combined light near infrared
spectrum using the resolved NICMOS photometry 
indicates that this system is composed of an L6.5 primary
and a T2 secondary \citep[see $\S$~5.3]{me0423}.
H$\alpha$ emission from T dwarfs is rare; 
only three other T dwarfs have been detected in 
emission \citep{me03opt}, one of which,
2MASS J12373919+6526148 \citep{me99}, is unusually active
and is speculated to be a very tight ($\rho \sim$ 0.1~R$_{\sun}$)
interacting binary system \citep{me00b,me02c}.
The H$\alpha$ line flux as measured from combined light optical spectral 
data (Kirkpatrick et al.\ in prep.), flux calibrated to
SDSS $i^{\prime}$ photometry (20.22$\pm$0.04; Geballe et al.\ 2002),
is $1.7{\times}10^{-17}$ ergs cm$^{-2}$ s$^{-1}$. Using the component 
bolometric luminosities listed in Table~\ref{tab_binparam}, we
derive $\log{L_{H\alpha}/L_{bol}}$ = -5.5 if the emission arises
from the L6.5 primary, and -5.2 if it arises from the T2 secondary.
Compared to similar-typed objects exhibiting H$\alpha$ emission --
e.g., DENIS-P J0205.4-1159 (L7; $\log{L_{H\alpha}/L_{bol}}$ $<$ -6.2; Mohanty \& Basri 2003) 
and SDSS 1254-0122 (T2; $\log{L_{H\alpha}/L_{bol}}$ = -5.8; Burgasser et al.\ 2003a) --
the emission flux from either component
is not necessarily extreme, but is nevertheless rare 
in this spectral type regime (cf.\ Figure 3 in Burgasser et al.\ 2002a).

Turning to the 6708~{\AA} \ion{Li}{1} line, 
\citet{liu05} have pointed out that the detection of this feature
in a substellar binary can be used as a powerful
constraint of systemic age in conjunction with theoretical
evolutionary models, particularly if the absorption
can be attributed to one or both
components.  SDSS 0423-0414 exhibits a prominent \ion{Li}{1}
line, and as luminous flux in this spectral region is dominated by 
the earlier-type primary (the optical classification of this
source is L7.5; Cruz et al.\ 2003) it is likely that this component
is responsible for the absorption.  This deduction is supported by the
fact that atomic Li gas is likely to be depleted in the photosphere of the
secondary (for which we derive {\teff} = 1260$\pm$70~K)
as Li is incorporated into LiCl and LiOH at {\teff}s $\lesssim$ 1500~K 
and pressures $\gtrsim$ 1-10 bar \citep{lod99}.  

Assuming then
that the observed absorption arises from the L6.5 primary, which
must then have M $\lesssim$ 0.065 M$_{\sun}$, an upper age
limit of 1.7~Gyr can be deduced using the theoretical evolutionary models
of \citet{bur97}; this is illustated in Figure~\ref{fig_0423evol}.
This age is on the young side for a field dwarf, and may explain
the presence of H$\alpha$ emission 
in one or both components; observations of fully-convective 
lower main sequence stars in the field       
and clusters shows that magnetic activity is commonly enhanced in young         
stars \citep{haw99,rei03}.
It is also consistent with the kinematics of this system,
as its tangential velocity, $V_{tan}$ = 24.0$\pm$0.7 km~s$^{-1}$,
is on the low end of the T Dwarf $V_{tan}$ distribution of \citet{vrb04}.
An estimated minimum age for this system of 0.5~Gyr can be
argued from the absence of low surface gravity features 
in its combined-light optical spectrum
(e.g., VO absorption and weakened alkali lines; see Kirkpatrick 2005).
Thus, 
the components of
SDSS 0423-0414 are among the few brown dwarfs with well-constrained
ages, distances and bolometric luminosities.

\subsection{2MASS 0518-2828}

The detection of a companion to 2MASS 0518-2828 appears to confirm
the binary hypothesis of \citet{cru04} for this source, 
put forth to explain its unusual
near infrared spectrum.  2MASS 0518-2828 exhibits
clear {\meth} absorption at 1.6~$\micron$ but no {\meth} band at 2.2~$\micron$.
This is in contrast with trends in the standard L/T spectral sequence,
where the 2.2~$\micron$ band is seen to develop in the latest-type L dwarfs
first, followed by 1.6~$\micron$ absorption at the start of the T sequence 
\citep{geb02}.  \citet{cru04} found that the combination of
L6 and T4 spectra, with the latter scaled to be 20\% brighter
at 1.27~$\micron$ 
adequately matches the observed spectral energy distribution for 2MASS 0518-2828. 
The small separation of this source, and the corresponding poor
determination of its relative photometry, makes it impossible to verify the
conjectured spectral types of the components based on these {\em HST} observations
(although our photometry are consistent with these types).
However, the fact that 2MASS 0518-2828 is resolved into
two components makes this scenario
likely. 

\subsection{SDSS 1021-0304}

Like SDSS 0423-0414 and 2MASS 0518-2828, SDSS 1021-0304 appears
to be a binary straddling the L/T transition, with 
component F110W-F170M colors indicating
spectral types of $\lesssim$T2 and T5.
However, this source is particularly interesting as the two
components have nearly equal magnitudes at F110W, while the
secondary is a full magnitude fainter at F170M. 
Because {\water} and {\meth} absorption bands encompassed 
by the F110W filter bandpass (cf.\ Fig.\ 1) become
stronger with later spectral types, the equivalent magnitudes
of the two components suggests that the secondary
has a brighter peak flux density, like 2MASS 0518-2828.

To explore this possibility,
we performed a spectral decomposition of the combined
light near infrared spectra of this source and SDSS 0423-0414
using a method similar to that
described in \citet{me0423} and \citet{rei2252}.
In brief, our technique involves the combination of 
various pairings of standard spectra (sources with well-defined classifications)
after scaling them to the relative fluxes of the binary system under
investigation.  The hybrid spectra were then 
compared to the combined light spectrum of each (unresolved) 
binary to determine the best match.
We performed our analysis on low resolution ({\ldl} $\sim$ 150)
near infrared data obtained with the SpeX spectrograph \citep{ray03} 
mounted on the 3.0m NASA Infrared Telescope Facility.
Details on the acquisition, reduction and
characteristics of these data are described in detail in \citet{me04wide3,me06a}.
The comparison basis set was composed of equivalent
spectra of late L and T dwarf standards from \citet{kir99,cru03}; and
\citet{me06a}; specifically:
2MASS J08354256-0819237 (L5), 
2MASS J04390101-2353083 (L6.5),
DENIS-P J0205.4-1159\footnote{DENIS-P J0205.4-1159 is either
a resolved, near-equal mass
binary \citep{koe99,leg02} or a triple \citep{bou05}, and is 
arguably a poor choice for this analysis. However, we did not have an
alternate L7 comparison source, and DENIS-P J0205.4-1159
is currently the L7 optical spectral standard on the \citet{kir99} scheme.} \citep[L7]{del97},
2MASS J16322911+1904407 (L8), SDSS 0151+1244 (T1), SDSS 1254-0122 (T2),
2MASS 1209-1004 (T3), 2MASS 2254+3123 (T4), 2MASS 1503+2525 (T5)
and SDSS 1624+0029 (T6).
Pairings of the standard spectra were scaled to the observed F110W flux
ratios, then added together and normalized.
The quality of agreement between the resulting hybrid spectra
and those of the binaries was quantitatively determined
by comparison of the {\water}, {\meth} and $K/J$ spectral
indices defined in \citet{me06a}, as well as the relative F170M
flux ratios between the spectral components.  

The best matches for SDSS 0423-0414 and SDSS 1021-0304
are shown in Figure~\ref{fig_1021specfit}. 
For the former, we confirm previous results by \citet{me0423},
finding a best fit to a combination of the L6.5 2MASS 0439-2353
and the T2 SDSS 1254-0122.
A hybrid spectrum of the 
T1 SDSS 0151+1244 and the T5 2MASS 1503+2525 provides
the best match for SDSS 1021-0304.
Note that the derived spectral types of the secondary components agree
with photometric classifications based on F110W-F170M colors.
In both cases the hybrid spectra
show remarkable agreement with the binary spectra,
both in terms of band strengths and the overall spectral
energy distribution, across the full 0.8-2.5 $\micron$ band.  The relative
F170M magnitudes as measured from the scaled
component spectra are in rough agreement with {\em HST} photometry,
differing by at most $\sim$0.1~mag, an offset attributable to 
the low resolution and calibration uncertainties in the spectral data.

Examining the relative fluxes of the best-fit 
component spectra for SDSS 1021-0304 in more detail, 
a remarkable fact is revealed.  The emergent flux density
of the T5 secondary of this system is 31\% brighter than that of the T1
primary at the peak of the spectral energy distribution (1.27~$\micron$),
and 24\% brighter at the 1.05 $\micron$ flux peak.  
This is despite the fact that the secondary is 11\% cooler 
and 37\% less luminous overall.  
That these spectral peaks differ significantly between the two components
while F110W magnitudes are roughly equivalent can be explained by
the redistribution of flux within the F110W bandpass.  The increased
brightening at 1.05 and 1.27 $\micron$ in the secondary component
is offset by deeper {\water}
and {\meth} bands at 1.1 and 1.35 $\micron$.  At the bottom of these molecular
features, and at shorter and longer
wavelengths, the primary component is brighter.  The components of
SDSS 0423-0414 do not exhibit this same
brightness inversion, although the peak flux densities
are rather similar (differing by less than 25\% at 1.27~$\micron$)
given the large differences in spectral type.
The brightening of the secondary component
of SDSS 1021-0304 is similar to that hypothesized for 
2MASS 0518-2828, and more recently observed for the
T1.5+T5.5 binary SDSS J153417.05+161546.1 
\citep[hereafter SDSS 1534+1615]{liu06}.
Indeed, the photometric and spectroscopic properties of
SDSS 1021-0304 and SDSS 1534+1615 are quite similar.
We discuss the observed $J$-band brightening, and its implications
on the transition between L dwarfs and T dwarfs, in $\S$~7.

\subsection{2MASS 1217-0311}

\citet{me03hst} reported a possible faint companion to this object
in {\em HST} Wide Field Planetary Camera 2 imaging, but our
NICMOS observations fail to reveal this source.  
Assuming that the relative flux at
F110W between 2MASS 1217-0311 and the putative companion is as bright
or brighter than that at F1042M ($\lambda_c$ = 1.02~$\micron$), 
it would have been easily detected
at the separation (0$\farcs$21) and flux ratio (2.4~mag)
previously observed.  It is likely that the prior detection was
an unfortunate combination of cosmic ray hits localized near the 
target source.

\subsection{2MASS 1553+1532}

2MASS 1553+1532 is the latest-type binary in our sample, and the best
resolved.
The F110W-F170M colors are consistent with very similar spectral
types (T6.5 and T7), effective temperatures (within 7\%) and
masses ($q = 0.90{\pm}0.02$).  With a separation of 0$\farcs$349$\pm$0$\farcs$005
the 2MASS 1553+1532 pair is resolvable by
ground-based imaging under the best seeing conditions,
and was previously reported as a possible
binary by \citet{me02aas} based on imaging observations
with the Keck 10m Near Infrared Camera 
\citep[hereafter NIRC]{mat94} on 2000 July 22 (UT).
We have revisited these data to determine whether the two components
share common proper motion, and to search for orbital motion.

Conditions during the Keck observations in 2000 were
particularly excellent, with clear
skies and seeing of 0$\farcs$3 (full width at half maximum) at $K_s$
during the observations.  2MASS 1553+1532 was observed in this
filter, with 10 dithered exposures of 20s each obtained sequentially.
Immediately following these observations, 10 dithered 20s exposures of 
the unresolved T dwarf 2MASS 2254+3123 were also obtained.  Despite the
large angular offset between the sources, we
used these observations for PSF calibration as no other sufficiently
bright and unresolved sources were detected in the 2MASS 1553+1532 fields.
Raw images for both datasets were pairwise subtracted
to eliminate first order background emission, and checked for linearity.
No further reduction of the data (e.g., flat fielding) 
was done as relative photometry was not a priority. 
Figure~\ref{fig_1553contour} shows the PSFs
of 2MASS 2254+3123 and 2MASS 1553+1532.  The latter is clearly extended along
a NNE/SSW axis, but the underlying components overlap substantially.
We extracted astrometric information
using a PSF fitting algorithm similar to that
described above, but in this case comparing $3{\arcsec}{\times}3\arcsec$
subsections of each  
pair-wise subtracted frame of 2MASS 1553+1532 to all 10 observations
of 2MASS 2254+3123, for a total of 100 separate fits.
Imposing the condition that 
average residuals be less than 3\% of the peak source flux, the 42 best
fits gave a mean separation $\rho$ = 0$\farcs$30$\pm$0$\farcs$03
(assuming a camera pixel scale of 0$\farcs$153) and position angle 
$\theta$ = 199$\pm$7$\degr$.

The difference in epoch between the {\em HST} and Keck images is 3.126 yr.
A preliminary proper motion of this object as measured by the USNO
infrared parallax program \citep{vrb04} of $\sim$0$\farcs$4 yr$^{-1}$
(F.\ Vrba 2006, private communication) implies 
a total motion of the system of $\sim$1$\farcs$3 over this period.  
Yet the change in the 
relative separation between the two components is only
0$\farcs$05$\pm$0$\farcs$03.  These observations therefore confirm
common proper motion for this pair, which are almost certainly
gravitationally bound.

As for orbital motion, both the separation and position angle
of this system have changed only slightly
between the Keck and {\em HST} observations.  
While only marginally significant ($\Delta{theta}$ = 9$\pm$7$\degr$),
Figure~\ref{fig_1553nircvshst} illustrates that these slight changes are perceptible. 
However, the small position angle
change is significantly less than that expected ($\sim$26$\degr$) 
for its estimated 45~yr orbital period.  This suggests that the true
orbital period
may be much longer, possibly due to smaller component masses in a younger
system, or a particularly eccentric orbit; or that the orbital inclination
is quite different from a face-on projection 
(consistent with the slight change in the separation of the two components).
Further high-resolution imaging 
may constrain these possibilities in a reasonably short time period
($\sim$5~yr), but mapping of the full orbit is clearly 
a long-term prospect.

\section{An Updated Assessment of Brown Dwarf Multiplicity}

\subsection{The Binary Fraction}

The fraction of resolved binaries in our sample is $\epsilon_b^{obs}$ = 5/22 = 
23$_{-6}^{+11}$\%, where the uncertainties take into
account the size of the sample \citep{me03hst}.  This
is similar to resolved binary fractions measured for other
large high resolution imaging samples of VLM field dwarfs 
\citep{rei01,rei06,bou03,clo03,giz03,sie05}.  However, as all of these
samples are largely magnitude-limited, this fraction is biased in
favor of unresolved, near-equal mass ratio binaries.  We can estimate 
the underlying (i.e., volume-limited) binary
fraction, $\epsilon_b$, as (cf. Eqns.\ 4 and 5 in \citet{me03hst})
\begin{equation}
{\epsilon}_b = \frac{{\epsilon}_b^{obs}}{{\alpha}(1-{\epsilon}_b^{obs})+{\epsilon}_b^{obs}},
\end{equation}
where
\begin{equation}
{\alpha} \equiv \frac{\int_0^1{(1+q^{2.64})^{3/2}f(q)dq}}{\int_0^1{f(q)dq}}
\end{equation}
is the fractional increase in volume sampled for binaries with
flux ratio $f_B/f_A \approx q^{2.64}$ (Eqn.\ 6)
and mass ratio ratio distribution
$f(q)$.   In \citet{me03hst}, both flat and delta-function forms
of $f(q)$ were considered; here, we explicitly calculate 
$\alpha$ = 2.50$_{-0.06}^{+0.04}$ using
a power-law mass ratio distribution as described below.
This yields $\epsilon_b$ = 11$_{-3}^{+7}$\%, a value consistent with previous determinations
of bias-corrected VLM binary fractions \citep{me03hst,clo03,sie05} 
and volume-limited
estimates \citep[$\sim$15\% and 12$_{-3}^{+7}$\%, respectively]{bou03,rei06}.  

It is important to stress that this
fraction applies only to those binary systems that are resolvable by direct imaging.
For our study, this limits the phase space sampled to $\rho \gtrsim$ 1~AU
and $q \gtrsim$ 0.35, or $\rho \gtrsim$ 5~AU and $q \gtrsim$ 0.2.  
The equivalent phase space of F-G and M dwarf binaries in the studies of \citet{duq91} and
\citet{rei97} yield binary fractions of 39\% (combining both mass ratio
and period distributions) and 24$^{+6}_{-4}$\% (for $M_V > 9$), respectively.
Hence, in equivalent phase spaces the binary fraction of brown dwarfs 
in our sample is less than that of more massive stars.

But does this mean that the overall binary fraction of brown dwarfs is less?
\citet{max05} have proposed that a substantial fraction (50-67\%)
of VLM binaries may be hiding in more closely separated systems ($\rho \lesssim$ 2.6~AU)
and can only be resolved as spectroscopic binaries.
This projected separation corresponds to an angular separation of 
$\lesssim$0$\farcs$13 at the average distance of sources in our
sample ($\sim$20 pc), only 3 NIC1 pixels.  Indeed, over 25\%
of resolved brown dwarf binaries identified to date have angular
separations below this limit, with the majority close to the resolution
limits of $HST$.  This supports the possibility that 
a significant number of 
more closely-separated and/or more distant systems remain unresolved.
Bayesian statistical analysis of high resolution imaging studies by \citet{rei06},
which takes into account the possible presence of unresolved systems, 
indicates an overall VLM binary fraction of 24\%,
twice that of the resolved fraction.\footnote{Note that this analysis
assumes a symmetric Guassian separation distribution; the fraction
may be higher (lower) if there is an excess (deficiency) of short period
systems (cf.\ Maxted \& Jeffries 2005).}  This still places the binary fraction
of VLM dwarfs at 1/3 that of solar-type stars, and 2/3 that of M dwarfs,
consistent with a decreasing binary fraction toward later spectral types
and lower masses.

\subsection{The Separation Distribution}

The projected separation distribution of 30 brown dwarf binaries
resolved by high resolution imaging surveys to date are
shown in Figure~\ref{fig_sepdist}. 
These include systems listed in
\citet{me06ppv}\footnote{This sample incorporates binaries
identified in 
\citet{mrt99,lei01,rei01,rei06,pot02,bou03,me03hst,giz03,mcc04,me05gl337cd,liu05,liu06};
and this study.}
that have estimated primary masses below 0.075~M$_{\sun}$.
In accordance with
previous studies, we find that this distribution peaks at
very close separations, $\sim$4~AU with a broad peak spanning 2--8~AU.
This is significantly lower than the 30~AU peak of the F-G and M 
binary separation distributions \citep{duq91,fis92}. Indeed, no
brown dwarf field binaries have been identified with separations
$\gtrsim 15$~AU.  
However, two wider brown dwarf binaries systems have been
recently identified in young cluster/associations:
2MASS J11011926-7732383 \citep{luh04}, a 240~AU binary
in the $\sim$2~Myr Chameleon I association;
and 2MASS J1207334-393254 \citep{cha04,cha05}, 
a 40~AU, very low mass (M$_{tot}$ $\sim$ 0.03~M$_{\sun}$)
brown dwarf binary
in the $\sim$8~Myr TW Hydrae association \citep{giz02}.
Both systems are very young, and it remains unclear
as to whether their configurations are stable long-term \citep{mug05}.
The widest VLM field binary so far identified, 
DENIS J055146.0-443412.2 \citep[220~AU]{bil05},
is likely composed of two VLM stars.  
Hence, the wide separation brown dwarf binary desert originally suggested
by \citet{mrt00b} --- not to be confused with the brown dwarf companion desert 
around solar-type stars \citep{mrc00} ---
remains a distinct characteristic of brown
dwarf binaries in the field.

At closer separations, imaging surveys are limited 
by angular resolution.  Hence, the true peak of the brown dwarf
separation distribution may be lower than that inferred from
Figure~\ref{fig_sepdist}.
Nevertheless, it is interesting to note that
the separations of binaries identified in this survey --- and indeed all T dwarf
binaries identified to date ---
are $\lesssim$~5~AU, such that most T dwarf
binaries have separations below the peak of the brown dwarf
distribution.  The apparent compactness of T dwarf binaries as compared to
warmer M- and L-type brown dwarf systems
is consistent with a maximum binary
separation that scales with total system mass
\citep{rei01,me03hst,clo03}, since cooler brown 
dwarfs have lower masses than warmer ones at a given age.
However, the smaller separations of T dwarf binaries 
may also be due to selection effects. T dwarfs are intrinsically
fainter and typically found at closer distances to the Sun than M and L dwarfs in
magnitude-limited surveys.  Hence, T dwarf binaries can generally be observed 
at higher linear resolution.  
A statistically robust
volume-limited sample of M-, L- and T-type brown dwarfs would provide
an adequate check for mass dependency in the separation distribution of 
substellar objects.

\subsection{The Mass Ratio Distribution}

The mass ratio distribution of brown dwarf binaries is shown in
Figure~\ref{fig_qdist}.
This distribution is clearly peaked at $q \sim 1$, 
with 50$\pm$9\% of all known systems having near-equal mass components. 
Again, because the majority of these systems were
originally selected from magnitude-limited surveys, there is an inherent bias
in the discovery of equal-mass systems that scales approximately as 
$(1+q^{2.64})^{3/2}$.  A bias-corrected distribution, also shown in
Figure~\ref{fig_qdist}, nevertheless shows that near-equal mass ratio systems
are predominant.   This result is robust even when sensitivity
limits are taken into account.
The bias-corrected
frequency of binaries drops by a factor of 7.8 from $q = 1$ to $q = 0.5$,
even though most imaging programs are complete for companions down to or
below this limit. 
A fit to the bias-corrected distribution
for $q > 0.5$ to a power-law, $f(q) \propto q^{\gamma}$,
yields $\gamma$ = 4.2$\pm$1.0, slightly flatter but nevertheless consistent 
with a Bayesian analysis of VLM binaries \citep{rei06}.

In summary, our sample supports prior results on brown dwarf multiplicity,
namely:
\begin{itemize}
\item The resolved binary fraction of brown dwarfs is lower than that
of stars, $\epsilon_b \approx 11$\% for $\rho \gtrsim$ 3 AU and $q \gtrsim$ 0.3;
\item The separation distribution of resolved brown dwarfs peaks around 4~AU;
the true peak may lie at lower separations due to resolution limits of imaging programs;
\item The maximum separations of field brown dwarf binaries appears to decrease for
later spectral types, consistent with a mass-dependent trend; and
\item Most brown dwarf pairs have near-equal mass ratios, with
a bias-corrected distribution of 
$f(q) \propto q^{(4.2{\pm}1.0)}$ indicated by current data.
\end{itemize} 
These characteristics of brown dwarf field binaries provide key
empirical constraints for the theoretical modeling of
brown dwarf formation and dynamical evolution, issues that are discussed in
detail in \citet{me06ppv,luh06}; and \citet{whi06}.

\section{Binaries and the L/T Transition}

\subsection{$J$-band Brightening - Intrinsic to the L/T Transition}

Three of the binaries in our sample -- 2MASS 0518-2828, SDSS 0423-0414
and SDSS 1021-0304 -- are composed of brown dwarfs that span
the transition between L dwarfs and T dwarfs.  This spectral type
range has been the focus of both observational and theoretical studies as it
encompasses dramatic changes in the atmospheric
properties (e.g., photospheric condensate dust depletion) and 
spectral energy distributions (e.g., the onset of {\meth} absorption)
of cool brown dwarfs.  This transition also
exhibits several unusual traits,
including an apparent brightening of absolute $J$-band magnitudes
from late-type L to mid-type T dwarfs \citep{dah02,tin03,vrb04}.
This so-called ``$J$-band bump'' 
has been attributed to dynamic atmospheric processes,
such as condensate cloud fragmentation \citep{me02b}, a sudden increase
in sedimentation efficiency \citep{kna04}
or a global collapse of the condensate cloud layer \citep{tsu05}.
However, \citet{tsu03}
have also argued that age and/or surface gravity
effects amongst disparate field sources may be responsible.

The component fluxes of the SDSS 1021-0304 binary
demonstrate that the last hypothesis can be largely ruled out for this system.
Under the reasonable
assumption of coevality, these brown dwarfs have similar ages, masses
and (presumably) radii, implying nearly identical surface gravities.
Yet the T5 secondary of this system is clearly brighter than the T1
primary at 1.05 and 1.27 $\micron$.  
Similar trends suggested in the 2MASS 0518-2828 system and
observed in SDSS 1534+1615 
demonstrate that SDSS 1021-0304 is not a unique case.  Hence, 
a brightening of surface fluxes at these wavelengths appears to be 
an {\em intrinsic feature} of the L/T transition.

\subsection{A ``Bump'' or a ``Plateau''?}

In their analysis of the SDSS 1534+1615 binary,
\citet{liu06} proposed that the $J$-band bump may be artificially
enhanced by a significant contribution of binaries amongst
mid-type T dwarfs (such ``crypto-binarity'' has also been suggested
by Burrows, Sudarsky \& Hubeny 2006).
To examine this hypothesis in detail,
Figure~\ref{fig_absmag} compares 
absolute MKO $J$- and $K$-band magnitudes to spectral
type for 50 L and T dwarf systems with measured parallaxes 
\citep[precision $>$20\%]{dah02,tin03,vrb04},
companions to nearby Hipparcos stars 
\citep{bec88,nak95,me00a,kir01,mcc04} and
resolved absolute magnitudes for SDSS 0423-0414, SDSS 1021-0304
and the T1 + T6 binary Epsilon Indi B \citep{mcc04}.
For consistency, spectral types for L dwarfs are based on optical data
and the \citet{kir99} classification scheme, while those for T dwarfs
are based on near infrared data and the \citet{me06a} scheme. 
MKO $J$ magnitudes for the SDSS 0423-0414 and SDSS 1021-0304 components
are based on their F110W-F170M colors and Eqn.\ 2; 
$K$-band component photometry
is derived from synthetic colors measured from the component spectral templates.

The $J$-band bump is seen clearly in these data in the T1-T5 spectral type
range, and is well traced by the absolute magnitude/spectral type relation
of \citet{tin03}.
Yet one of these data points is the unresolved SDSS 1021-0304
systems, and its individual
component fluxes are only slightly brighter 
($J = 14.33$ and 14.29) 
than the latest-type L dwarfs ($J \sim 14.7$).  The same holds true for the
T2 secondary of SDSS 0423-0414 ($J = 14.38$) and the T1 primary
of Epsilon Indi B ($J = 14.30$).
Indeed, {\em all} of the resolved components spanning types T1 to T5, have nearly identical absolute $J$-band magnitudes.

Does this mean that the $J$-band bump is largely an artifact of multiplicity?
Possibly, but only if the T2 SDSS 1254-0122,
the T3.5 SDSS 1750+1759 (both unresolved in this study) and the T4.5 
2MASS J05591914-1404488
\citep[unresolved in \citet{me03hst}]{me00c} are all closely-separated multiples.
This is not out of the realm of possibility, for as discussed in $\S$6.2
the separations of brown dwarf binaries likely extend below
imaging resolution limits. One or all of these 
systems may also have been imaged at an unfortunate orbital angle, 
as was the case initially for Kelu~1 \citep{mrt99c,liu05,gel06}.
Furthermore, the fact that SDSS 1254-0122 and 2MASS 0559-1404
are $\sim$0.5 and $\sim$0.8 mag brighter at $J$-band than
the T2 and T5 secondaries of SDSS 0423-0414 and SDSS 1021-0304, 
respectively, suggests that the former are unresolved multiples.
If the primaries of these hypothetical systems 
are constrained to have $M_J \approx 14.3$, then the secondaries of SDSS 1254-0122
and SDSS 1750+1759 would have $M_J \approx$ 15.3-15.5 and be $\sim$T6 dwarfs.
2MASS 0559-1404 would be required to have near
equal-magnitudes components or be a higher multiple system. 
High resolution radial velocity monitoring observations are needed to test
these possibilities.

It is important to point out that 
the absolute $J$-band magnitudes of the early-type T dwarf 
resolved binary components examined here are still $\sim$0.4 mag
brighter than the latest-type L dwarfs (but $\sim$0.5 mag fainter at $K$-band).
Hence, some broad-band brightening may still be present across the L/T transition.
Furthermore, even if the $J$-band ``bump'' is a more modest ``plateau'', 
there remains a significant ($\sim$30\%) brightening
at 1.05 and 1.27 $\micron$ due to flux redistribution within the $J$-band spectral region,
a feature not yet reproduced self-consistently by
current atmosphere models.


\subsection{The Origin of $J$-band Brightening}

That the observed brightening is concentrated in the 
1.05 and 1.27~$\micron$ flux peaks is an important clue to its origin.
The photospheric atomic and molecular gas opacities of low-temperature
brown dwarfs show distinct minima at these wavelengths,
shaped by strong {\water} and {\meth} bands and bracketed by
pressure-broadened \ion{K}{1} at shorter wavelengths 
and collision-induced H$_2$ absorption at longer wavelengths.
Condensate opacities, for grain sizes ($\sim$40-80~$\micron$) 
computed in a self-consistent manner \citep{ack01}, are roughly constant
across the near infrared band.  In the L dwarf regime, condensates
are a dominant source of opacity at the $J$-, $H$- and $K$-band 
spectral peaks.  However, theoretical atmosphere models incorporating
condensate clouds indicate
that the photospheric opacity from these species 
are dominant only in the 1.05 and 1.27~$\micron$
flux peaks for {\teff} $\lesssim$ 1300-1500~K; i.e., at the L/T transition
(cf.\ Figure 16 in Burrows, Sudarsky \& Hubeny 2006).
If these condensates are suddenly removed, the total opacity at these 
wavelengths decreases, allowing brighter emission from deeper and hotter
layers.  

This is the underlying thesis for dynamical atmospheric
explanations for the $J$-band brightening \citep{me02b,kna04}.
However, one must also consider whether 
higher gas opacities at longer wavelengths,
with the increased photospheric abundances of {\water} and {\meth}
molecules below {\teff} $\approx$ 1300-1800~K \citep{bur99,lod02}
and stronger H$_2$ absorption, might
lead to a redistribution of flux into the 1.05 and 1.27~$\micron$ flux peaks.
Current cloud models that assume constant 
sedimentation efficiency \citep{mar02} or particle
size distributions \citep{bur06} do not show this to be the case.
Hence, a dynamic mechanism for clearing out photospheric condensate
dust may still be necessary to explain the evolution of brown
dwarf atmospheres across the L/T transition.

\subsection{The Frequency of L/T Binaries}

Is there evidence from the distribution of binary frequency as a function
of spectral type that binaries play a special role through the L/T
transition?  To address this, we
have compiled results from high resolution imaging
of L and T dwarfs by \citet{koe99,mrt99,rei01,rei06,clo03,bou03,me03hst,giz03}; and
this study.\footnote{We also include the recently identified 
Kelu~1 binary \citep{liu05,gel06}, a target of prior searches.}
We did not include individual binary discoveries made serendipitously 
\citep{got02,pot02,mcc04,me05gl337cd} or those identified 
as part of as yet unpublished surveys \citep{liu06,stu06} in order
to make a fair assessment of the observed binary fraction.
Care was taken to identify duplicate
sources in each of the imaging studies, and classifications were verified
through published optical (L dwarfs, on the \citet{kir99} scheme) 
and near infrared data (T dwarfs, on the \citet{me06a} scheme).
Only those resolved pairs
that had a high probability of companionship, based on either
common proper motion confirmation, resolved spectroscopy and/or photometric
colors,
or very low probability of coincidence with an unrelated background source,
were considered as bona-fide binaries.
The complete sample includes 129 L dwarfs
and 34 T dwarfs, of which 33 are binary.

Figure~\ref{fig_ltbinfrac} plots the observed binary fraction of these sources
as a function of spectral type, binned by individual subclasses and
into subclass groups of L0-L2 (62 sources), L2.5-L4.5 (28 sources), 
L5-L6.5 (27 sources), L7-L9.5 (12 sources), T0-T3.5 (7 sources),
T4-T5.5 (12 sources) and T6-T8 (15 sources).
Note that these fractions have not been corrected for 
selection bias (resulting in an overestimate from equal-brightness systems)
or sensitivity/resolution limits (resulting in an underestimate by missing
closely separated or low mass ratio systems).  
This sample may also be subject to more subtle biases, such as the smaller
typical distances of later-type, intrinsically fainter brown dwarfs, resulting
in greater linear resolution for these objects (although this effect may be offset
by the apparent decrease in separations for lower-mass brown dwarfs).
It nevertheless serves to illustrate possible trends.

There
is clearly significant structure in the binary fraction distribution
for individual
subclasses, although this could be attributed 
to small number statistics.  
By binning the subclasses (reducing
statistical uncertainties), a remarkable
result emerges.  For most of the sample, binary
fractions are consistent with the overall fraction, 
$\epsilon_b^{obs}$ = 20$\pm$4\%.
Yet the L7-L9.5 and T0-T3.5 subclass groups -- the L/T transition objects ---
have fractions that are twice as high, 
42$_{-10}^{+12}$\% combining all 19 systems in this spectral 
type range.  This deviation is significant at the 98\% confidence level
compared to the sample mean.

Why would the observed binary fraction of L/T transition objects
be so high?  We posit the following scenario.  
Analysis of the 
SDSS 0423-0414 and SDSS 1021-0304 components, and prior results from
\citet{kir00,me02a,dah02,vrb04} and \citet{gol04}, all indicate that the L/T transition
spans a relatively narrow range of effective 
temperatures, $\Delta${\teff} $\approx$ 200--300~K.
However, the cooling rate of brown dwarfs is largely insensitive to 
changes in the photospheric opacity \citep{cha00}, such as the removal
of condensates or emergence of {\meth} absorption. Brown dwarfs
must therefore progress through the L/T transition relatively rapidly, implying
fewer such sources per spectral subtype for a given field sample.  
On the other hand, the analysis of $\S$5.3, and similar results
by \citet{cru04,me0423,rei2252}; and \citet{liu06}, all demonstrate
that early-type T dwarf spectral
features can be reproduced from the combined light 
of a late-type L and mid-type T dwarf binary.  
It is therefore possible that such hybrid binaries, if unrecognized, 
could significantly contaminate a spectral sample 
of early-type T dwarfs.  

To illustrate this point,
consider the following example.  
Assuming that L5-L8 dwarfs have 
{\teff} $\approx$ 1700-1300~K and
L8-T5 dwarfs have {\teff} $\approx$ 1300-1100~K 
(Table 5 and \citet{gol04}), 
the mass function simulations of \citet{me04mfxn} 
predict a relative space density
of $N_{L/T}/N_{L} \approx 0.9$ 
between these two groups, largely independent of the shape
of the underlying mass function.  
However, 
because mid- to late-type L dwarfs are roughly twice as bright
as L/T transition objects, the relative number observed in a 
magnitude-limited sample (the best approximation for current
imaging samples) is 
$(N_{L/T}/N_{L})^{obs}$ $\approx 0.3$.
Now consider that all brown dwarfs in a magnitude-limited
sample have a resolvable binary
fraction of $\sim$25\%, and that 20\% of all late-type L dwarf
binaries have T dwarf secondaries (this is roughly 
consistent with the mass ratio
distribution of Figure~\ref{fig_qdist}).  
These binaries would
exhibit a combined light
spectrum similar to a late-type L/early-type T dwarf, 
and would be identified as such in an unresolved spectroscopic sample.
Hence, the observed binary fraction among late-type L dwarfs
in this scenario would be $\sim$20\%, while the fraction
of L/T transition binaries would be
\begin{equation}
\epsilon_{L/T}^{obs} = \frac{0.2{\times}0.25N_L+0.25N_{L/T}}{0.2{\times}0.25N_L+N_{L/T}} = \frac{0.05+0.075}{0.05+0.3} \approx 36\%;
\end{equation}
i.e., nearly twice the underlying binary fraction. 
This numerical example serves to illustrate that the binary
hypothesis provides both a qualitative and quantitative
explanation for the peak in the binary fraction of L/T transition objects.
More complete modeling of this effect will be presented in a 
forthcoming publication.

We therefore conclude that multiplicity does play an important role in the L/T 
transition, contaminating samples of ``true'' transition objects and leading
to a greater $J$-band brightening than that inferred for resolved
systems.  These binaries also provide a detailed 
and intriguing picture of this still poorly-understood transition,
and a list of all currently known L/T binaries 
is given in Table~\ref{tab_ltbinaries}.
Further study of these source will provide improved
understanding of the physical mechanisms governing this transition,
including the depletion of photospheric condensates, the emergence
of {\meth} gas and the possible role of atmospheric dynamics
in brown dwarf spectral evolution.

\section{Summary}

We have identified 5 binaries in of a sample of 22 T dwarfs imaged
with {\em HST} NICMOS.  
Of these, three are well-resolved, permitting determination
of their component spectral types, relative
bolometric luminosities and {\teff}s, and systemic mass ratios.
The identification of 2MASS 0518-2828 as a closely-separated
binary confirms previous suspicions of multiplicity 
based on this object's unusual near infrared spectrum.
The bias-corrected resolved
binary fraction of this sample ($\epsilon_b = 11_{-3}^{+7}$\%), 
the near-unity mass ratios of the components of these 
systems ($q \gtrsim 0.7$) and their small
projected separations ($\rho \lesssim 5$~AU) are all consistent 
with previously identified trends amongst VLM dwarfs,
indicating that they are salient properties of brown dwarf field binaries.

Three of the binaries in our sample, SDSS 0423-0414, 2MASS 0518-2828 and 
SDSS 1021-0304, are composed of sources spanning the L to T transition,
and spectral decomposition analysis of SDSS 1021-0304
reveals that its T5 secondary is 25--30\% brighter at 1.05 and
1.27~$\micron$ than its T1 primary, despite being 35\% less luminous overall.
The properties of these sources, as well as the recently
discovered binary SDSS 1534+1615, indicate that
the $J$-band brightening previously
observed amongst late-type L and mid-type field T dwarfs
is an intrinsic feature
of the L/T transition and not the result of age, surface gravity
or metallicity effects.
In support of the results of \citet{bur06} and \citet{liu06}, we find
that the $J$-band bump may be more of a $J$-band ``plateau'',
with T1-T5 dwarfs having $M_J \approx 14.3$, enhanced by the
presence of unresolved binaries in this spectral type range.  Indeed,
we find that the frequency of L/T transition binaries
is twice as high as those of all other L and T dwarfs, a statistically
significant deviation that can be explained if spectroscopic samples
of L/T transition objects are significantly contaminated by
binaries composed of earlier-type and later-type components.
Taken together, the properties of L/T binary systems
provide the most conclusive evidence to date that
the L/T transition occurs relatively rapidly,
driven by the removal of photospheric
condensates that is likely to be facilitated by dynamic atmospheric processes.  
Further parallax and multiplicity measurements
will better 
constrain the flux evolution and relative numbers
of L/T transition objects, important constraints for understanding
the physical mechanism of photospheric condensate depletion 
and the atmospheric evolution of brown dwarfs
as they cool below {\teff} $\approx$ 1500~K.

\acknowledgments

The authors would like to thank Santiago Arribas and
Patricia Royle at STScI for their technical support of {\em HST} 
program GO-9833, and David Sprayberry and Meg Whittle at Keck for 
their support during the NIRC observations of 2MASS 1553+1532.
A.~J.~B.\ acknowledges useful discussions with Antonin Bouchez,
Randy Campbell, Michael Liu (who gave our
original manuscript a careful read), Mark McCaughrean and Frederick Vrba. 
We also thank our referee, Brian Patten, for his prompt critique of our 
manuscript and helpful suggestions.
K.~L.~C.\ acknowledges support by a National Science Foundation 
Astronomy and Astrophysics
Postdoctoral Fellowship under AST-0401418.
A.~B.\ acknowledges support under NASA grant
NNG04GL22G and through the NASA Astrobiology
Institute under Cooperative Agreement No.\ CAN-02-OSS-02
issued through the Office of Space Science.
This work is based in part on observations made with the
NASA/ESA Hubble Space Telescope, obtained at the Space Telescope
Science Institute, which is operated by the Association of
Universities for Research in Astronomy, Inc., under NASA contract NAS 5-26555.
These observations are associated with proposals GO-9833 and GO-10247.
This publication makes use of data from the Two Micron All Sky Survey, 
which is a
joint project of the University of Massachusetts and the Infrared
Processing and Analysis Center, and funded by the National
Aeronautics and Space Administration and the National Science
Foundation. 2MASS data were obtained from the NASA/IPAC Infrared
Science Archive, which is operated by the Jet Propulsion
Laboratory, California Institute of Technology, under contract
with the National Aeronautics and Space Administration.
This publication has benefitted from the M, L, and T 
dwarf compendium housed at DwarfArchives.org and 
maintained by Chris Gelino, Davy Kirkpatrick, and Adam Burgasser;
and the VLM Binary Archive maintained by N.\ Siegler at 
\url{http://paperclip.as.arizona.edu/$\sim$nsiegler/VLM\_binaries/}. 
The authors wish to recognize and acknowledge the 
very significant cultural role and reverence that 
the summit of Mauna Kea has always had within the 
indigenous Hawaiian community.  We are most fortunate 
to have the opportunity to conduct observations from this mountain.

Facilities: \facility{Hubble Space Telescope(NICMOS)}; 
\facility{IRTF(SpeX)}; \facility{Keck(NIRC)}

\clearpage
\pagestyle{empty}
\begin{deluxetable}{llcccccccccl}
\tabletypesize{\scriptsize}
\tablecaption{T Dwarf Targets \label{tab_allt}}
\tablewidth{0pt}
\tablehead{
 & & \multicolumn{3}{c}{J2000 Coordinates\tablenotemark{b}} &
 \multicolumn{3}{c}{2MASS Photometry} \\
\cline{3-5} \cline{6-8}
\colhead{Name} &
\colhead{SpT\tablenotemark{a}} &
\colhead{$\alpha$} &
\colhead{$\delta$} &
\colhead{Epoch} &
\colhead{$J$} &
\colhead{$H$} &
\colhead{$K_s$} &
\colhead{$\pi$} &
\colhead{$\mu$} &
\colhead{$\phi$} &
\colhead{Ref\tablenotemark{c}} \\
 & & & & &
\colhead{(mag)} &
\colhead{(mag)} &
\colhead{(mag)} &
\colhead{(arcs)} &
\colhead{(arcs yr$^{-1}$)} &
\colhead{($\degr$)} \\
\colhead{(1)} &
\colhead{(2)} &
\colhead{(3)} &
\colhead{(4)} &
\colhead{(5)} &
\colhead{(6)} &
\colhead{(7)} &
\colhead{(8)} &
\colhead{(9)} &
\colhead{(10)} &
\colhead{(11)} &
\colhead{(12)} \\
}
\startdata
SDSS J015141.69+124429.6 & T1 & 01$^h$51$^m$41$\fs$55 & +12$\degr$44$\arcmin$30$\farcs$0 & 1997.70 & 16.57$\pm$0.13 & 15.60$\pm$0.11 & 15.18$\pm$0.19 & 0.047$\pm$0.003 & 0.743$\pm$0.004 &  93 & {\bf 1},9 \\
SDSS J020742.48+000056.2 & T4.5 & 02$^h$07$^m$42$\fs$84 & +00$\degr$00$\arcmin$56$\farcs$4 & 2000.63 & 16.63$\pm$0.05\tablenotemark{d} & 16.66$\pm$0.05\tablenotemark{d} & 16.62$\pm$0.05\tablenotemark{d} & 0.035$\pm$0.010 & 0.156$\pm$0.011 &  96 & {\bf 1},9 \\
2MASS J02431371$-$2453298 & T6 & 02$^h$43$^m$13$\fs$71 & $-$24$\degr$53$\arcmin$29$\farcs$8 & 1998.87 & 15.38$\pm$0.05 & 15.14$\pm$0.11 & 15.22$\pm$0.17 & 0.094$\pm$0.004 & 0.355$\pm$0.004 & 234 & {\bf 2},9 \\
2MASS J03480772$-$6022270 & T7 & 03$^h$48$^m$07$\fs$72 & $-$60$\degr$22$\arcmin$27$\farcs$0 & 1999.89 & 15.32$\pm$0.05 & 15.56$\pm$0.14 & 15.60$\pm$0.23 & \nodata & 0.77$\pm$0.04 & 201 & {\bf 3} \\
2MASS J04151954$-$0935066 & T8 & 04$^h$15$^m$19$\fs$54 & $-$09$\degr$35$\arcmin$06$\farcs$6 & 1998.87 & 15.70$\pm$0.06 & 15.54$\pm$0.11 & 15.43$\pm$0.20 & 0.174$\pm$0.003 & 2.255$\pm$0.003 &  76 & {\bf 2},9 \\
SDSS J042348.57$-$041403.5 & T0 & 04$^h$23$^m$48$\fs$58 & $-$04$\degr$14$\arcmin$03$\farcs$5 & 1998.73 & 14.47$\pm$0.03 & 13.46$\pm$0.04 & 12.93$\pm$0.03 & 0.0659$\pm$0.0017 & 0.333$\pm$0.003 & 284 & {\bf 1},9 \\
2MASS J05160945$-$0445499 & T5.5 & 05$^h$16$^m$09$\fs$45 & $-$04$\degr$45$\arcmin$49$\farcs$9 & 1998.72 & 15.98$\pm$0.08 & 15.72$\pm$0.17 & 15.49$\pm$0.20 & \nodata & 0.34$\pm$0.03 & 232 & {\bf 3} \\
2MASS J05185995$-$2828372 & T1p & 05$^h$18$^m$59$\fs$95 & $-$28$\degr$28$\arcmin$37$\farcs$2 & 1999.01 & 15.98$\pm$0.10 & 14.83$\pm$0.07 & 14.16$\pm$0.07 & \nodata & \nodata & \nodata & {\bf 4} \\
2MASS J07271824+1710012 & T7 & 07$^h$27$^m$18$\fs$24 & +17$\degr$10$\arcmin$01$\farcs$2 & 1997.83 & 15.60$\pm$0.06 & 15.76$\pm$0.17 & 15.56$\pm$0.19 & 0.110$\pm$0.002 & 1.297$\pm$0.005 & 126 & {\bf 2},9 \\
2MASS J07554795+2212169 & T5 & 07$^h$55$^m$47$\fs$95 & +22$\degr$12$\arcmin$16$\farcs$9 & 1998.83 & 15.73$\pm$0.06 & 15.67$\pm$0.15 & 15.75$\pm$0.21 & \nodata & \nodata & \nodata & {\bf 2} \\
SDSS J083717.22$-$000018.3 & T1 & 08$^h$37$^m$17$\fs$21 & $-$00$\degr$00$\arcmin$18$\farcs$0 & 2000.11 & 16.90$\pm$0.05\tablenotemark{d} & 16.21$\pm$0.05\tablenotemark{d} & 15.98$\pm$0.05\tablenotemark{d} & 0.034$\pm$0.014 & 0.173$\pm$0.017 & 185 & {\bf 5},9 \\
SDSS J092615.38+584720.9 & T4.5 & 09$^h$26$^m$15$\fs$37 & +58$\degr$47$\arcmin$21$\farcs$2 & 2000.22 & 15.90$\pm$0.07 & 15.31$\pm$0.10 & 15.45$\pm$0.19 & \nodata & $<$ 0.3 & \nodata & {\bf 1},7 \\
SDSS J102109.69$-$030420.1 & T3 & 10$^h$21$^m$09$\fs$69 & $-$03$\degr$04$\arcmin$19$\farcs$7 & 1998.94 & 16.25$\pm$0.09 & 15.35$\pm$0.10 & 15.13$\pm$0.17 & 0.034$\pm$0.005 & 0.183$\pm$0.003 & 249 & {\bf 5},10 \\
SDSS J111010.01+011613.1 & T5.5 & 11$^h$10$^m$10$\fs$01 & +01$\degr$16$\arcmin$13$\farcs$0 & 2000.12 & 16.34$\pm$0.12 & 15.92$\pm$0.14 & $>$ 15.1 & \nodata & 0.34$\pm$0.10 & 110 & {\bf 1},11 \\
2MASS J1217110$-$0311131 & T7.5 & 12$^h$17$^m$11$\fs$10 & $-$03$\degr$11$\arcmin$13$\farcs$1 & 1999.08 & 15.86$\pm$0.06 & 15.75$\pm$0.12 & $>$ 15.9 & 0.091$\pm$0.002 & 1.0571$\pm$0.0017 & 274 & {\bf 6},10 \\
SDSS J125453.90$-$012247.4 & T2 & 12$^h$54$^m$53$\fs$93 & $-$01$\degr$22$\arcmin$47$\farcs$4 & 1999.07 & 14.89$\pm$0.04 & 14.09$\pm$0.03 & 13.84$\pm$0.05 & 0.0732$\pm$0.0019 & 0.491$\pm$0.003 & 285 & {\bf 5},10 \\
2MASS J15031961+2525196 & T5 & 15$^h$03$^m$19$\fs$61 & +25$\degr$25$\arcmin$19$\farcs$6 & 1999.39 & 13.94$\pm$0.02 & 13.86$\pm$0.03 & 13.96$\pm$0.06 & \nodata & \nodata & \nodata & {\bf 7} \\
2MASS J15530228+1532369 & T7 & 15$^h$53$^m$02$\fs$28 & +15$\degr$32$\arcmin$36$\farcs$9 & 1998.15 & 15.83$\pm$0.07 & 15.94$\pm$0.16 & 15.51$\pm$0.18 & \nodata & \nodata & \nodata & {\bf 2} \\
SDSS J162414.37+002915.6 & T6 & 16$^h$24$^m$14$\fs$36 & +00$\degr$29$\arcmin$15$\farcs$8 & 1999.31 & 15.49$\pm$0.05 & 15.52$\pm$0.10 & $>$ 15.5 & 0.092$\pm$0.002 & 0.3832$\pm$0.0019 & 270 & {\bf 8},12 \\
SDSS J175032.96+175903.9 & T3.5 & 17$^h$50$^m$32$\fs$93 & +17$\degr$59$\arcmin$04$\farcs$2 & 1999.23 & 16.34$\pm$0.10 & 15.95$\pm$0.13 & 15.48$\pm$0.19 & 0.036$\pm$0.005 & 0.204$\pm$0.008 &  61 & {\bf 1},9 \\
2MASS J22282889$-$4310262 & T6 & 22$^h$28$^m$28$\fs$89 & $-$43$\degr$10$\arcmin$26$\farcs$2 & 1998.90 & 15.66$\pm$0.07 & 15.36$\pm$0.12 & 15.30$\pm$0.21 & \nodata & 0.31$\pm$0.03 & 175 & {\bf 3} \\
2MASS J22541892+3123498 & T4 & 22$^h$54$^m$18$\fs$92 & +31$\degr$23$\arcmin$49$\farcs$8 & 1998.48 & 15.26$\pm$0.05 & 15.02$\pm$0.08 & 14.90$\pm$0.15 & \nodata & \nodata & \nodata & {\bf 2} \\
2MASS J23391025+1352284 & T5 & 23$^h$39$^m$10$\fs$25 & +13$\degr$52$\arcmin$28$\farcs$4 & 2000.91 & 16.24$\pm$0.11 & 15.82$\pm$0.15 & 16.15$\pm$0.31 & \nodata & 0.83$\pm$0.11 & 159 & {\bf 2},7 \\
\enddata
\tablenotetext{a}{Near infrared spectral types from \citet{me06a}.}
\tablenotetext{b}{Coordinates from the
2MASS All Sky Point Source Catalog \citep{skr06}.}
\tablenotetext{c}{Discovery reference in boldface type, followed by references for additional photometric
and astrometric data.}
\tablenotetext{d}{MKO $JHK$ from \citet{leg02} or \citet{kna04}.}
\tablerefs{
(1) \citet{geb02}; (2) \citet{me02a}; (3) \citet{me03wide2};
(4) \citet{cru04}; (5) \citet{leg00}; (6) \citet{me99}; 
(7) \citet{me03wide1}; (8) \citet{str99}; (9) \citet{vrb04};
(10) \citet{tin03}; (11) \citet{tin05}; (12) \citet{dah02}.}
\end{deluxetable}
\clearpage

\begin{deluxetable}{llccccccc}
\tabletypesize{\scriptsize}
\tablecaption{Log of {\em HST} Observations for Programs GO-9833 and GO-10247.}
\tablewidth{0pt}
\tablehead{
 & & \multicolumn{2}{c}{F090M} &
\multicolumn{2}{c}{F110W} &
\multicolumn{2}{c}{F170M} & \\
\cline{3-4} \cline{5-6}  \cline{7-8}
\colhead{Object} &
\colhead{UT Date} &
\colhead{$t$} &
\colhead{$m_{lim}$\tablenotemark{a}} &
\colhead{$t$} &
\colhead{$m_{lim}$\tablenotemark{a}} &
\colhead{$t$} &
\colhead{$m_{lim}$\tablenotemark{a}} &
\colhead{Roll Angle\tablenotemark{b}} \\
\colhead{} &
\colhead{} &
\colhead{(s)} &
\colhead{(mag)} &
\colhead{(s)} &
\colhead{(mag)} &
\colhead{(s)} &
\colhead{(mag)} &
\colhead{($\degr$)} \\
\colhead{(1)} &
\colhead{(2)} &
\colhead{(3)} &
\colhead{(4)} &
\colhead{(5)} &
\colhead{(6)} &
\colhead{(7)} &
\colhead{(8)} &
\colhead{(9)} \\
}
\startdata
SDSS 0151+1244 & 2003 Nov 3 & 56  & 19.0 &  864 & 22.9 & 1519 & 21.2 & 212.0 \\
SDSS 0207+0000 & 2004 Feb 3 & 48  & 18.4  &  864 & 22.1 & 1519 & 20.9 & 208.1 \\
2MASS 0243$-$2453 & 2004 Feb 15 & 56 & 19.5 & 864 & 23.6 & 1519 & 21.5 & 220.6 \\
2MASS 0348$-$6022 & 2004 May 26 & 88 & 20.2 & 1599 & 22.7 & 3071 & 21.4 & 321.1 \\
2MASS 0415$-$0935 & 2004 Feb 20 & 56 & 19.0 &  864 & 22.6 & 1519 & 21.1 & 214.1 \\
SDSS 0423$-$0414 & 2004 Jul 22 & 72  & 19.3 &  816 & 22.6 & 1519 & 21.3 & 15.3 \\
2MASS 0516$-$0445 & 2004 Jul 25 & 48  & 18.6 &  864 & 22.7 & 1519 & 21.1 & 15.9 \\
2MASS 0518$-$2828\tablenotemark{c} & 2004 Sep 7 & 960  & 21.0 &  64 & 19.9 & 416 & 19.9 & 38.9 \\
2MASS 0727+1710 & 2004 Mar 27 & 56  & 19.5 &  864 & 23.6 & 1519 & 21.1 & 232.7 \\
2MASS 0755+2212 & 2004 Mar 29 & 56  & 18.6 &  864 & 22.4 & 1519 & 20.9 & 236.9 \\
SDSS 0837$-$0000 &  2004 Feb 18 & 48  & 18.6 &  864 & 23.3 & 1519 & 21.7 & 196.5 \\
SDSS 0926+5847 &  2004 Feb 5 & 48  & 19.1 &  960 & 22.7 & 1600 & 21.2 & 319.5 \\
SDSS 1021$-$0304 &  2004 May 22 & 48  & 18.9 &  864 & 22.9 & 1519 & 21.0 & 245.9 \\
SDSS 1110+0116 & 2004 May 22 & 48  & 18.9 &  864 & 23.3 & 1519 & 21.5 & 247.1 \\
2MASS 1217$-$0311 & 2004 Apr 26 & 48 & 18.7 &  864 & 22.7 & 1519 & 21.3 & 255.7 \\
SDSS 1254$-$0122 & 2004 Feb 13 & \nodata & \nodata &  896 & 22.7 & 1519 & 21.1 & 64.9 \\
2MASS 1503+2525 & 2003 Sep 9 &  80  & 18.9 &  791 & 21.9 & 1519 & 20.7 & 216.0 \\
2MASS 1553+1532 & 2003 Sep 7 & 56  & 19.5 &  864 & 23.5 & 1519 & 21.7 & 226.9 \\
SDSS 1624+0029 & 2003 Sep 9 & 48  & 18.8 &  864 & 22.7 & 1519 & 20.9 & 230.8 \\
SDSS 1750+1759 & 2003 Sep 12 & 56  & 18.6 &  864 & 22.1 & 1519 & 20.9 & 231.8 \\
2MASS 2228$-$4310 & 2004 May 26 & 72  & 19.2 &  912 & 23.0 & 1519 & 21.3 & 37.3 \\
2MASS 2254+3123 & 2003 Sep 15 & 64  & 19.4 &  864 & 22.8 & 1519 & 21.2 & 299.7 \\
2MASS 2339+1352 & 2003 Nov 1 & 56  & 18.8 &  864 & 22.6 & 1519 & 20.9 & 218.5 \\
\enddata
\tablenotetext{a}{{L}imiting magnitude estimated for a 5$\sigma$ flux peak detection.}
\tablenotetext{b}{Telescope roll angle, East from North.}
\tablenotetext{c}{Also observed at F145M (320 s) and F160W (160 s).}
\end{deluxetable}

\begin{deluxetable}{cccc}
\tabletypesize{\scriptsize}
\tablecaption{NICMOS NIC1 Aperture Corrections. \label{tab:apcor}}
\tablewidth{0pt}
\tablehead{
\colhead{Aperture} & & & \\
\colhead{Radius} &
\colhead{F090M\tablenotemark{a}} &
\colhead{F110W} &
\colhead{F170M}  \\
\colhead{(pixels)} &
\colhead{(mag)} &
\colhead{(mag)} &
\colhead{(mag)} \\
}
\startdata
2.0 & -0.71$\pm$0.05 & -0.79$\pm$0.02 & -1.096$\pm$0.009 \\ 
2.5 & -0.60$\pm$0.04 & -0.695$\pm$0.017 & -0.869$\pm$0.012 \\ 
3.0 & -0.47$\pm$0.04 & -0.637$\pm$0.014 & -0.757$\pm$0.008 \\ 
3.5 & -0.35$\pm$0.04 & -0.557$\pm$0.011 & -0.714$\pm$0.008 \\ 
4.0 & -0.26$\pm$0.04 & -0.451$\pm$0.008 & -0.694$\pm$0.007 \\ 
4.5 & -0.22$\pm$0.04 & -0.349$\pm$0.007 & -0.658$\pm$0.007 \\ 
5.0\tablenotemark{b} & -0.20$\pm$0.05 & -0.275$\pm$0.005 & -0.593$\pm$0.006 \\ 
5.5 & -0.18$\pm$0.05 & -0.232$\pm$0.005 & -0.503$\pm$0.006 \\ 
6.0 & -0.17$\pm$0.04 & -0.211$\pm$0.004 & -0.410$\pm$0.006 \\ 
6.5 & -0.16$\pm$0.04 & -0.199$\pm$0.003 & -0.333$\pm$0.006 \\ 
7.0 & -0.15$\pm$0.03 & -0.190$\pm$0.003 & -0.279$\pm$0.005 \\ 
7.5 & -0.14$\pm$0.03 & -0.181$\pm$0.002 & -0.247$\pm$0.004 \\ 
8.0 & -0.14$\pm$0.03 & -0.1736$\pm$0.0019 & -0.232$\pm$0.004 \\ 
8.5 & -0.14$\pm$0.04 & -0.1681$\pm$0.0015 & -0.225$\pm$0.004 \\ 
9.0 & -0.13$\pm$0.04 & -0.1638$\pm$0.0012 & -0.219$\pm$0.003 \\ 
9.5 & -0.12$\pm$0.04 & -0.1597$\pm$0.0010 & -0.212$\pm$0.002 \\ 
10.0 & -0.12$\pm$0.04 & -0.1549$\pm$0.0007 & -0.2039$\pm$0.0017 \\ 
10.5 & -0.11$\pm$0.02 & -0.1498$\pm$0.0006 & -0.1969$\pm$0.0013 \\ 
11.0 & -0.115$\pm$0.012 & -0.1448$\pm$0.0006 & -0.1919$\pm$0.0009 \\ 
\enddata
\tablecomments{Values include corrections from a
reference aperture (11.5 pixels) to an infinitely sized aperture 
of -0.1136, -0.1393 and -0.1888
mag as given in \citet{sch05}.}
\tablenotetext{a}{Measured from observations of 2MASS 0518-2828.}
\tablenotetext{b}{Adopted aperture radius for single sources.}
\end{deluxetable}

\begin{deluxetable}{llcccc}
\tabletypesize{\scriptsize}
\tablecaption{NICMOS Photometry.}
\tablewidth{0pt}
\tablehead{
\colhead{Object} &
\colhead{SpT} &
\colhead{F090M} &
\colhead{F110W} &
\colhead{F170M} &
\colhead{F110W-F170M} \\
 & & \colhead{(mag)} &
\colhead{(mag)} &
\colhead{(mag)} &
\colhead{(mag)} \\
\colhead{(1)} &
\colhead{(2)} &
\colhead{(3)} &
\colhead{(4)} &
\colhead{(5)} &
\colhead{(6)} \\
}
\startdata
SDSS 0151+1244 & T1 & 18.53$\pm$0.09 & 17.26$\pm$0.05 & 15.54$\pm$0.05 & 1.72$\pm$0.04 \\
SDSS 0207+0000 & T4.5 & 19.9$\pm$0.2 & 17.84$\pm$0.05 & 16.87$\pm$0.05 & 0.97$\pm$0.04 \\
2MASS 0243$-$2453 &  T6 & 19.5$\pm$0.2 & 16.23$\pm$0.05 & 15.74$\pm$0.06 & 0.49$\pm$0.04 \\
2MASS 0348$-$6022 & T7 &  18.51$\pm$0.09 & 16.11$\pm$0.06 & 16.16$\pm$0.05 & -0.05$\pm$0.05 \\
2MASS 0415$-$0935 & T8 &  19.04$\pm$0.10 & 16.47$\pm$0.05 & 16.67$\pm$0.06 & -0.20$\pm$0.04 \\
SDSS 0423$-$0414 & T0 &  16.68$\pm$0.10\tablenotemark{a} & 15.28$\pm$0.05\tablenotemark{a} & 13.62$\pm$0.05\tablenotemark{a} & 1.66$\pm$0.04\tablenotemark{a} \\
2MASS 0516$-$0445 & T5.5 & 18.96$\pm$0.09 & 16.66$\pm$0.05 & 16.14$\pm$0.05 & 0.52$\pm$0.03 \\
2MASS 0518$-$2828\tablenotemark{b} & T1p & 18.55$\pm$0.08 & 16.69$\pm$0.07 & 14.92$\pm$0.05 & 1.77$\pm$0.06 \\
2MASS 0755+2212 & T5 & 18.89$\pm$0.09 & 16.58$\pm$0.05 & 16.08$\pm$0.05 & 0.50$\pm$0.03 \\
SDSS 0837$-$0000 &  T1 & $>$18.6 & 17.91$\pm$0.05 & 16.21$\pm$0.05 & 1.70$\pm$0.04 \\
SDSS 0926+5847 & T4.5 & 18.66$\pm$0.09 & 16.57$\pm$0.05 & 15.64$\pm$0.05 & 0.93$\pm$0.03 \\
SDSS 1021$-$0304 &  T3 & 19.34$\pm$0.15\tablenotemark{a} & 17.09$\pm$0.05\tablenotemark{a} & 15.83$\pm$0.05\tablenotemark{a} & 1.26$\pm$0.04\tablenotemark{a} \\
SDSS 1110+0116 & T5.5 & 19.15$\pm$0.12 & 17.27$\pm$0.05 & 16.49$\pm$0.05 & 0.78$\pm$0.04 \\
2MASS 1217$-$0311 & T7.5 &  18.91$\pm$0.09 & 16.73$\pm$0.05 & 16.77$\pm$0.05 & -0.04$\pm$0.04 \\
SDSS 1254$-$0122 & T2 &  \nodata & 15.66$\pm$0.05 & 14.10$\pm$0.05 & 1.56$\pm$0.03 \\
2MASS 1503+2525 & T5 &  16.84$\pm$0.08 & 14.65$\pm$0.05 & 14.18$\pm$0.05 & 0.47$\pm$0.03 \\
2MASS 1553+1532A & T7 &  19.16$\pm$0.16 & 17.16$\pm$0.05 & 17.01$\pm$0.05 & 0.15$\pm$0.04 \\
2MASS 1553+1532B & T7: &  $>$19.5 & 17.46$\pm$0.05 & 17.40$\pm$0.05 & 0.06$\pm$0.04 \\
SDSS 1624+0029 & T6 &  18.46$\pm$0.09 & 16.28$\pm$0.05 & 15.96$\pm$0.05 & 0.32$\pm$0.03 \\
SDSS 1750+1759 & T3.5 &  18.80$\pm$0.09 & 17.18$\pm$0.05 & 16.04$\pm$0.05 & 1.14$\pm$0.04 \\
2MASS 2228$-$4310 & T6 &  18.77$\pm$0.09 & 16.36$\pm$0.05 & 16.05$\pm$0.05 & 0.31$\pm$0.03 \\
2MASS 2254+3123 & T4 &  18.12$\pm$0.10 & 16.05$\pm$0.05 & 15.09$\pm$0.05 & 0.96$\pm$0.04 \\
2MASS 2339+1352 & T5 &  18.86$\pm$0.09 & 16.91$\pm$0.05 & 16.27$\pm$0.05 & 0.64$\pm$0.03 \\
\enddata
\tablecomments{Aperture photometry derived using an 5-pixel aperture and aperture corrections from Table~2, unless otherwise noted. 
Values are given on the Arizona Vega system ($M_{Vega}$ = 0.02).
Uncertainties include RMS scatter in count rates between individual exposures,
aperture correction uncertainties, 5\% flux calibration (3\% in F110W-F170M
color) and 1\% sensitivity variation (zeropoint drift).}
\tablenotetext{a}{Photometry measured using a 15-pixel aperture
and no aperture correction to incorporate both components.}
\tablenotetext{b}{Additional photometry for 2MASS 0518-2828: F145M = 15.86$\pm$0.05, F160W = 15.20$\pm$0.05.}
\end{deluxetable}

\begin{deluxetable}{lcccccc}
\tabletypesize{\scriptsize}
\tablecaption{Binary Properties. \label{tab_binparam}}
\tablewidth{0pt}
\tablehead{
\colhead{Parameter} &
\colhead{SDSS 0423-0414} &
\colhead{2MASS 0518-2828} &
\colhead{SDSS 0926+5847} &
\colhead{SDSS 1021-0304} &
\multicolumn{2}{c}{2MASS 1553+1532} \\
}
\startdata
UT Date &  2004 Jul 22 & 2004 Sep 7 & 2004 Feb 5 & 2004 May 22 & 2003 Sep 7 & 2000 Jul 22 \\
Instrument & {\em HST} NICMOS & {\em HST} NICMOS & {\em HST} NICMOS & {\em HST} NICMOS & {\em HST} NICMOS & Keck NIRC \\
$\rho$~($\arcsec$) & 0$\farcs$164$\pm$0$\farcs$005 & 0$\farcs$051$\pm$0$\farcs$012 & 0$\farcs$070$\pm$0$\farcs$006  & 0$\farcs$172$\pm$0$\farcs$005 & 0$\farcs$349$\pm$0$\farcs$005 & 0$\farcs$30$\pm$0$\farcs$02 \\
\phm{$\rho$}~(AU) & 2.49$\pm$0.07 & 1.8$\pm$0.5 & 2.6$\pm$0.5 & 5.0$\pm$0.7 & 4.2$\pm$0.7 & 3.6$\pm$0.7 \\
$\theta$ ($\degr$) & 19$\fdg$2$\pm$0$\fdg$8  & 189$\degr$$\pm$8$\degr$ & 132$\fdg$9$\pm$1$\fdg$9 & 244$\fdg$6$\pm$0$\fdg$8 & 189$\fdg$9$\pm$0$\fdg$4 & 199$\degr$$\pm$7$\degr$ \\
$\Delta$F090M & 0.88$\pm$0.03 & 1.6$\pm$0.4 & \nodata & 0.21$\pm$0.10 & \nodata & \nodata \\
$\Delta$F110W & 0.526$\pm$0.015 & 0.8$\pm$0.5\tablenotemark{a} & 0.4$\pm$0.2\tablenotemark{a} & 0.06$\pm$0.04 & 0.31$\pm$0.04 & \nodata \\
$\Delta$F170M & 0.820$\pm$0.013 & 0.9$\pm$0.6\tablenotemark{a} & 0.4$\pm$0.3\tablenotemark{a} & 1.030$\pm$0.019 & 0.461$\pm$0.016 & \nodata \\
F110W~(A) & 15.80$\pm$0.05 & 17.11$\pm$0.18 & 17.14$\pm$0.10 & 17.81$\pm$0.05 & 17.15$\pm$0.05 & \nodata \\
\phm{F110W}~(B) & 16.33$\pm$0.05 & 17.9$\pm$0.3 & 17.54$\pm$0.13 & 17.87$\pm$0.05 & 17.47$\pm$0.05 & \nodata \\
F170M~(A) & 14.04$\pm$0.05 & 15.31$\pm$0.19 & 16.21$\pm$0.13 & 16.19$\pm$0.05 & 16.98$\pm$0.05 & \nodata \\
\phm{F170W}~(B) & 14.86$\pm$0.05 & 16.2$\pm$0.4 & 16.61$\pm$0.18 & 17.22$\pm$0.05 & 17.44$\pm$0.05 & \nodata \\
F110W-F170M~(A) & 1.76$\pm$0.07 & 1.8$\pm$0.3 & 0.93$\pm$0.16 & 1.63$\pm$0.07 & 0.17$\pm$0.07 & \nodata \\
\phm{F110W-F170M}~(B) & 1.47$\pm$0.07 & 1.7$\pm$0.5 & 0.9$\pm$0.2 & 0.66$\pm$0.07 & 0.03$\pm$0.07 & \nodata \\
Est.\ SpT & L6.5+T2\tablenotemark{b} & L6:+T4:\tablenotemark{b} & T4:+T4: & T1+T5\tablenotemark{b} & \multicolumn{2}{c}{T6.5+T7} \\
${\Delta}M_{bol}$ & 0.72$\pm$0.13 & 0.9$\pm$0.6 & 0.4$\pm$0.2 & 0.49$\pm$0.13 & \multicolumn{2}{c}{0.31$\pm$0.12}  \\
$M_{bol}$~(A) & 15.77$\pm$0.16 & \nodata & \nodata & 16.5$\pm$0.7 &  \multicolumn{2}{c}{\nodata} \\
\phm{$M_{bol}$}~(B) & 16.50$\pm$0.16 & \nodata & \nodata & 17.0$\pm$0.7 &  \multicolumn{2}{c}{\nodata} \\
T$_2$/T$_1$ & 0.847$\pm$0.013 & 0.81$\pm$0.11 & 0.91$\pm$0.05 & 0.894$\pm$0.015 & \multicolumn{2}{c}{0.931$\pm$0.014} \\
{\teff}\tablenotemark{c} (K)~(A) & 1490$\pm$100 & $\sim$1600 & $\sim$1330 & 1260$\pm$210 & \multicolumn{2}{c}{$\sim$980} \\
\phm{{\teff}$^b$ (K)}~(B) & 1250$\pm$80 & $\sim$1330 & $\sim$1330 & 1130$\pm$190 & \multicolumn{2}{c}{$\sim$890} \\
$q$ & 0.78$\pm$0.02 & 0.74$\pm$0.15 & 0.87$\pm$0.07 & 0.84$\pm$0.02 & \multicolumn{2}{c}{0.90$\pm$0.02} \\
Est.\ Mass\tablenotemark{d} (M$_{\sun}$)~(A) & 0.039--0.062 & 0.042--0.077 & 0.029--0.073 & 0.025--0.076 & \multicolumn{2}{c}{0.019--0.065} \\
\phm{Est.\ Mass$^c$ (M$_{\sun}$)}~(B) & 0.029--0.051 & 0.031--0.074 & 0.029--0.073 & 0.021--0.074 & \multicolumn{2}{c}{0.016--0.061} \\
Est.\ Period\tablenotemark{e} (yr) & $\sim$19  & $\sim$10  & $\sim$18 & $\sim$50 & \multicolumn{2}{c}{$\sim$45} \\
\enddata
\tablenotetext{a}{Estimated flux ratios including systematic effects; see $\S$4.1.}
\tablenotetext{b}{Based on spectral decomposition; see $\S$5.3.}
\tablenotetext{d}{Based on \citet{gol04} {\teff}/spectral type relation,
with the exception of SDSS 0423-0413 and SDSS 1021-0304, where {\teff}s are derived
from measured $M_{bol}$s and assumed radii of 0.095$\pm$0.10 R$_{\sun}$.}
\tablenotetext{d}{Based on ages of 0.5--5.0~Gyr (with the exception of 
SDSS 0423-0414, where 0.5-1.7~Gyr is assumed), estimated {\teff}s (except SDSS 0423-0414 and SDSS 1021-0304,
where $M_{bol}$ is used) and the evolutionary models of \citet{bur97}.}
\tablenotetext{e}{Assuming semimajor axes $a$ = 1.26$\rho$ \citep{fis92}.}
\end{deluxetable}

\begin{deluxetable}{lcccccc}
\tabletypesize{\scriptsize}
\tablecaption{Faint Source Detection Limits for Unresolved Targets. \label{tab_senslimits}}
\tablewidth{0pt}
\tablehead{
 &  & \multicolumn{2}{c}{$0{\farcs}04 \lesssim \rho \lesssim 0{\farcs}2$}
 & \multicolumn{2}{c}{$0{\farcs}2 \lesssim \rho \lesssim 1{\farcs}0$} & \\
 \cline{3-4} \cline{5-6}
\colhead{Object} &
\colhead{Distance\tablenotemark{a}} &
\colhead{$\Delta$F110W} &
\colhead{$q$} &
\colhead{$\Delta$F110W} &
\colhead{$q$} &
\colhead{$\rho_{min}$} \\
\colhead{} &
\colhead{(pc)} &
\colhead{(mag)} &
\colhead{} &
\colhead{(mag)} &
\colhead{} &
\colhead{(AU)} \\
\colhead{(1)} &
\colhead{(2)} &
\colhead{(3)} &
\colhead{(4)} &
\colhead{(5)} &
\colhead{(6)} &
\colhead{(7)} \\
}
\startdata
SDSS 0151+1244 & 21.3$\pm$1.4 & 3.1 & 0.34 &  5.3 & 0.16 & 0.9 \\
SDSS 0207+0000 & 29$\pm$8 & 2.8 & 0.38 &  3.8 & 0.27 & 1.2 \\
2MASS 0243$-$2453 & 10.6$\pm$0.5 & 3.1 & 0.34 & 4.6 & 0.20 & 0.5 \\
2MASS 0348$-$6022 & $\sim$7 & 3.1 & 0.34 & 5.4 & 0.15 & 0.3 \\
2MASS 0415$-$0935 & 5.75$\pm$0.10 & 3.2 & 0.33 & 6.0 & 0.12 & 0.2 \\
2MASS 0516$-$0445 & $\sim$19 & 3.3 & 0.32 & 5.9 & 0.13 & 0.8 \\
2MASS 0755+2212 & $\sim$20 & 3.5 & 0.29 & 5.8 & 0.13 & 0.9 \\
SDSS 0837$-$0000 & 29$\pm$12 & 3.4 & 0.31 & 3.7 & 0.28 & 1.2 \\
SDSS 1110+0116 & $\sim$23 & 3.0 & 0.35 & 5.3 & 0.16 & 1.0 \\
2MASS 1217$-$0311 & 11.0$\pm$0.2 & 3.1 & 0.34 & 5.3 & 0.16 & 0.5 \\
SDSS 1254$-$0122 & 13.7$\pm$0.4 & 3.5 & 0.29 & 6.5 & 0.10 & 0.6 \\
2MASS 1503+2525 & $\sim$9 & 3.1 & 0.34 &  6.5 & 0.10 & 0.4 \\
SDSS 1624+0029 & 10.9$\pm$0.2 & 3.1 & 0.34 &  6.0 & 0.12 & 0.5 \\
SDSS 1750+1759 & 28$\pm$4 & 3.0 & 0.35 &  4.6 & 0.20 & 1.2 \\
2MASS 2228$-$4310 & $\sim$13 & 3.2 & 0.33 & 5.7 & 0.14 & 0.6 \\
2MASS 2254+3123 & $\sim$19 & 2.9 & 0.36 & 5.3 & 0.16 & 0.8 \\
2MASS 2339+1352 & $\sim$26 & 3.5 & 0.29 & 5.6 & 0.14 & 1.1 \\
\enddata
\tablecomments{Given values are 3$\sigma$ detection limits for a single PSF
subtraction from individual calibration images,
separated into the core/first Airy ring region ($\rho \lesssim 0{\farcs}2$)
and the background-dominated region ($\rho \gtrsim 0{\farcs}4$).
Mass ratio ($q$) limits assume $\Delta$F110W $\sim$ ${\Delta}M_{bol}$
and Eqn.~6.}
\tablenotetext{a}{Distance estimates for objects without parallax measurements
are based on apparent 2MASS $J$-band
magnitudes and the $M_J$/spectral type relation of \citet{tin03}.}
\end{deluxetable}

\begin{deluxetable}{lccccccc}
\tabletypesize{\scriptsize}
\tablecaption{L/T Transition Binaries. \label{tab_ltbinaries}}
\tablewidth{0pt}
\tablehead{
\colhead{Name} &
\multicolumn{2}{c}{Spectral Types} &
\colhead{Distance\tablenotemark{a}} &
\multicolumn{2}{c}{Separation} &
\colhead{Period} &
\colhead{Note} \\
\cline{5-6}
\colhead{} &
\colhead{(A)} &
\colhead{(B)} &
\colhead{(pc)} &
\colhead{($\arcsec$)} &
\colhead{(AU)} &
\colhead{(yr)} &
\colhead{} \\
\colhead{(1)} &
\colhead{(2)} &
\colhead{(3)} &
\colhead{(4)} &
\colhead{(5)} &
\colhead{(6)} &
\colhead{(7)} &
\colhead{(8)} \\
}
\startdata
SDSS J042348.57$-$041403.5 & L6 & T2 & 15.2$\pm$0.4 & 0$\farcs$164$\pm$0$\farcs$005 & 2.49$\pm$0.07 &  $\sim$19 & 1,2 \\  
2MASS J05185995$-$2828372 & L6: & T4: & $\sim$34 & 0$\farcs$051$\pm$0$\farcs$012 & 1.8$\pm$0.5 & $\sim$10 & 3,4 \\
2MASS J08503593+1057156 & L6 & T: & 26$\pm$2 & 0$\farcs$16$\pm$0$\farcs$010 & 4.4$\pm$0.4 & $\sim$43  &  5 \\
2MASS J09201223+3517429 & L6.5 & T: & $\sim$21 & 0$\farcs$07$\pm$0$\farcs$010 & 1.5$\pm$0.5 & $\sim$6  &  5 \\
Gliese 337C & L8 & T: & 20.5$\pm$0.4 & 0$\farcs$53$\pm$0$\farcs$03 & 10.9$\pm$0.7  & $\sim$150 & 6,7 \\
SDSS J102109.69$-$030420.1 & T1 & T5 & 29$\pm$4 & 0$\farcs$172$\pm$0$\farcs$005 & 5.0$\pm$0.7 & $\sim$48 & 2,4 \\  
Epsilon Indi B & T1 & T6 & 3.626$\pm$0.013 & 0$\farcs$732$\pm$0$\farcs$002  & 2.654$\pm$0.012  & $\sim$15 & 7,8 \\
SDSS J153417.05+161546.1 & T1.5 & T5.5 & $\sim$36 & 0$\farcs$110$\pm$0$\farcs$005  & 3.9$\pm$0.6 & $\sim$28 & 9 \\
2MASS J17281150+3948593 & L7 & T: & $\sim$23 & 0$\farcs$131$\pm$0$\farcs$003  & 3.0$\pm$0.5 & $\sim$35 & 10 \\
2MASS J22521073-1730134 & L6 & T2: & 14$\pm$3 & 0$\farcs$130$\pm$0$\farcs$002 & 1.8$\pm$0.4  & $\sim$9 &  11 \\
\enddata
\tablenotetext{a}{Parallax distance measurements from \citet{esa97} and \citet{vrb04}
are given with uncertainties, all others are spectrophotometric distance estimates
from the discovery references.}
\tablerefs{(1) \citet{me0423}; (2) \citet{vrb04}; (3) \citet{cru04}; 
(4) This paper; (5) \citet{rei01};
(6) \citet{me05gl337cd}; (7) \citet{esa97}; (8) \citet{mcc04}; (9) \citet{liu06}; 
(10) \citet{giz03}; (11) \citet{rei2252}.}
\end{deluxetable}

\clearpage

\begin{figure}
\centering
\plotone{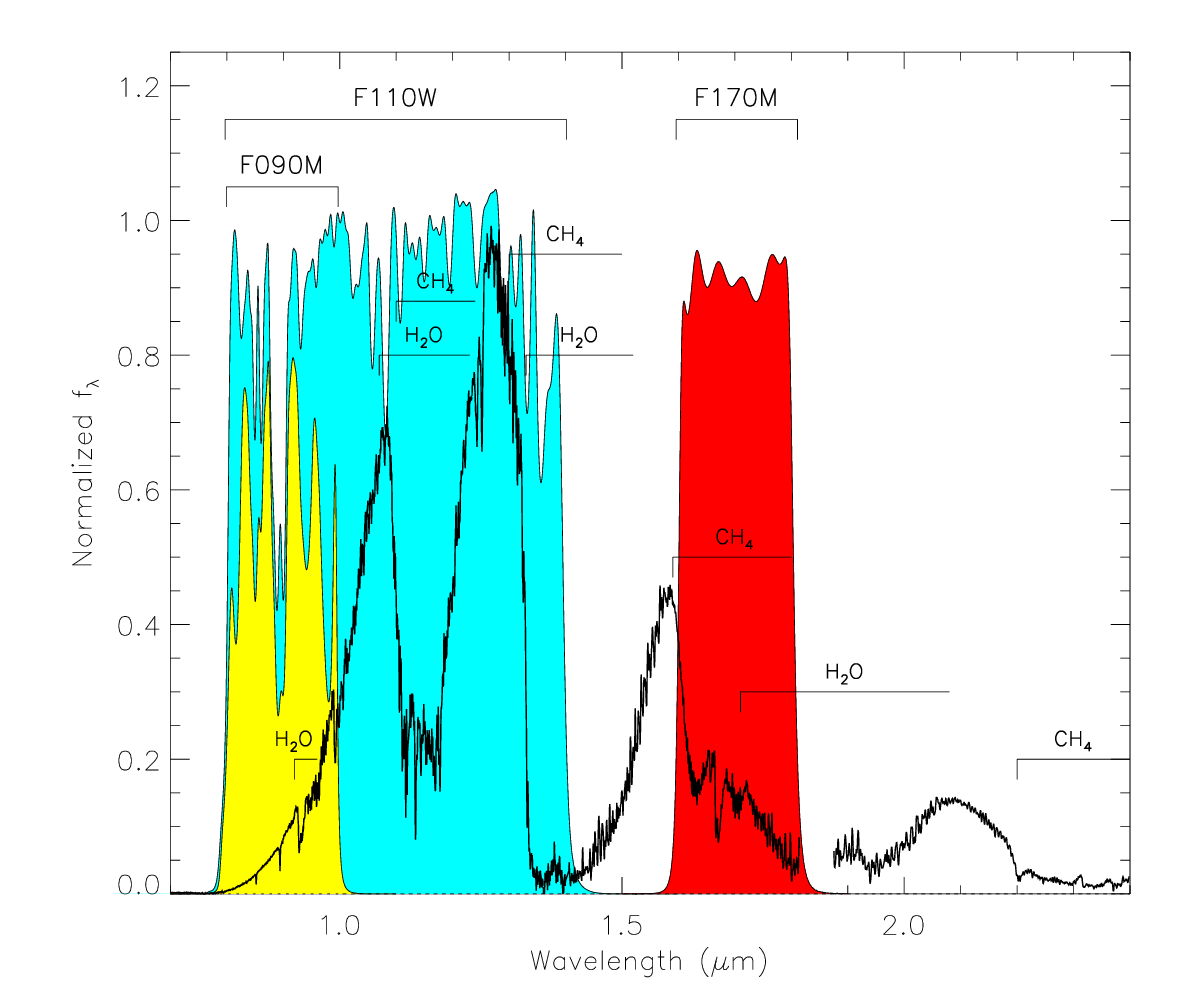}
\caption{NIC1 F090M (yellow), F110W (blue) and F170M (red) filter transmission
profiles overlaid 
on the red optical and near infrared spectrum of 2MASS 1503+2525 \citep{me03opt,me04wide3}.
Spectral data are normalized at 1.27~$\micron$.  Filter transmission profiles
are preflight measurements and do not include the NIC1 detector quantum efficiency
or optical element response curves.  Key {\water} and {\meth} bands present
in the spectra of T dwarfs are indicated.}
\end{figure}

\begin{figure}
\centering
\includegraphics[width=0.8\textwidth]{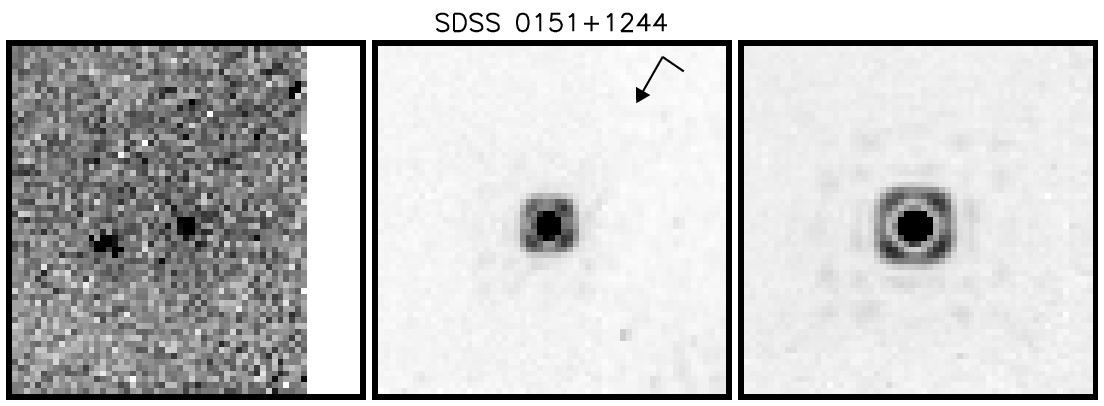}
\includegraphics[width=0.8\textwidth]{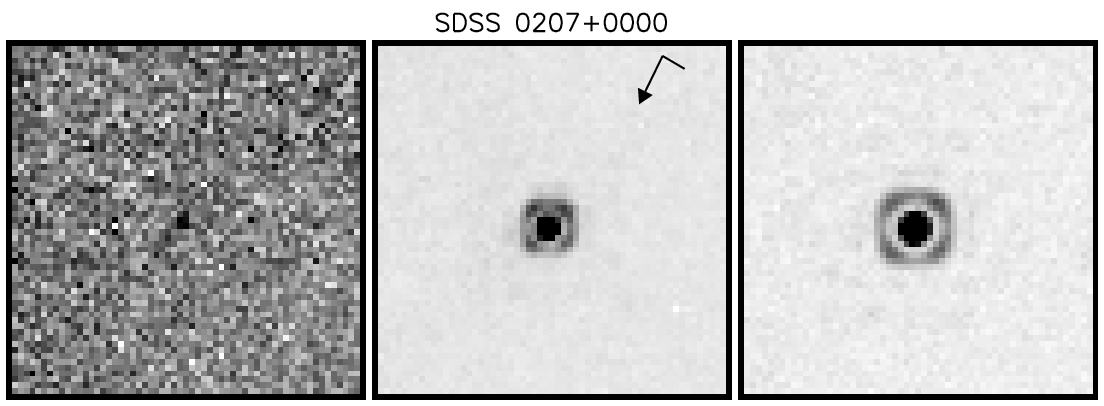}
\includegraphics[width=0.8\textwidth]{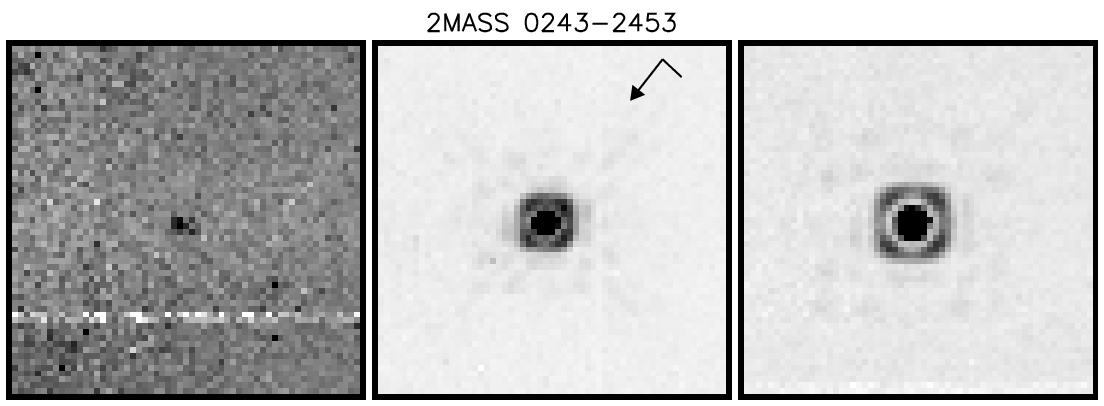}
\includegraphics[width=0.8\textwidth]{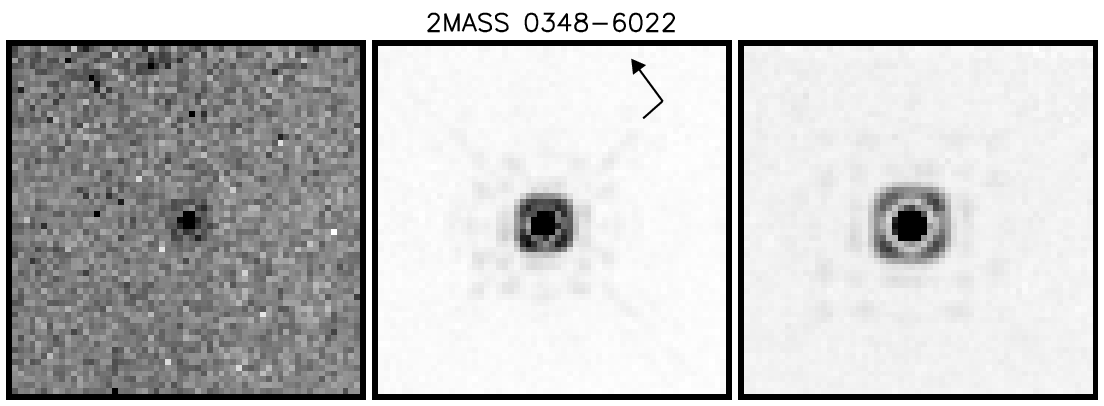}
\caption{NICMOS F090M (left), F110W (center) and F170M (right)
images of T dwarfs observed in this study.  
Subsections of 2$\farcs$5$\times$2$\farcs$5
in size are shown on a logarithmic scale.
Orientations North (arrow) and East (line) are indicated
in the center panel. \label{fig_image1}}
\end{figure}
\clearpage
\centering
\includegraphics[width=0.8\textwidth]{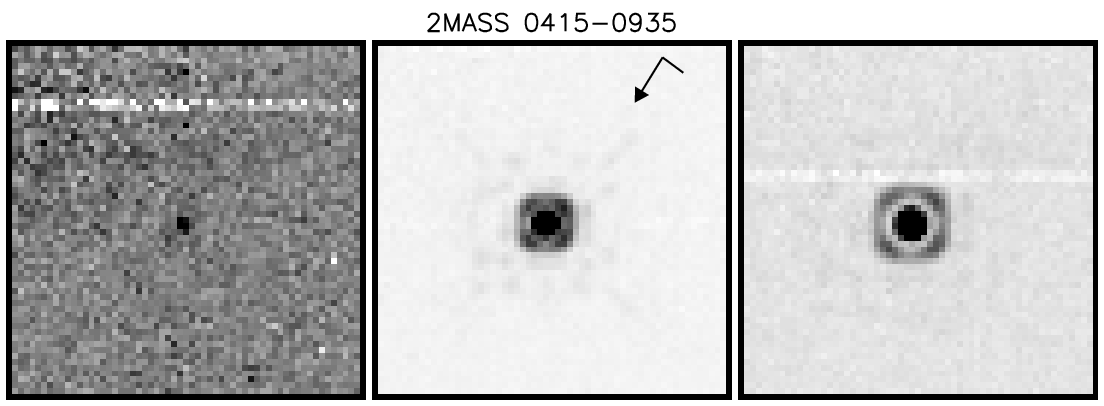}
\includegraphics[width=0.8\textwidth]{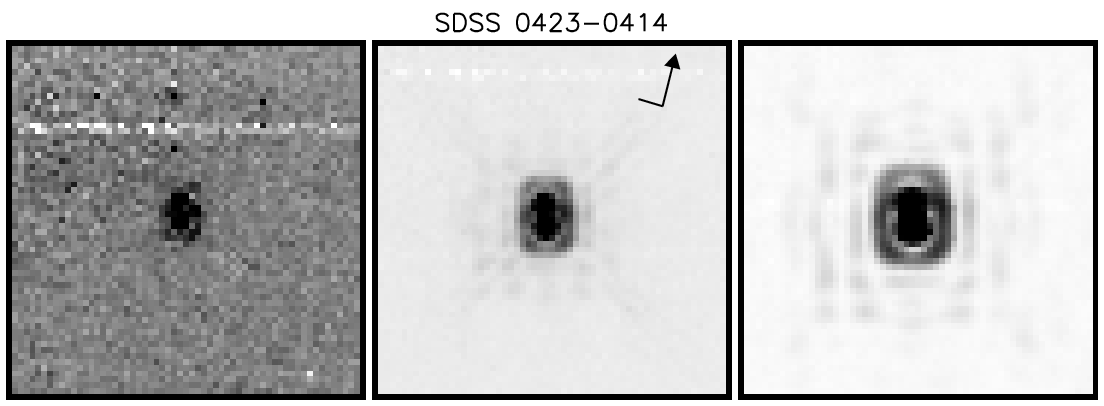}
\includegraphics[width=0.8\textwidth]{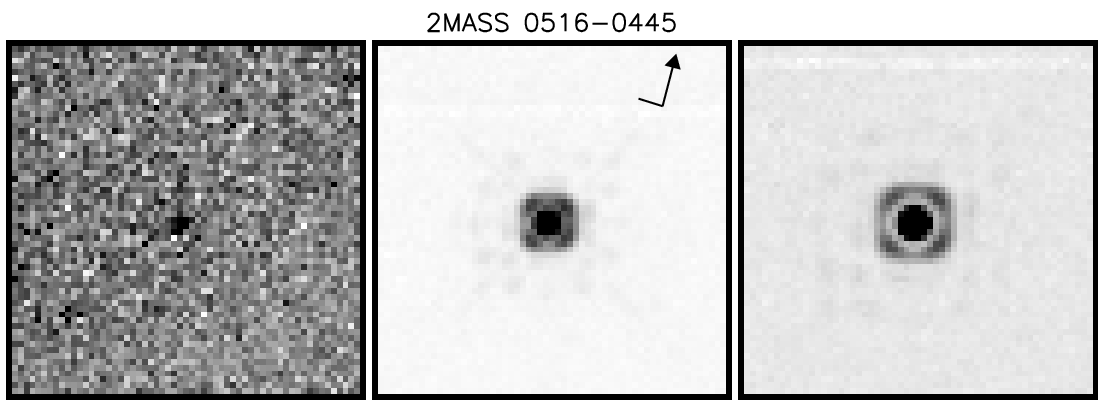}
\includegraphics[width=0.8\textwidth]{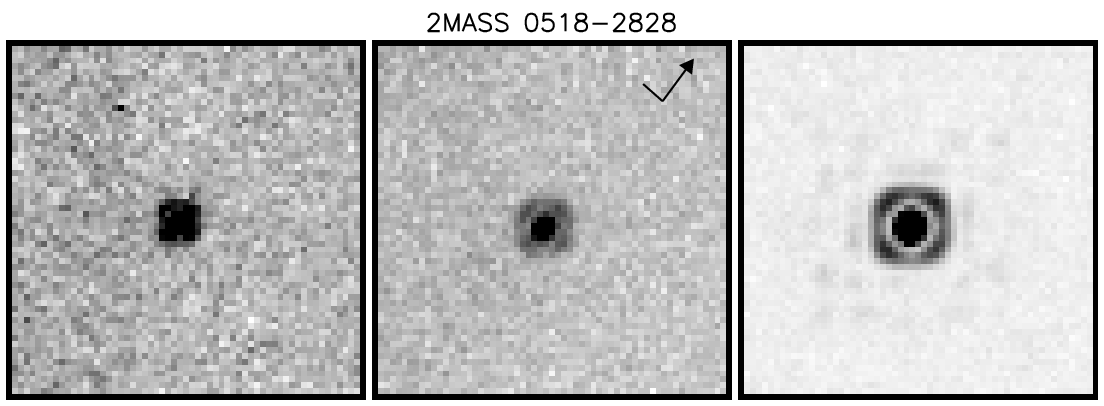}
\centerline{Fig. 2. --- Continued.}
\clearpage
\centering
\includegraphics[width=0.8\textwidth]{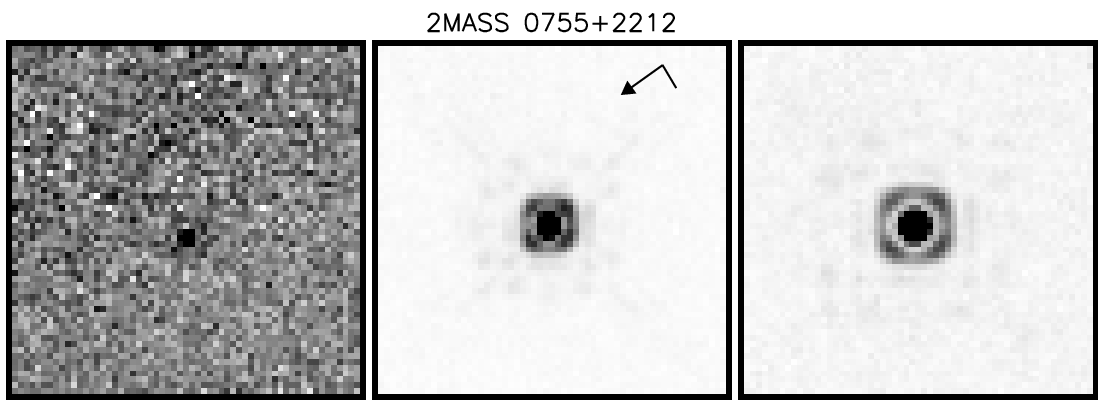}
\includegraphics[width=0.8\textwidth]{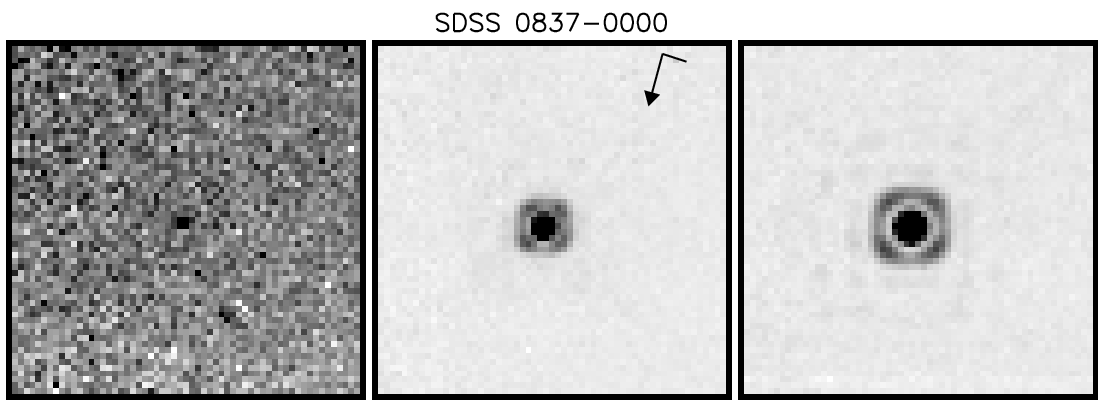}
\includegraphics[width=0.8\textwidth]{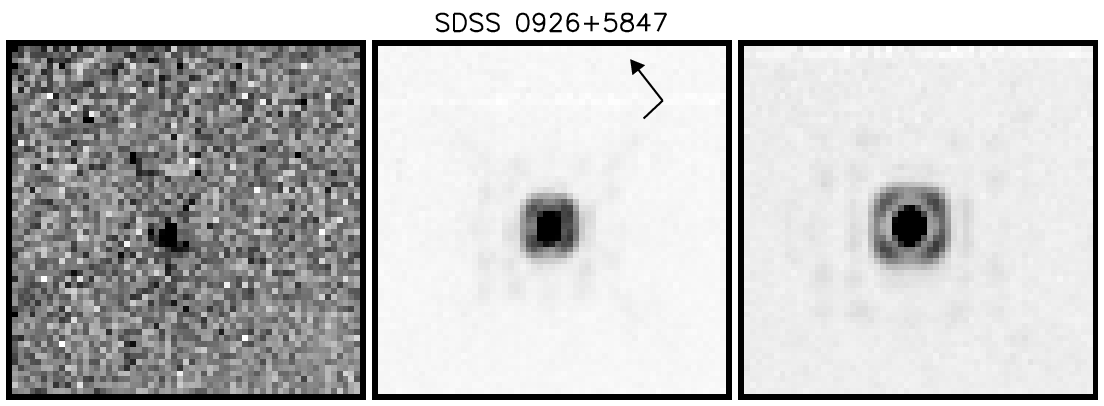}
\includegraphics[width=0.8\textwidth]{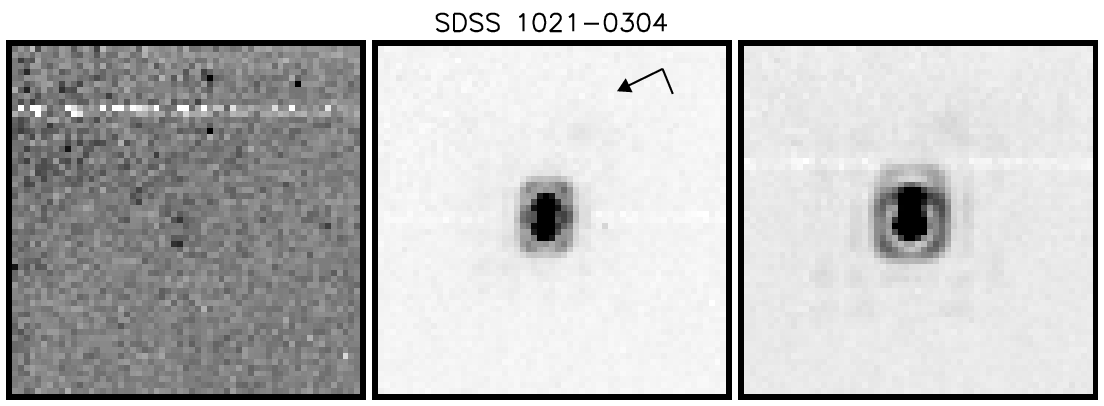}
\centerline{Fig. 2. --- Continued.}
\clearpage
\centering
\includegraphics[width=0.8\textwidth]{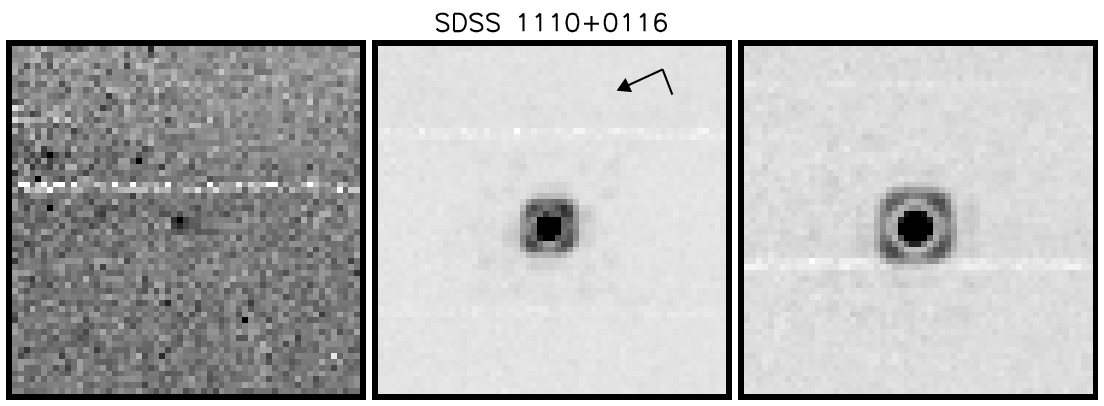}
\includegraphics[width=0.8\textwidth]{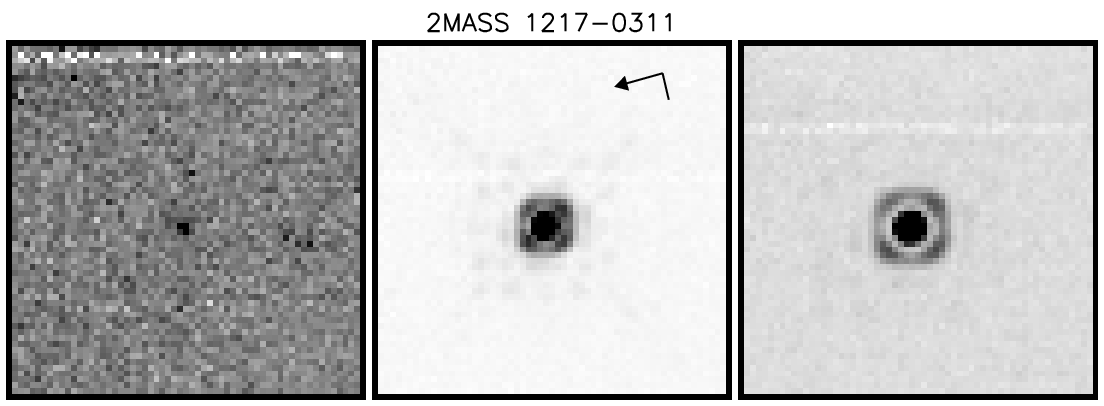}
\includegraphics[width=0.8\textwidth]{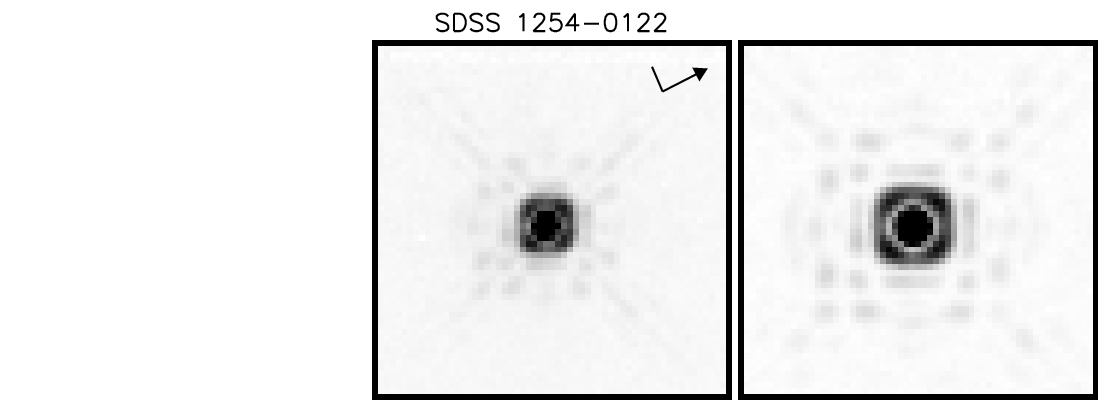}
\includegraphics[width=0.8\textwidth]{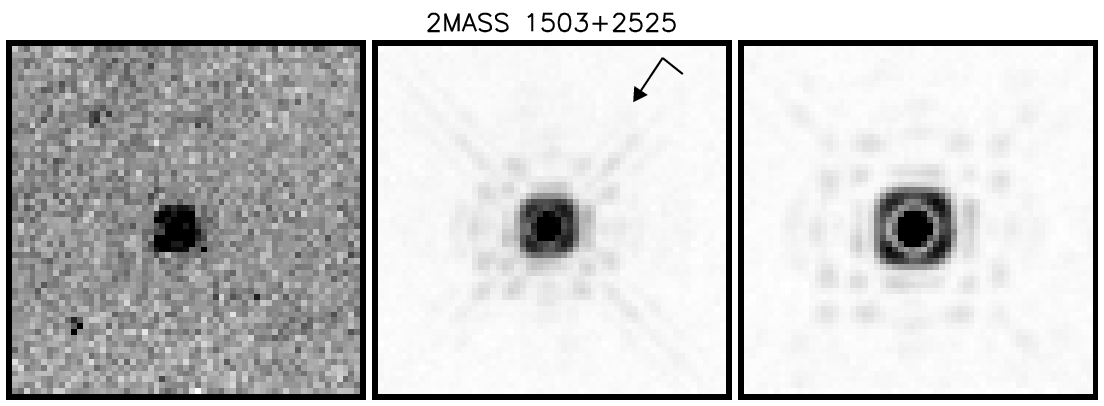}
\centerline{Fig. 2. --- Continued.}
\clearpage
\centering
\includegraphics[width=0.8\textwidth]{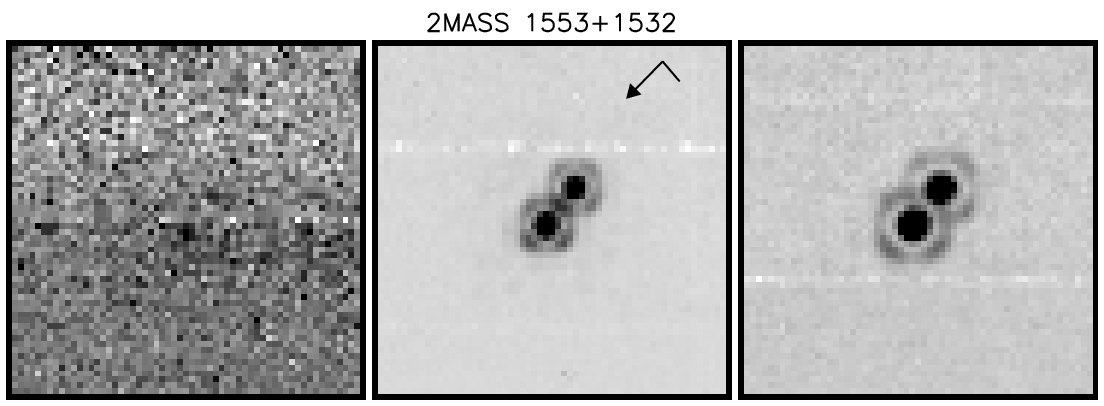}
\includegraphics[width=0.8\textwidth]{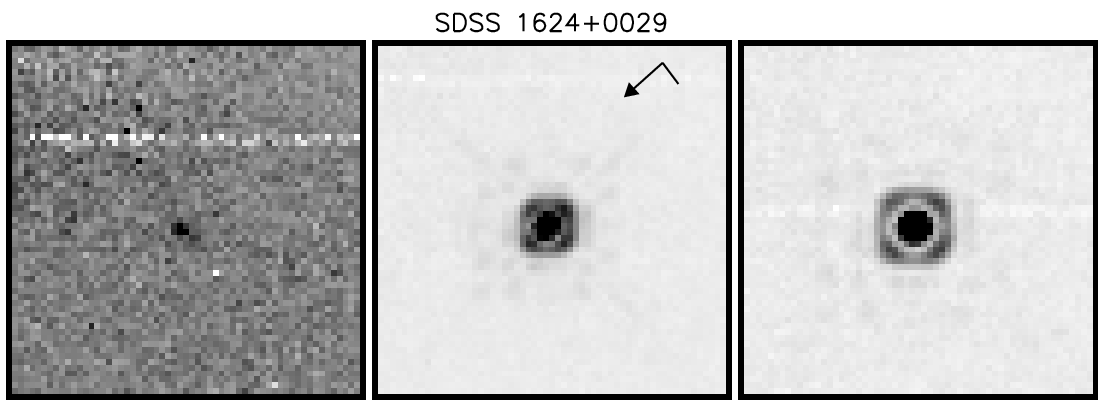}
\includegraphics[width=0.8\textwidth]{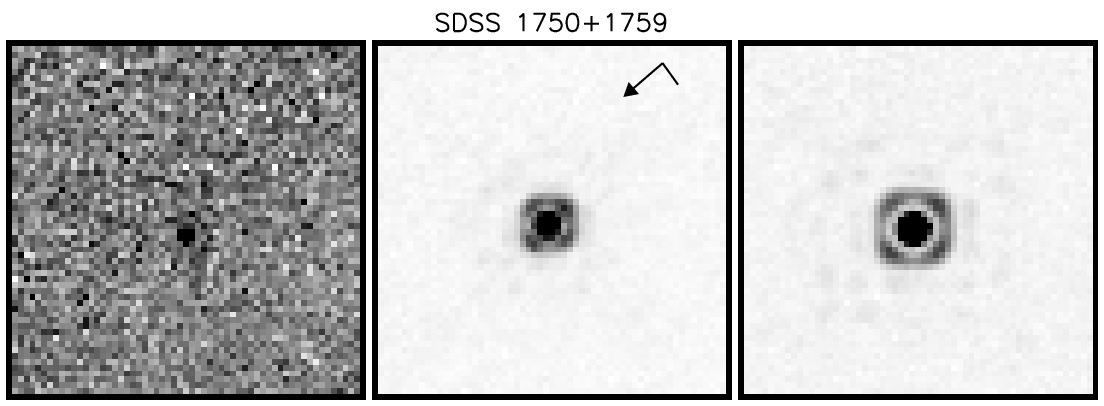}
\includegraphics[width=0.8\textwidth]{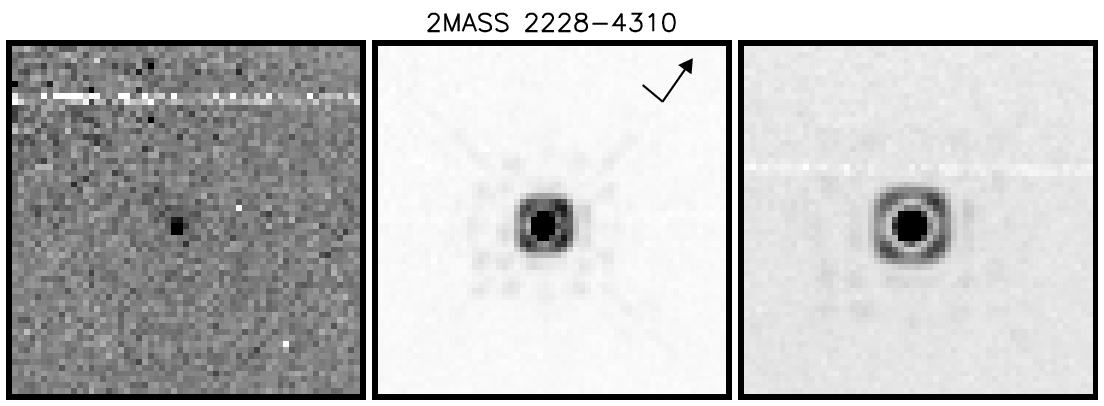}
\centerline{Fig. 2. --- Continued.}
\clearpage
\centering
\includegraphics[width=0.8\textwidth]{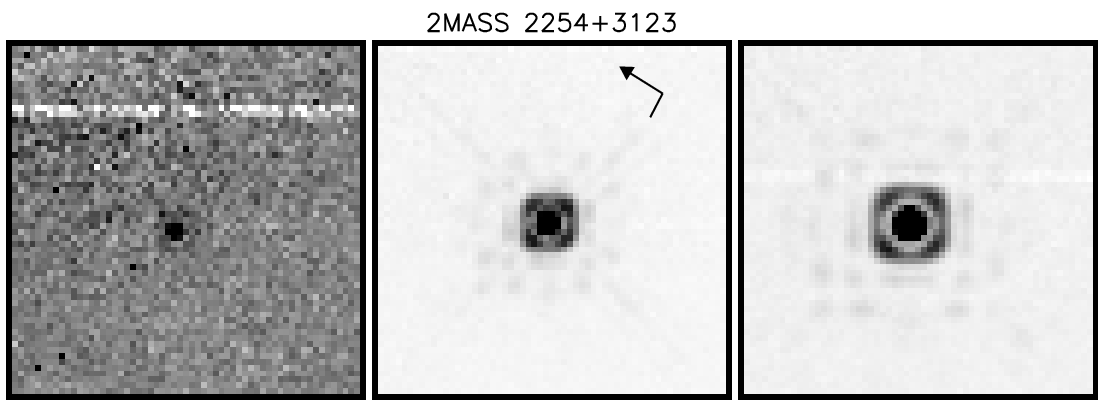}
\includegraphics[width=0.8\textwidth]{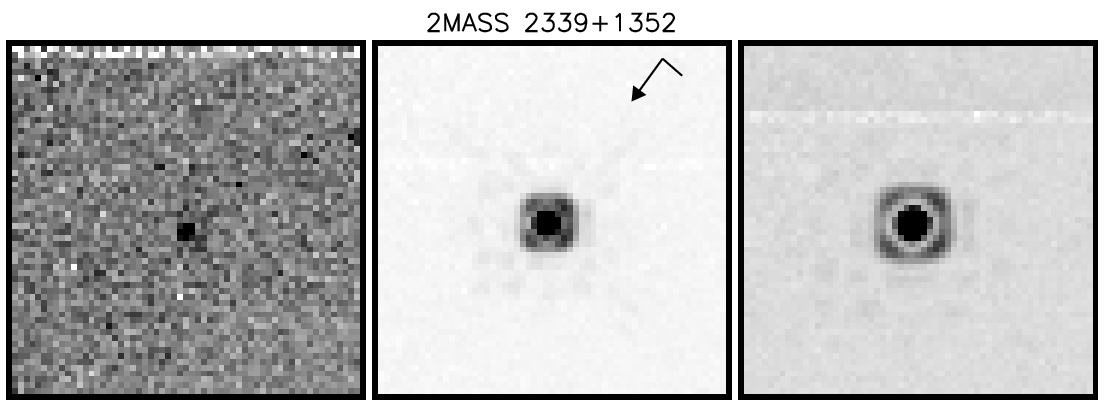}
\centerline{Fig. 2. --- Continued.}
\clearpage
\begin{figure}
\centering
\plotone{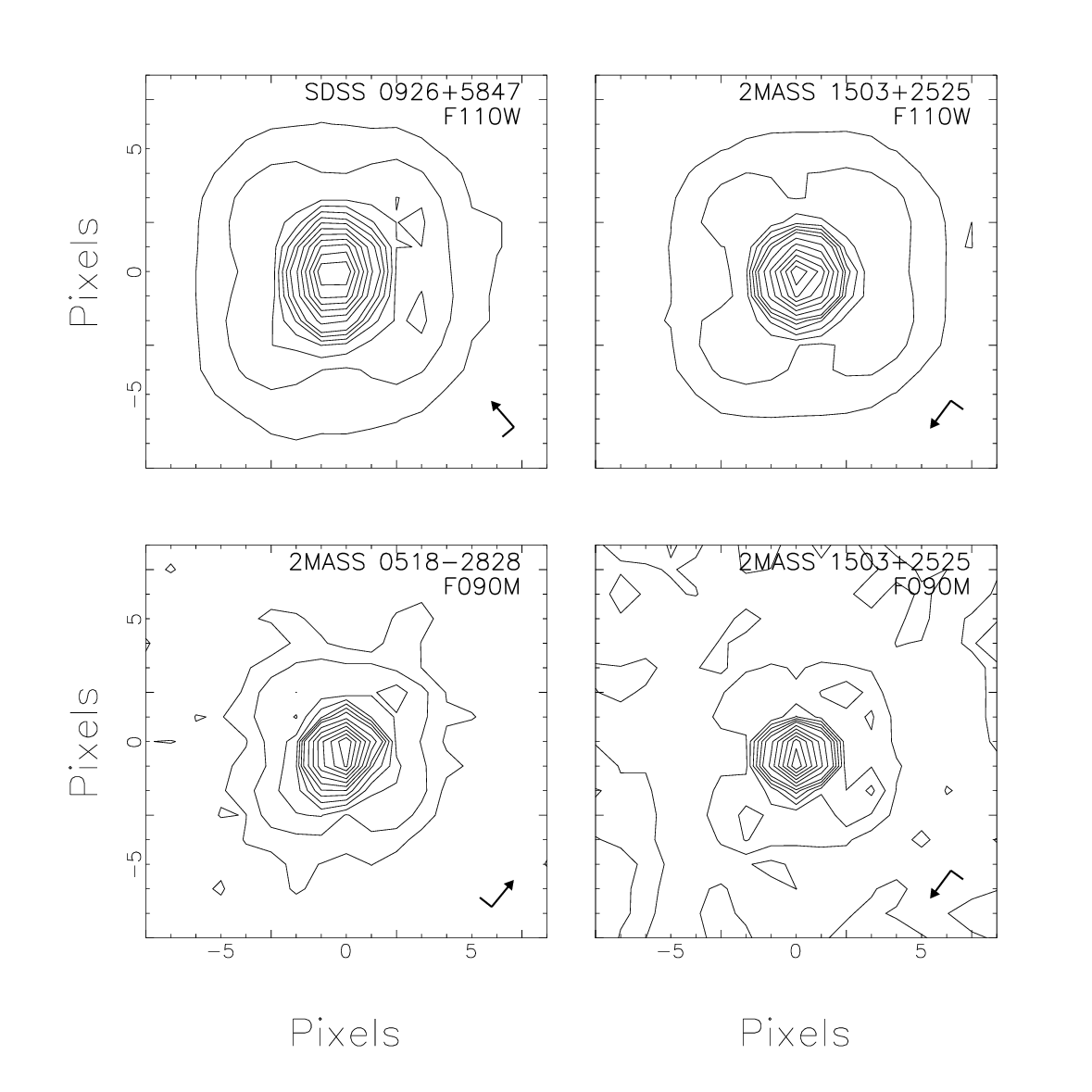}
\caption{Contour plots of F110W images of the central 0$\farcs$9$\times$0$\farcs$9 regions
around SDSS 0926+5847 (top left) and 2MASS 1503+2525 (top right), 
and F090M images of 2MASS 0518-2828 (bottom left)
and 2MASS 1503+2525 (bottom right). 
Lines indicate isofluxes of 0.01, 0.05, 0.1, 0.15, 0.2, 0.25, 0.3, 0.4, 0.5, 0.6, 0.7, 0.8, 0.8, 0.9 and 1.0 times the peak source flux.  Orientations (North and East)
for each image are indicated in the lower right corners.
\label{fig_contour}}
\end{figure}


\begin{figure}
\centering
\plotone{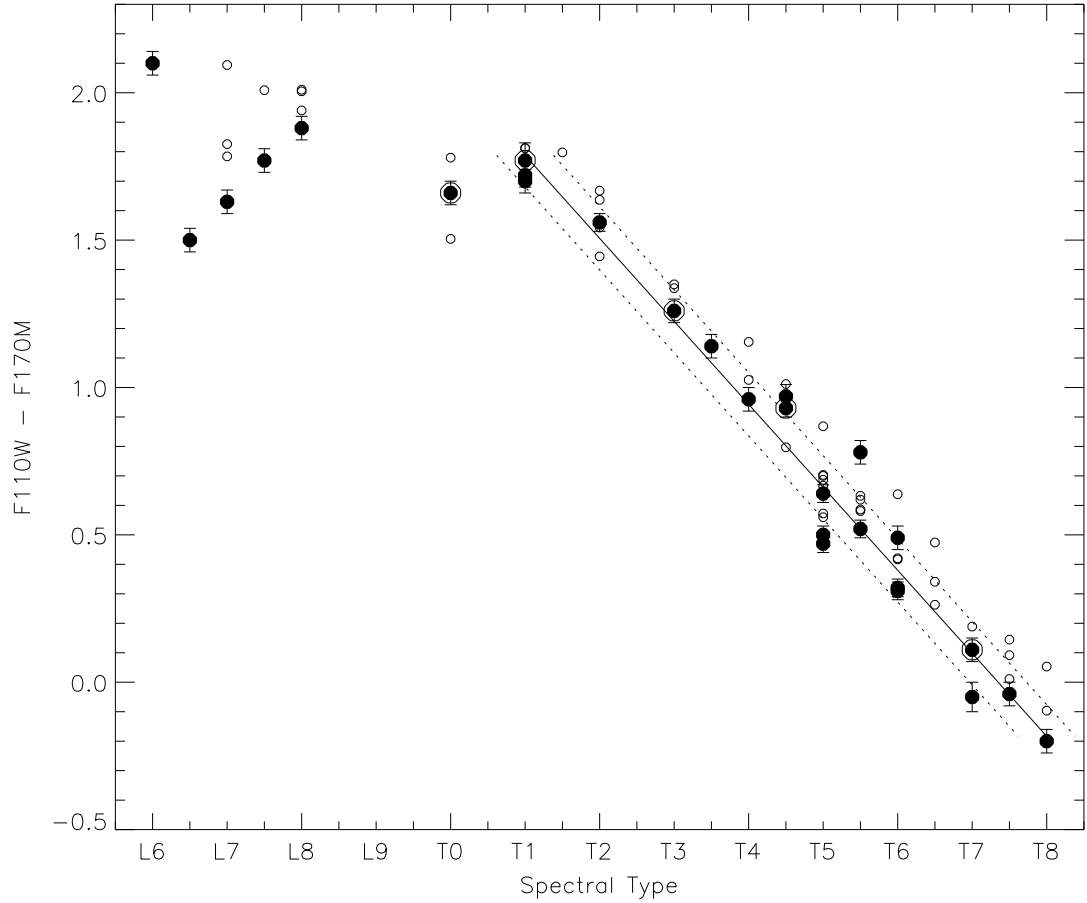}
\caption{F110W-F170M colors versus spectral type for 
subtypes L6 to T8.  L dwarf classifications are based on
optical spectra (e.g., Kirkpatrick et al.\ 1999); T dwarf
classifications are based on near infrared spectra (e.g., Burgasser et al.\ 2006).
Photometry from this program and \citet{rei06} are shown as
solid circles with error bars; multiple sources are encircled.
Spectrophotometric colors measured from low resolution near
infrared data from \citet{me06a} are shown as small open circles.
A linear fit to the photometric data of unresolved objects
is indicated by the solid line; $\pm$1$\sigma$ scatter about this
relation are indicated by the dotted lines. \label{fig_colorvsspt1}}
\end{figure}

\begin{figure}
\centering
\plotone{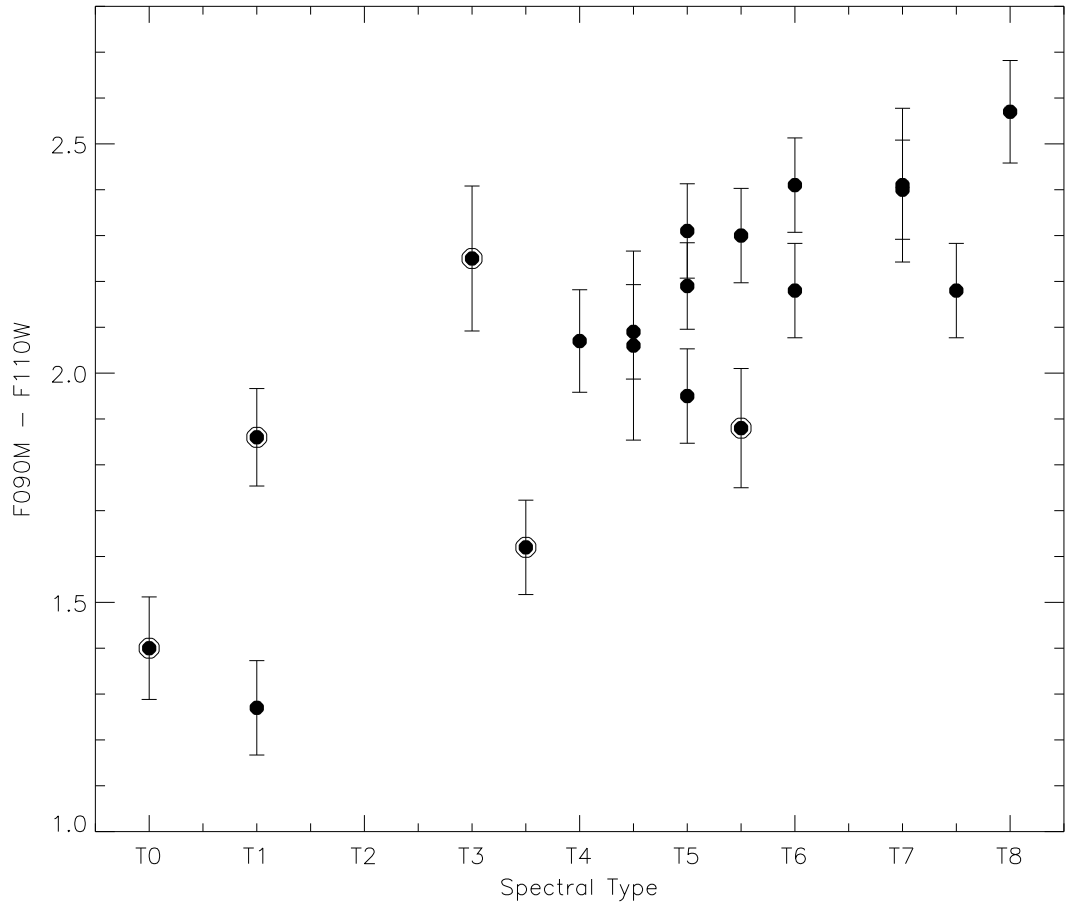}
\caption{F090M-F110W colors versus spectral type for T dwarfs in our sample.
Symbols are the same as in Figure~\ref{fig_colorvsspt1}. \label{fig_colorvsspt2}}
\end{figure}

\begin{figure}
\centering
\epsscale{0.7}
\plotone{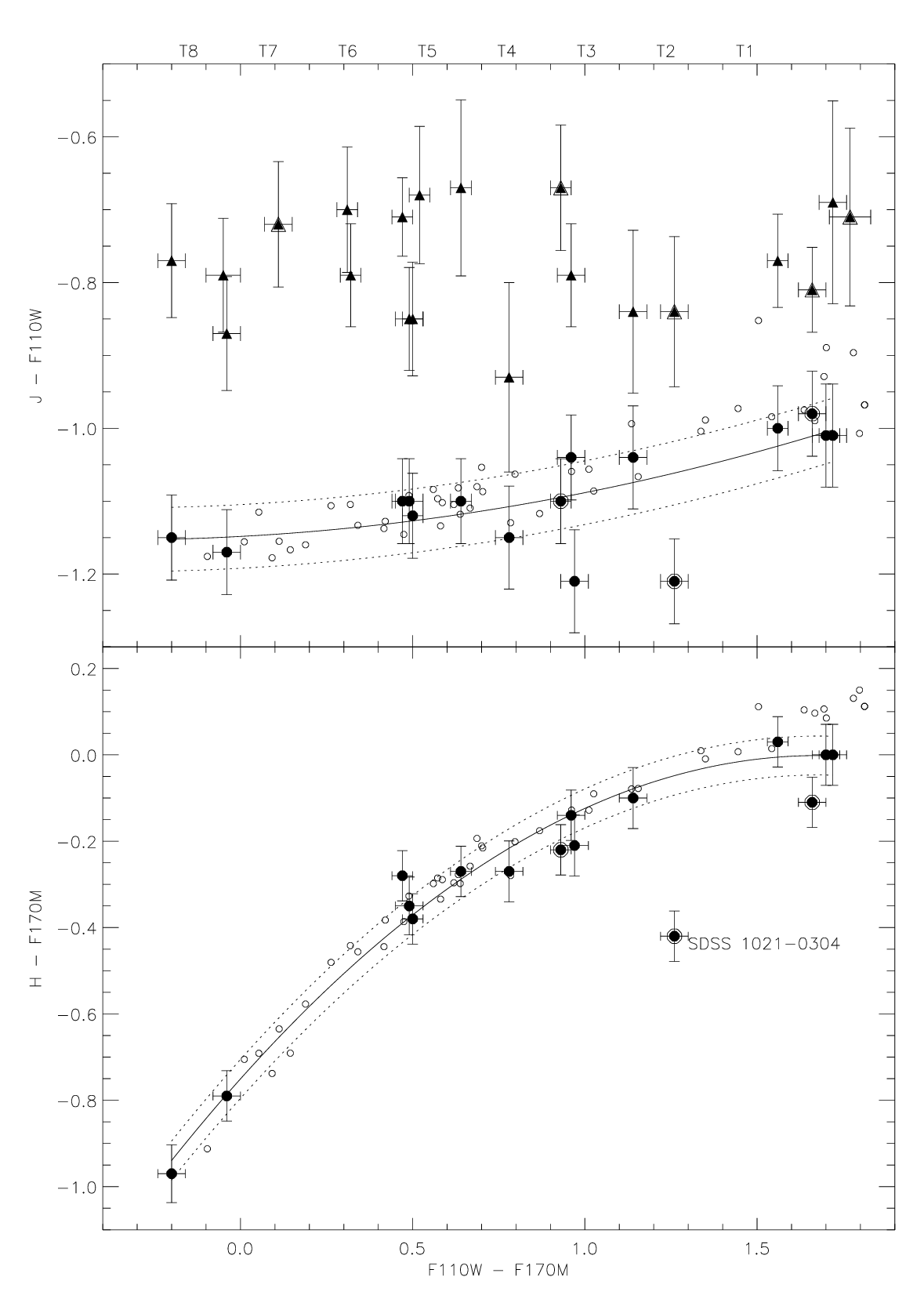}
\caption{{\em Top}: $J-F110W$ versus $F110W-F170M$ color
for sources in our sample.  Solid circles indicate colors on the MKO
system, solid triangles indicate colors on the 2MASS system.
Multiple sources are encircled.  Spectrophotometric colors 
on the MKO system measured from low resolution near
infrared data are shown as small open circles.  A polynomial
fit to the MKO photometry of unresolved sources is
indicated by the solid line; $\pm$1$\sigma$ scatter about this
relation are indicated by the dotted lines.
{\em Bottom}: $H-F170M$ color versus $F110W-F170M$ color
for sources in our sample.  Symbols are the same as in the top
panel, and only MKO photometry are shown.  A polynomial
fit to photometry for unresolved sources is
indicated by the solid line; $\pm$1$\sigma$ scatter about this
relation are indicated by the dotted lines.
The discrepant (multiple) source SDSS 1021-0304 is labeled. 
\label{fig_mkovshstvscolor}}
\end{figure}

\begin{figure}
\centering
\plotone{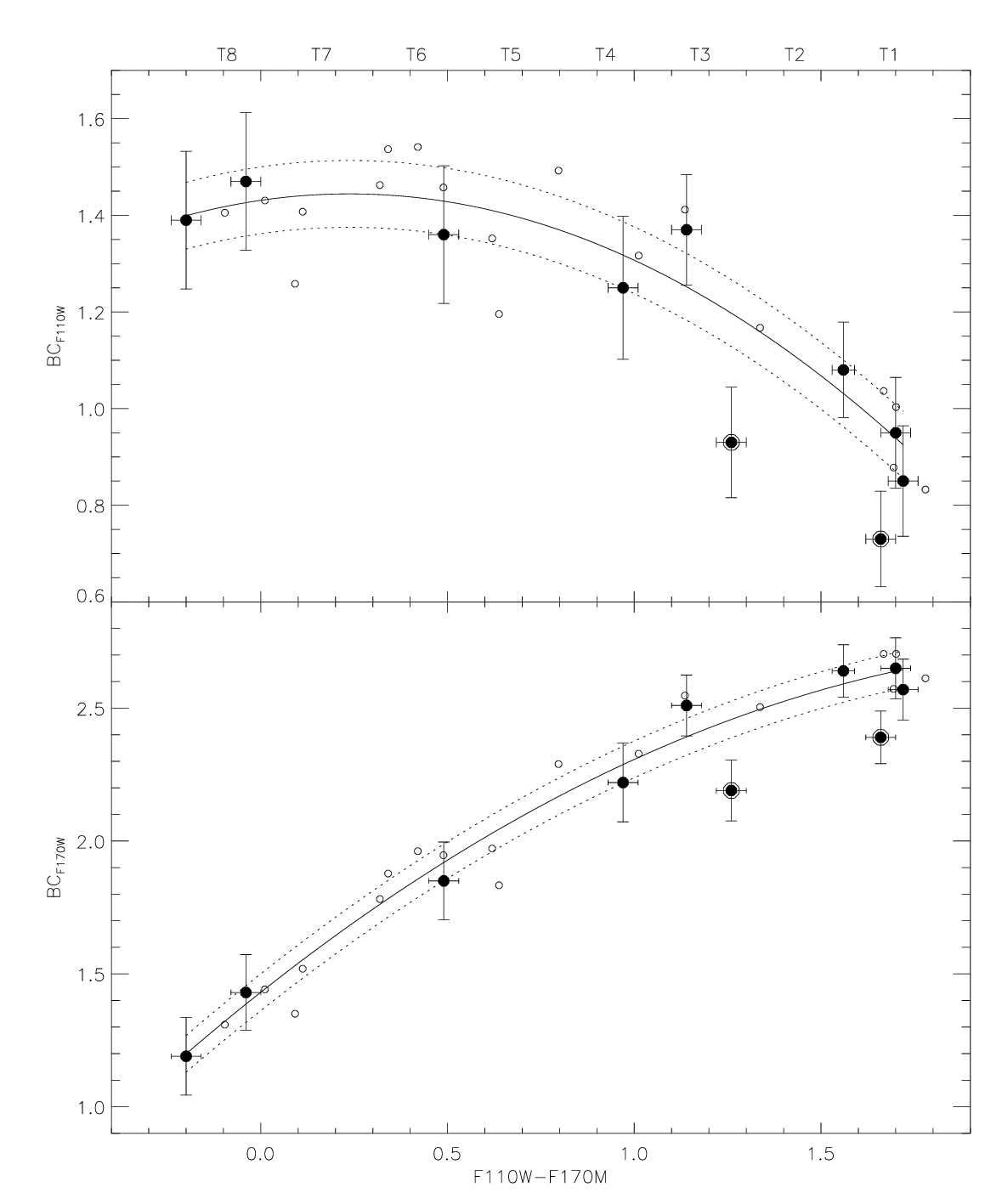}
\caption{F110W and F170M bolometric corrections (BC) as
a function of $F110W-F170M$ color.  BC values are based
on $K$-band bolometric corrections from \citet{gol04} and
MKO $K$-band photometry for sources in our sample.  
Spectrophotometric BCs 
measured from low resolution near
infrared data from \citet{me06a} are shown as small open circles.
Polynomial fits to the BCs of unresolved sources
observed in this program are
indicated by the solid lines; $\pm$1$\sigma$ scatter about these
relations are indicated by the dotted lines. \label{fig_bc}}
\end{figure}

\begin{figure}
\centering
\plotone{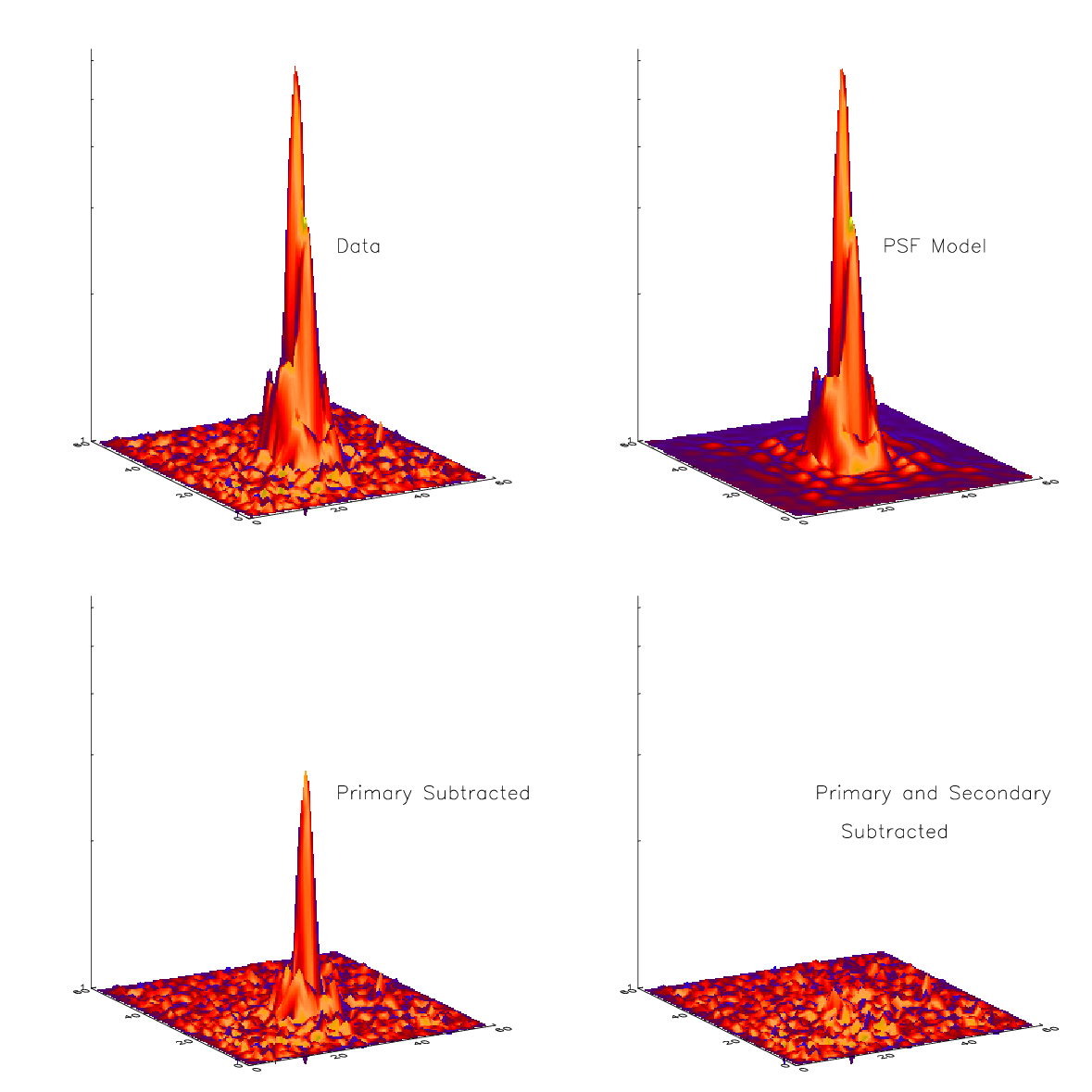}
\caption{Surface plots illustrating PSF fits to F170M imaging data
of the SDSS 1021-0304 binary.  Shown clockwise from upper left
are 2$\farcs$5$\times$2$\farcs$5 subsections of 
imaging data, the best fitting PSF model,
subtraction of the primary PSF model from the data,
and subtraction of the full PSF model from the data.
All four plots are normalized to a common logarithmic scale;
average residuals in the final subtraction are 0.4\% of the peak 
source flux. \label{fig_psffit}}
\end{figure}

\clearpage

\begin{figure}
\centering
\includegraphics[width=0.8\textwidth]{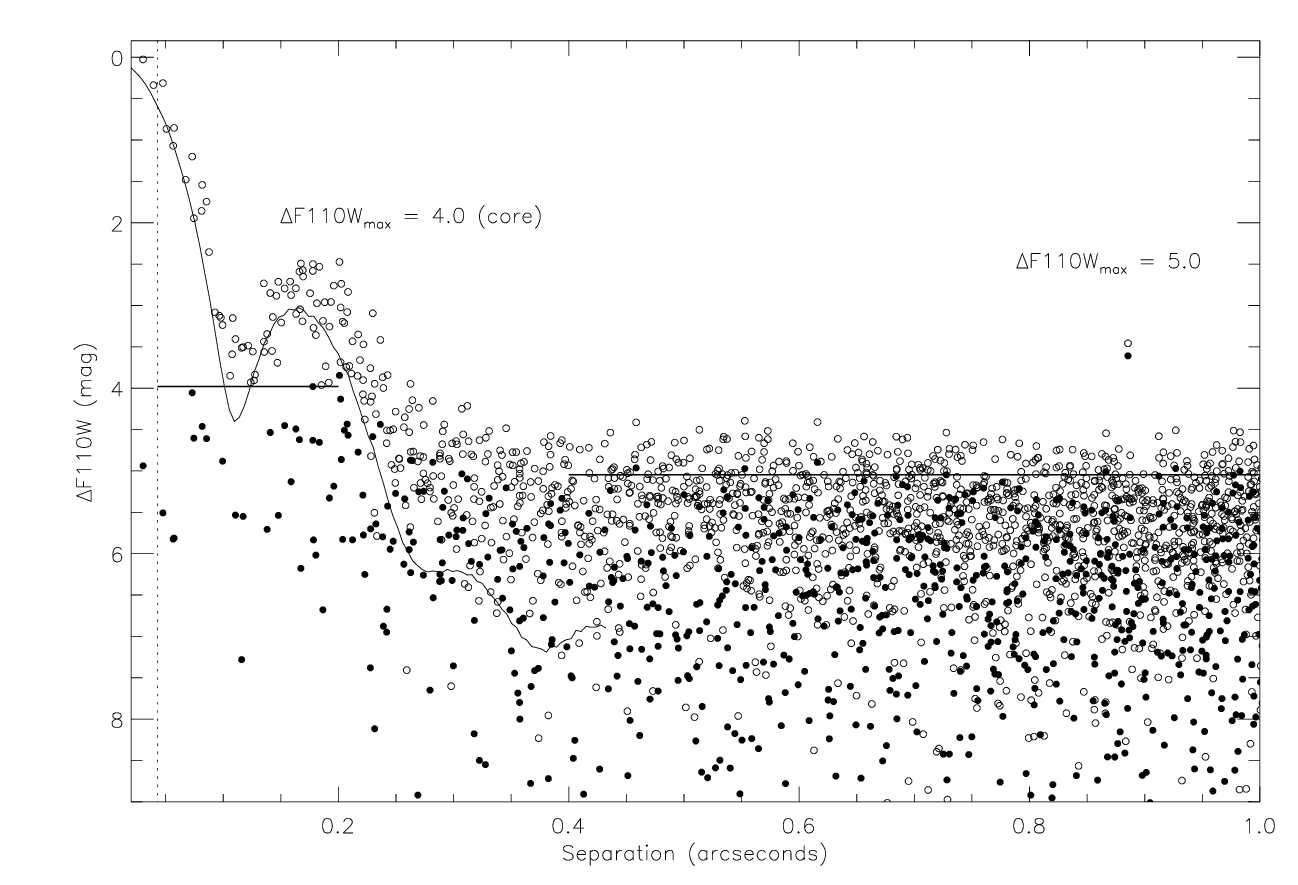}
\includegraphics[width=0.8\textwidth]{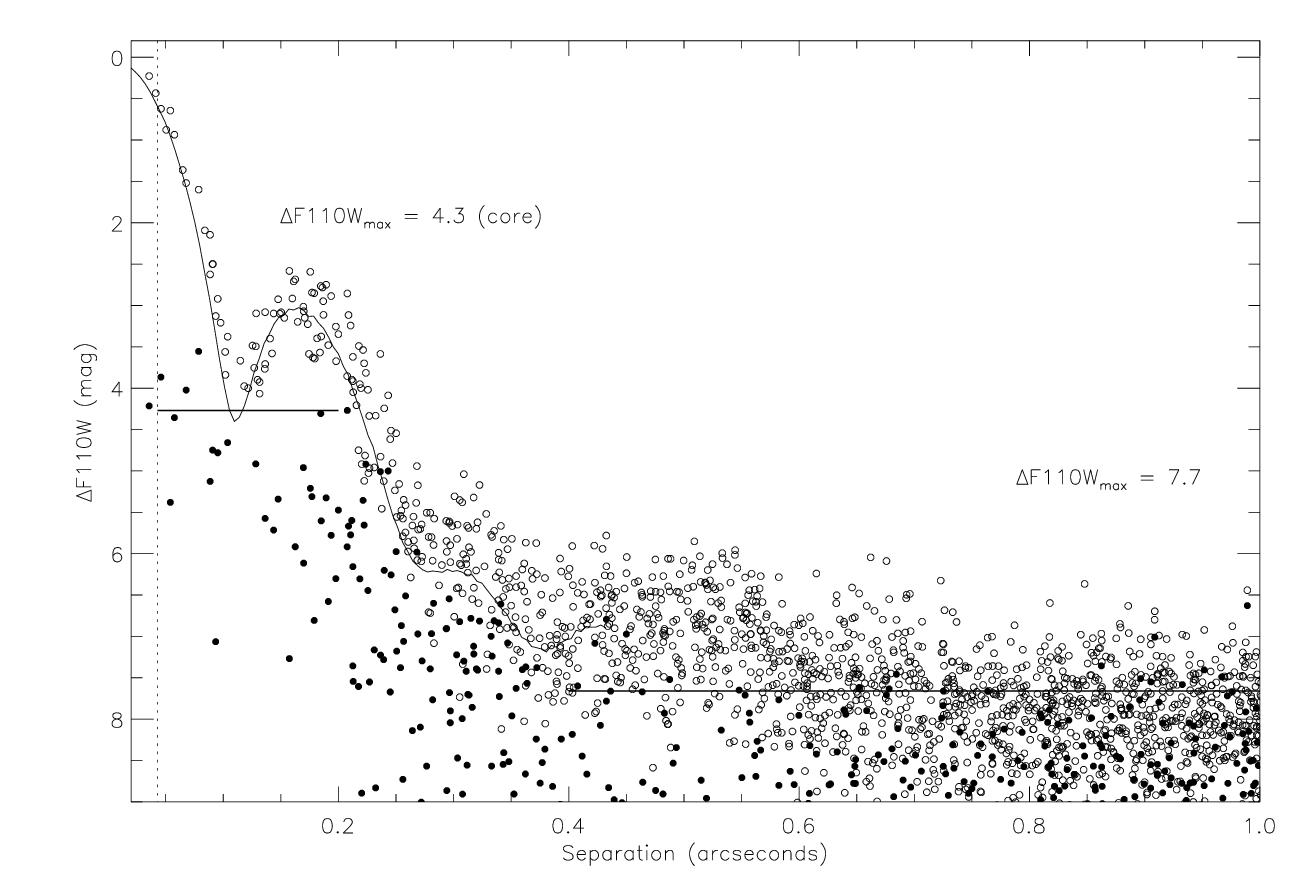}
\caption{Sensitivity limits for faint companions around 
the faintest and brightest unresolved sources, SDSS 0207+0000 (top) and 2MASS 1503+2525 
(bottom).  Open circles trace the relative brightness profile
with respect to the peak PSF in the original NICMOS image; filled circles trace
the brightness profile after PSF subtraction. The solid lines
trace the radial profile of an oversampled model PSF from Tiny Tim. 
Residuals in the PSF core (0$\farcs$04--0$\farcs$2) after model subtraction
are consistently $\sim$4 mag below the peak primary flux.  The far
wing sensitivity is dominated by background noise.
\label{fig_psfsub}}
\end{figure}

\begin{figure}
\centering
\plotone{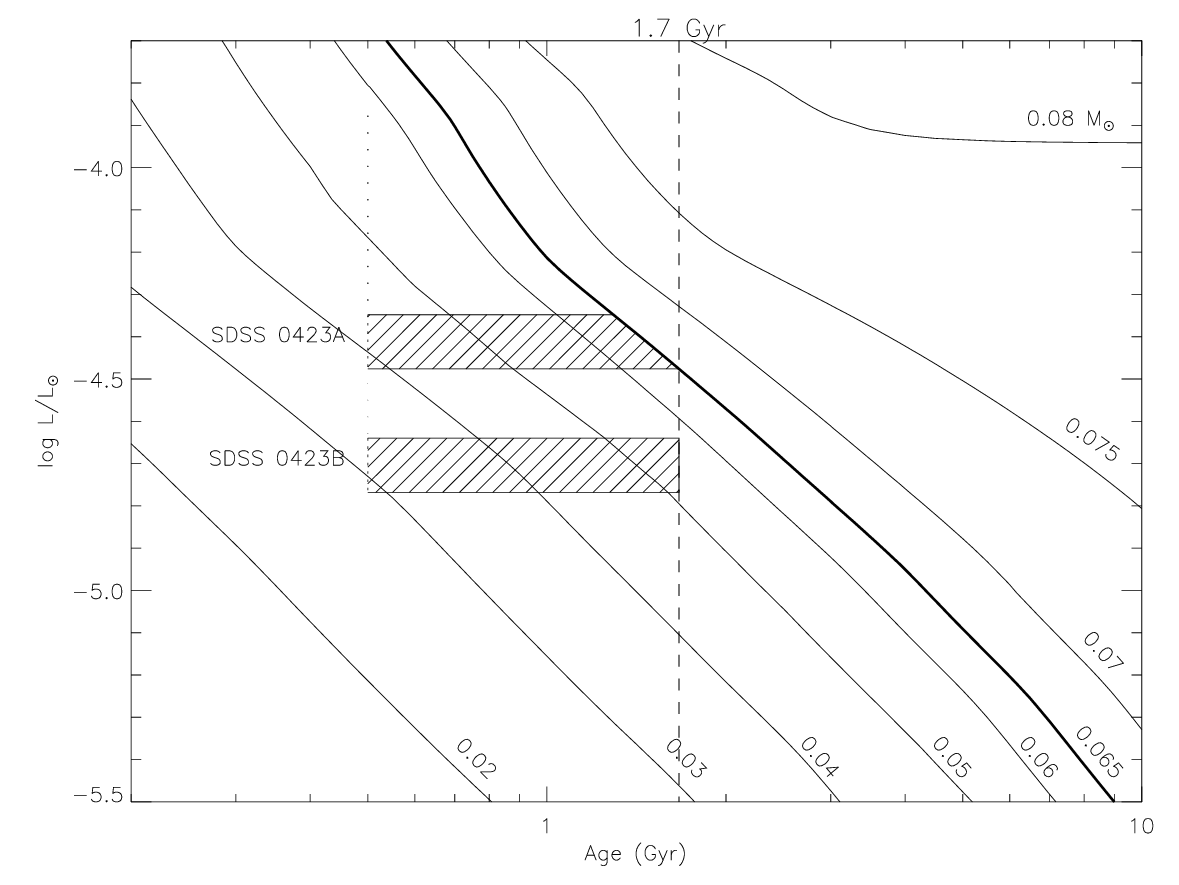}
\caption{Comparison of component bolometric luminosites for SDSS 0423-0414
to theoretical evolutionary models.  Isomass evolutionary tracks are indicated
by lines and labeled (in Solar masses).  The 1\% Li depletion limit is
indicated by the thick line, and effectively lies along the 0.065~M$_{\sun}$
track for the ages shown.  A minimum age for the system of $\sim$0.5 Gyr is based
on the absence of notable low surface gravity features in the optical and/or near
infrared spectrum of this object.  A maximum age of 1.7~Gyr is constrained
by the detection of strong \ion{Li}{1} absorption, most likely
from the primary component. \label{fig_0423evol}}
\end{figure}

\begin{figure}
\centering
\includegraphics[width=0.45\textwidth]{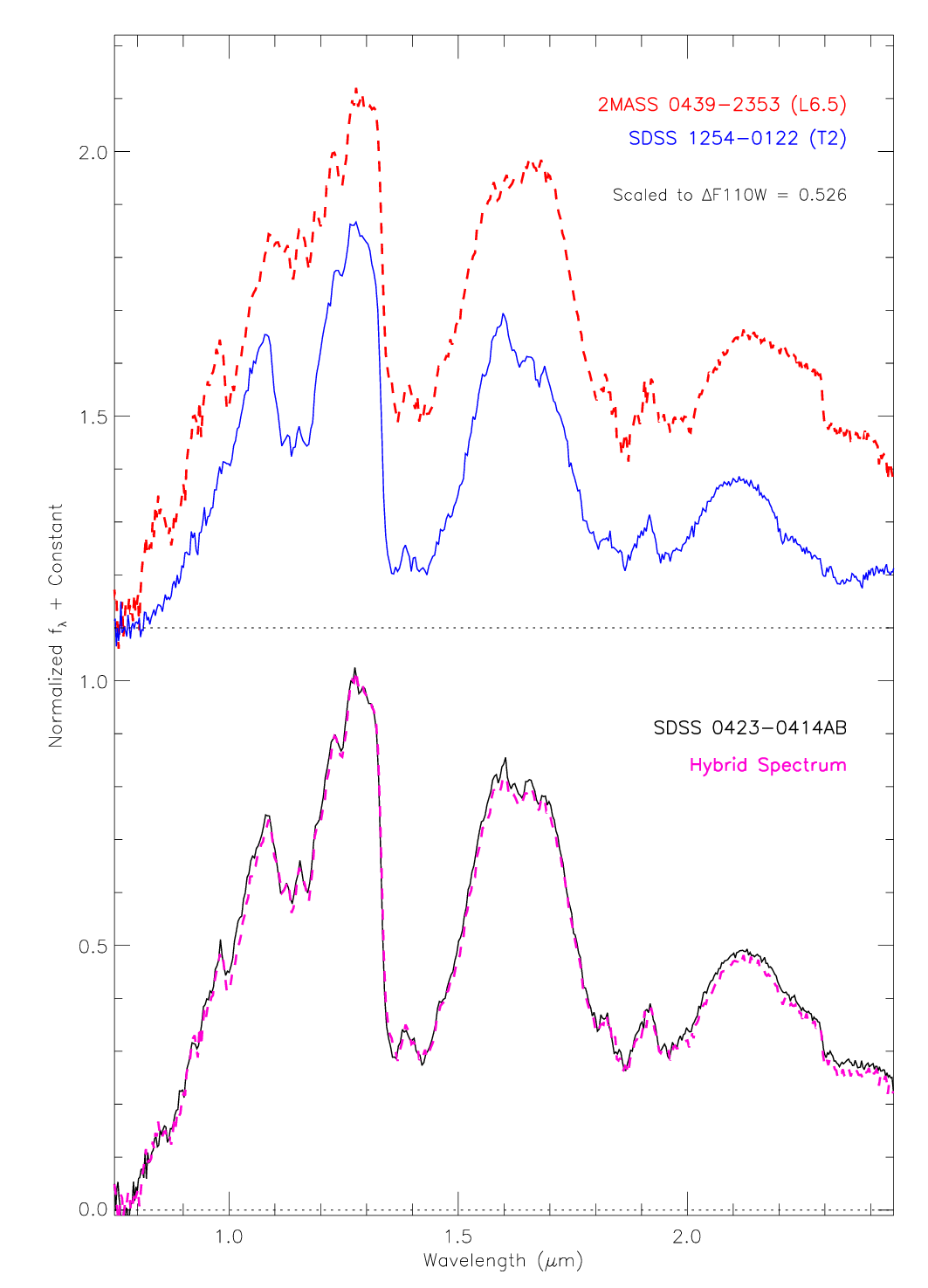}
\includegraphics[width=0.45\textwidth]{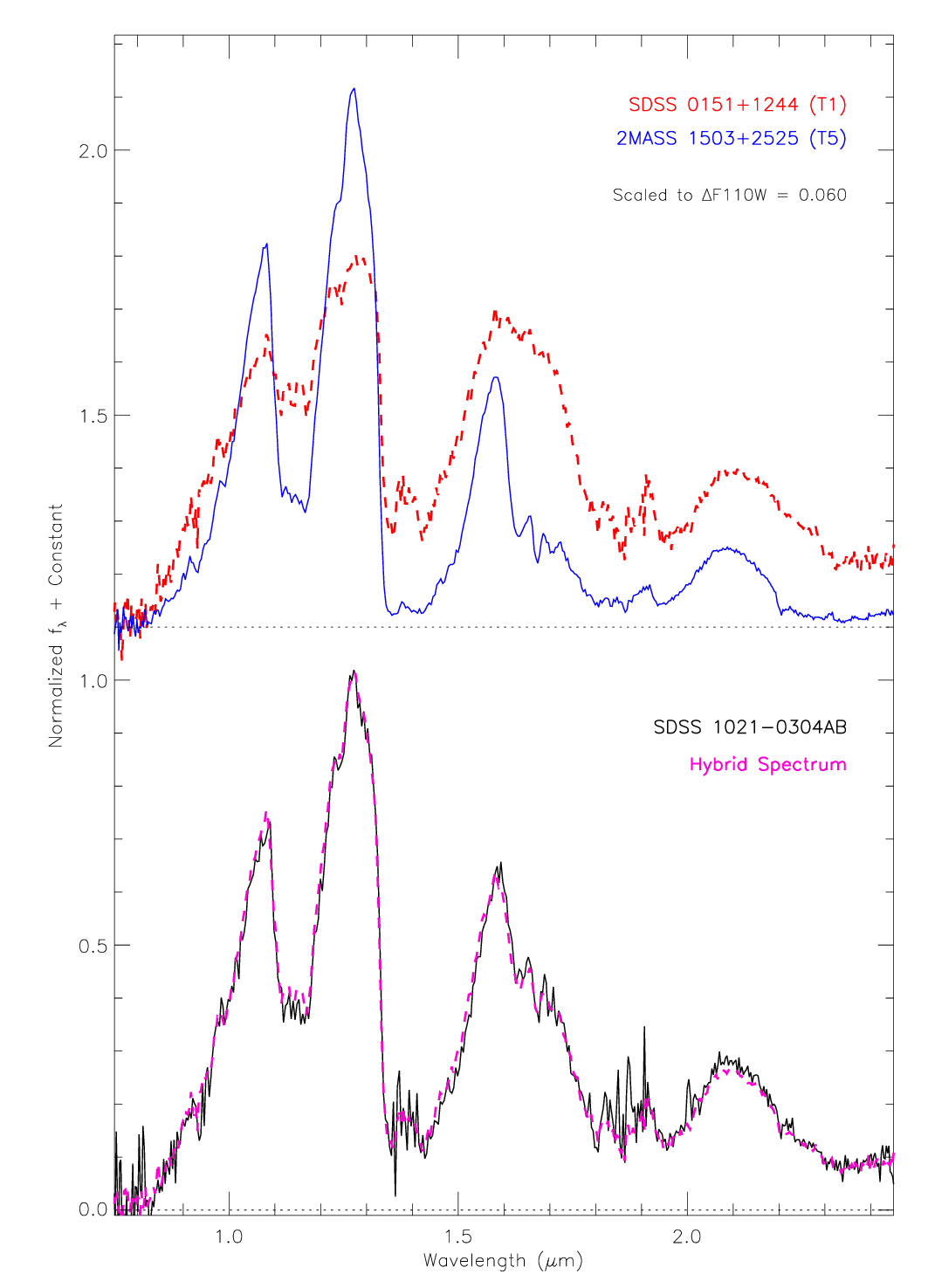}
\caption{Spectral decomposition of SDSS 0423-0414 (left) and SDSS 1021-0304 
(right).  The top of each panel shows the best fit spectral
standards to the primary (red dashed line)
and secondary (blue solid line) of each binary, 
normalized and scaled to the relative F110W magnitudes as measured
with NICMOS.  The bottom of each panel compares the
combined light spectrum of the binary (black solid line) to the hybrid
spectrum of the two spectral components (purple dashed line), both normalized at 1.27~$\micron$.
The agreement between the combined light and hybrid spectra is overall
quite remarkable. 
\label{fig_1021specfit}}
\end{figure}

\begin{figure}
\centering
\plotone{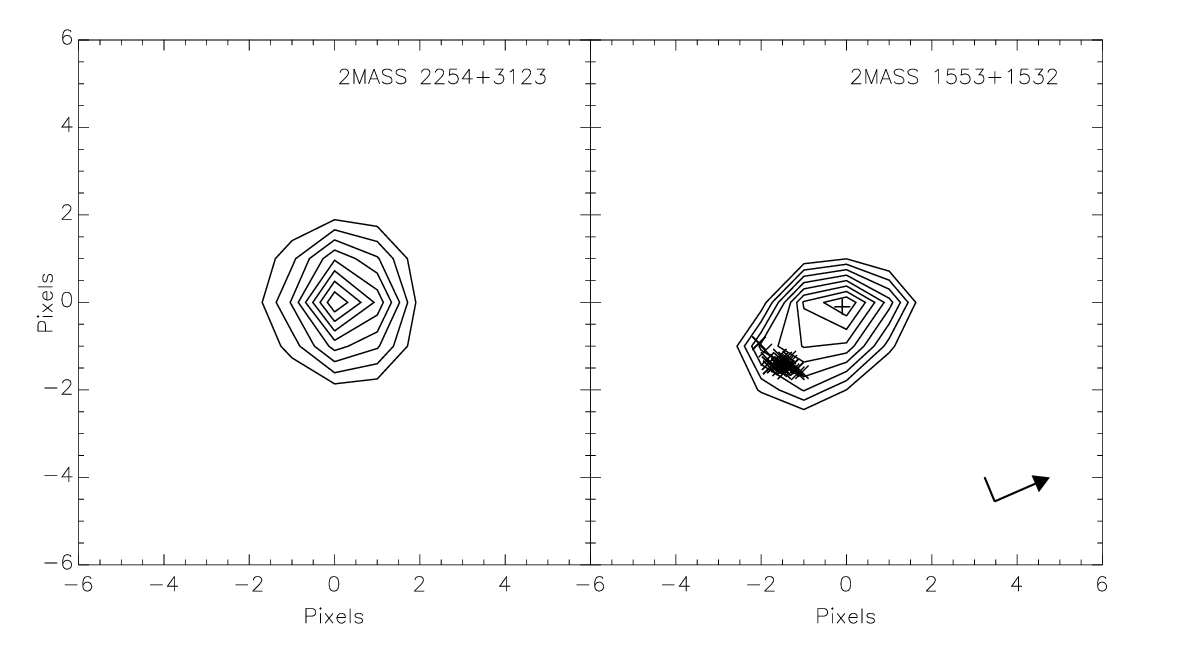}
\caption{Contour plots for one set of NIRC $K_s$-band images of 
2MASS 2254+3123 (left) and 2MASS 1553+1532 (right). 
The displayed boxes correspond to a 1$\farcs$8$\times$1$\farcs$8 area
on the sky, and 
orientation for both images is indicated by the arrow (pointing north) in the right
panel. Contour levels corresponding to 0.2, 0.3, 0.4, 0.5, 0.6, 0.7, 0.8 and
0.9 times the peak source flux are shown.  Relative separations for the
42 best PSF fits to 2MASS 1553+1532 are indicated by the plus symbol 
(corresponding to the primary) and crosses (corresponding to the secondaries),
and are consistent with a mean separation of 0$\farcs$30$\pm$0$\farcs$02
and position angle of 199$\pm$7$\degr$.  \label{fig_1553contour}}
\end{figure}

\begin{figure}
\centering
\plotone{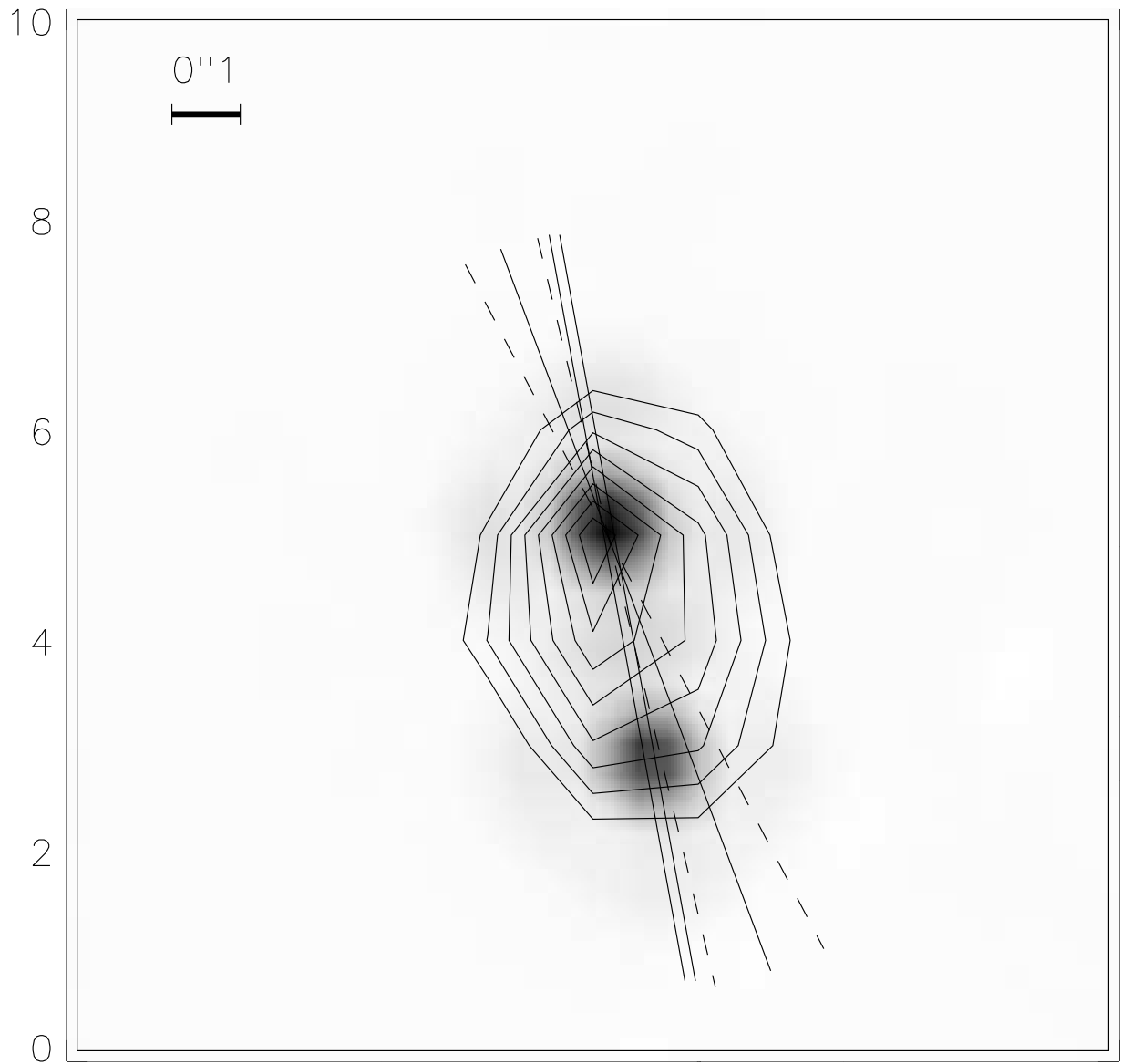}
\caption{Comparison between NIRC $K_s$-band (contours) and {\em HST} F110W (greyscale) 
images of 2MASS 1553+1532.
Both images are oriented with north up and east to the left, and 
angular scale is indicated
in the upper left corner.  The position angles between the primary and secondary
as determined by PSF fitting are overlaid; the single line and dashed lines
correspond to mean and 1$\sigma$ uncertainty as measured 
from the NIRC images, the double line corresponds to
the best fit from the {\em HST} images.  A slight rotation and expansion
of the system between the 
two imaging epochs can be discerned. \label{fig_1553nircvshst}}
\end{figure}

\begin{figure}
\centering
\epsscale{0.7}
\plotone{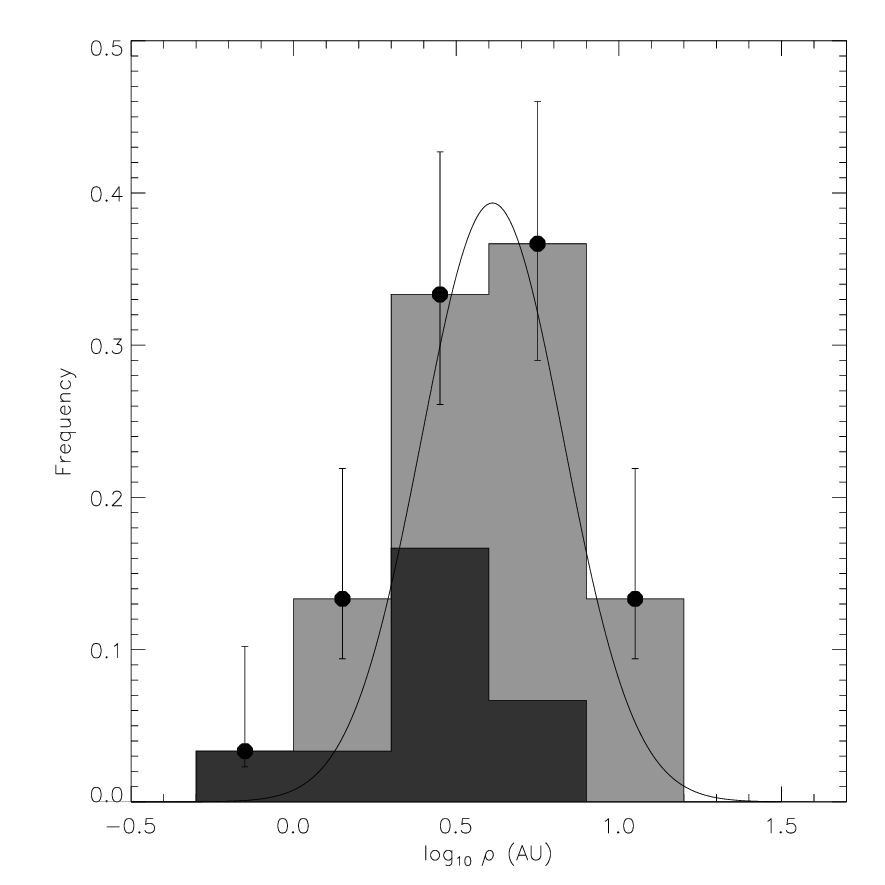}
\caption{Projected separation
distribution (light grey histogram) of 30 brown dwarf binaries identified in 
high resolution imaging studies by
\citet{mrt99,lei01,rei01,rei06,pot02,bou03,me03hst,giz03,mcc04,me05gl337cd,liu05,liu06};
and this study.  Uncertainties based on counting statistics are
indicated by error bars.
The distribution exhibits a peak at
$\rho \approx$ 4~AU ($\log_{10}{\rho} = 0.6$) as derived from a Gaussian fit 
(solid line),
although the decline at smaller separations
may be the result of resolution limits in the imaging studies.  Note that T dwarf binaries (dark
grey histogram) have smaller projected separations on average as compared
to all resolved brown dwarf binaries.
\label{fig_sepdist}}
\end{figure}

\begin{figure}
\centering
\plotone{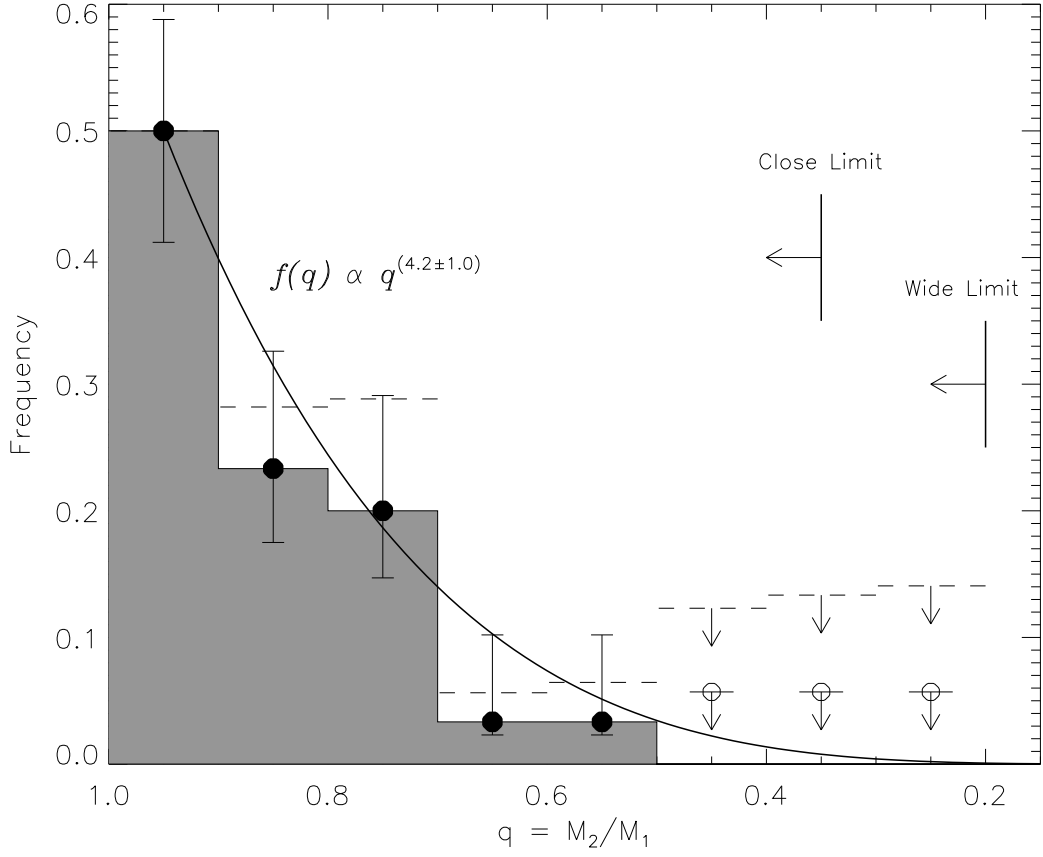}
\caption{Mass ratio distribution of the 30 brown dwarf binaries from
Figure~\ref{fig_sepdist}.  Uncertainties based on counting statistics are
indicated by error bars, upper limits are indicated by open circles.
Bias corrected values are shown by dashed lines.
A power law fit to the bias-corrected values,
$f(q) \propto q^{(4.2{\pm}1.0)}$,
is shown by the thick solid line. 
Sensitivity limits at close ($\rho \gtrsim 0{\farcs}04$)
and wide ($\rho \gtrsim 0{\farcs}2$) separations are indicated.
  \label{fig_qdist}}
\end{figure}

\begin{figure}
\centering
\includegraphics[width=0.45\textwidth]{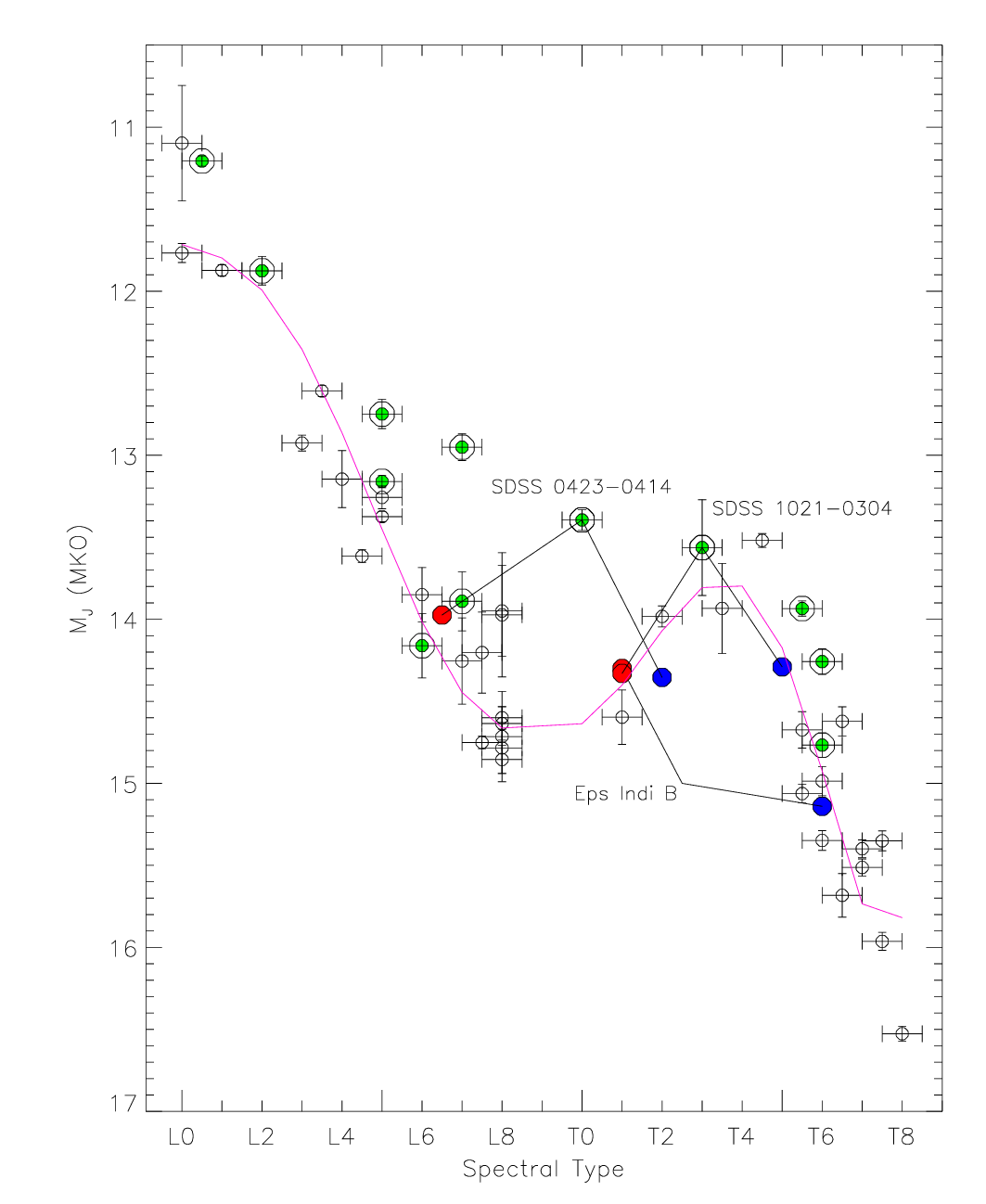}
\includegraphics[width=0.45\textwidth]{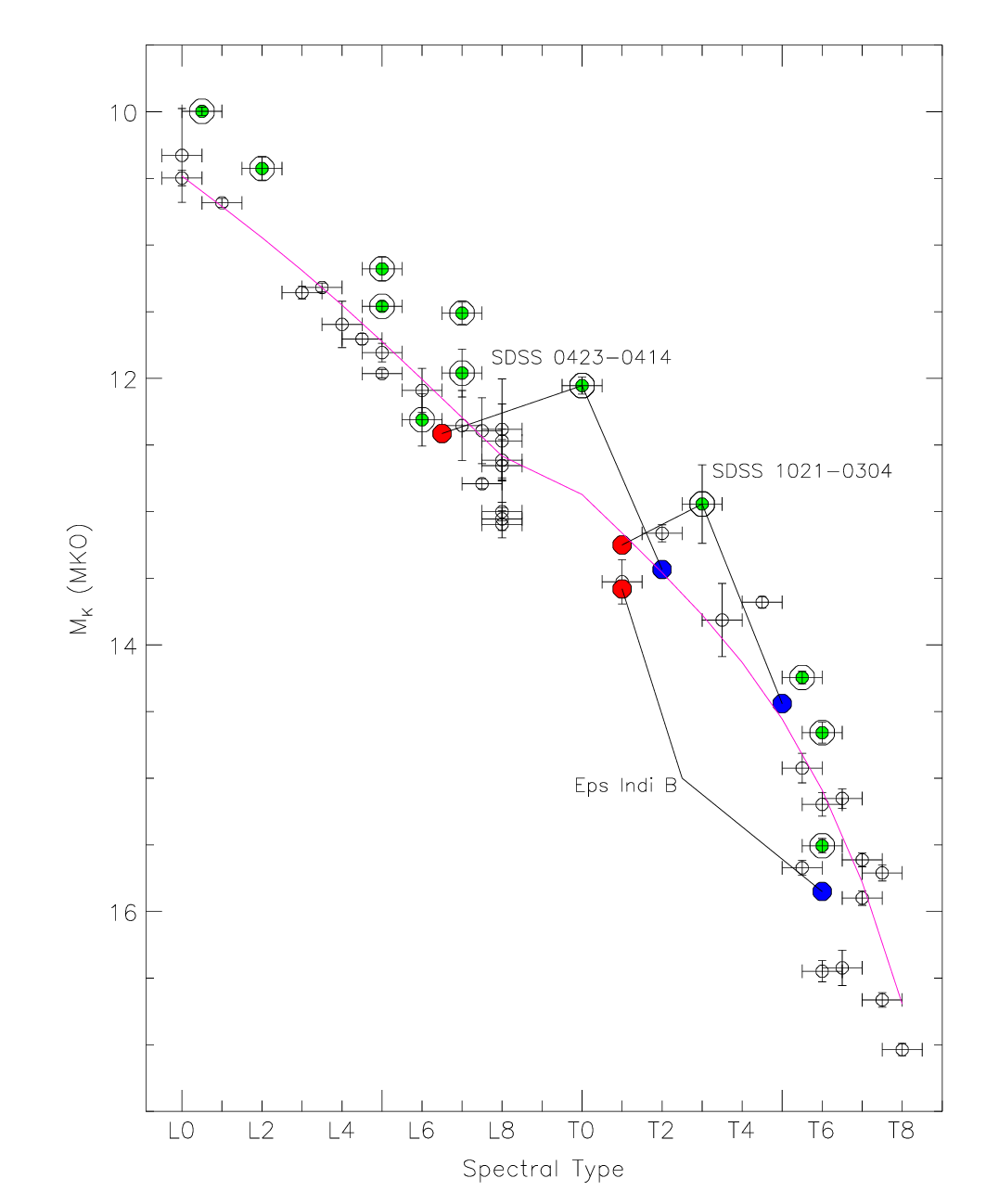}
\caption{Absolute MKO $J$- (left) and $K$-band (right) magnitudes versus spectral type
for field L and T dwarfs with parallax measurements 
\citep{dah02,tin03,vrb04} and companions to nearby Hipparcos
stars \citep{bec88,nak95,me00a,kir01,mcc04}.  Spectral types are based on optical
data for the L dwarfs and near infrared data for the T dwarfs.
Combined light photometry for known binaries are encircled. 
Primary (red circles) and secondary (blue circles) 
spectral types and magnitudes for the SDSS 0423-0414, SDSS 1021-0304 and 
Epsilon Indi B binaries are indicated.
Absolute MKO magnitude/spectral type relations from \citet{tin03}
are shown by the solid curves.  
\label{fig_absmag}}
\end{figure}

\begin{figure}
\centering
\includegraphics[width=0.45\textwidth]{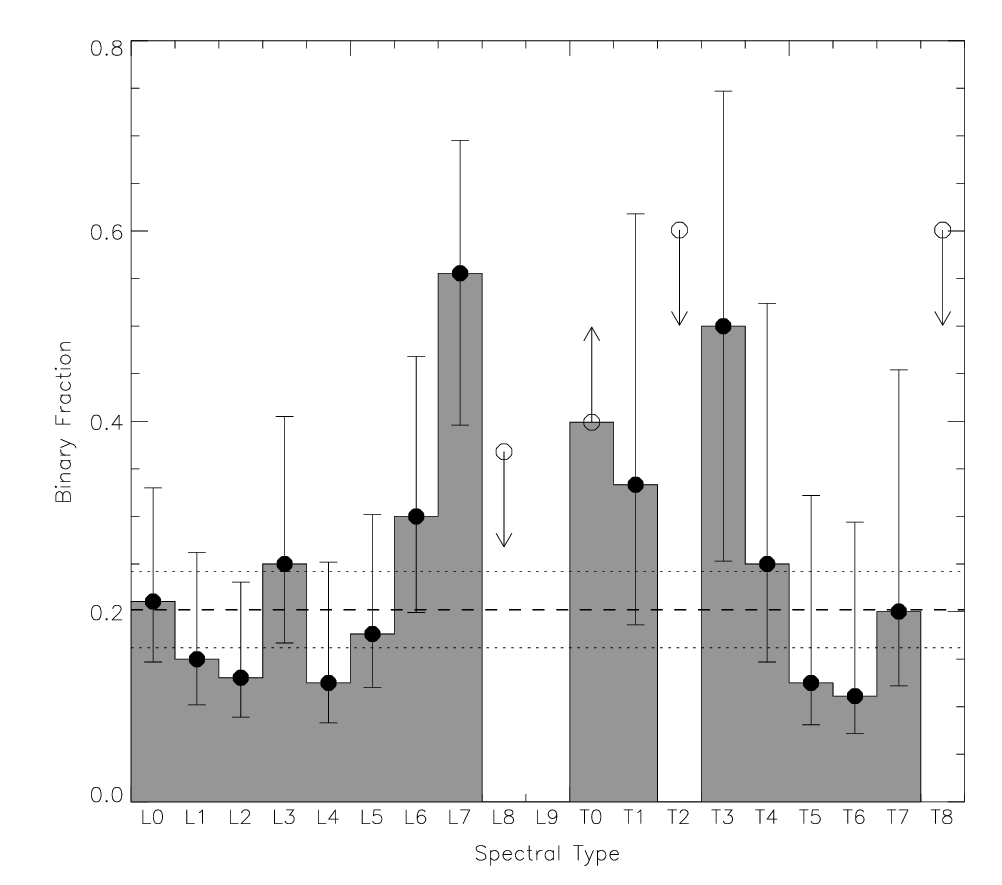}
\includegraphics[width=0.45\textwidth]{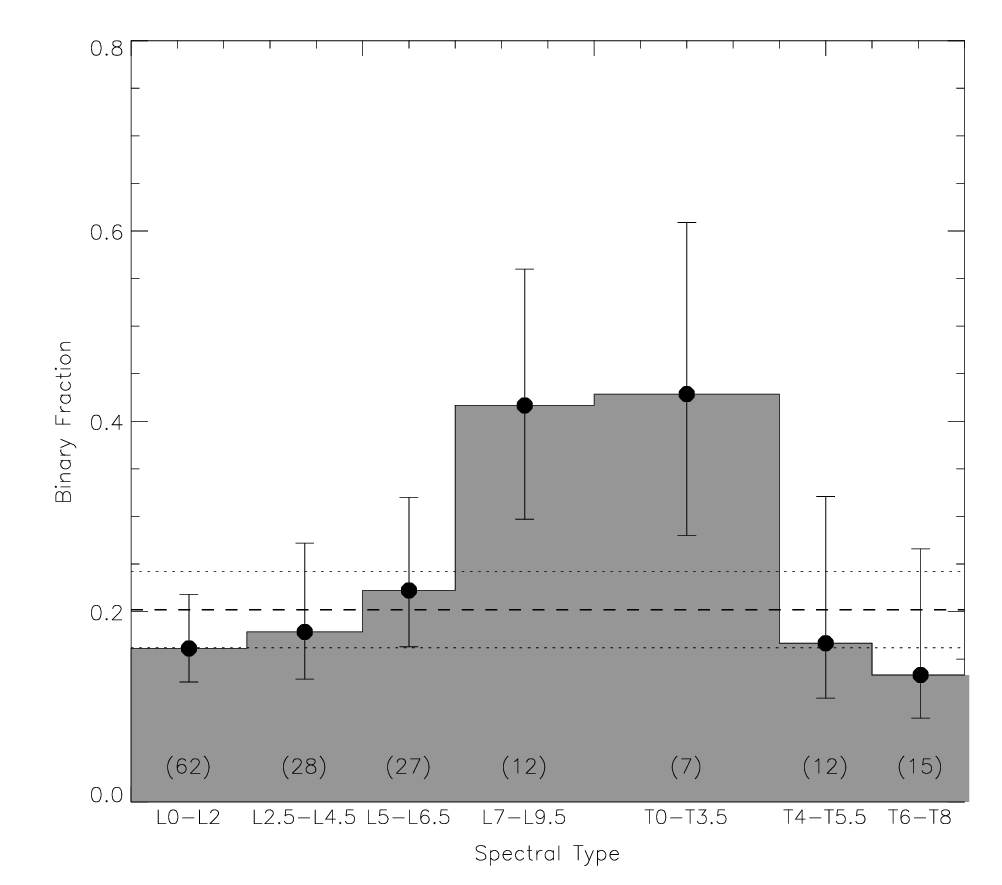}
\caption{Observed binary fractions of L and T dwarfs 
as a function of spectral type.
Data were compiled from the imaging surveys of  \citet{koe99,mrt99,rei01,rei06,clo03,bou03,me03hst,giz03}; and
this study.
Counting uncertainties are indicated in both panels.  The left plot
shows binary fractions broken down by individual subclasses; upper
(zero binaries) and lower limits
(all binaries) are indicated by arrows. The right
plot groups sources into spectral class bins of L0-L2, L2.5-L4.5, L5-L6.5,
L7-L9.5, T0-T3.5, T4-T5.5 and T6-T8, with the number of source in each bin labeled.  
The overall observed binary fraction,
$\epsilon_b^{obs}$ = 20$\pm$4\%, is indicated by the dashed and dotted lines.
\label{fig_ltbinfrac}}
\end{figure}

\end{document}